\documentclass[a4paper,12pt]{article}
\pdfoutput=1 

\usepackage{jheppub} 
\usepackage{amsmath}
\usepackage[T1]{fontenc} 

\title{\boldmath The connection between holographic entanglement and complexity of purification }

\author[a,c]{Mahdis Ghodrati,}
\author[a,c]{Xiao-Mei Kuang,}
\author[a,c]{Bin Wang,}
\author[b]{Cheng-Yong Zhang,}
\author[a]{Yu-Ting Zhou}

\affiliation[a]{Center for Gravitation and Cosmology, College of Physical Science and Technology,\\
Yangzhou University, Yangzhou 225009, China}
\affiliation[b]{Department of Physics and Center for Field Theory and Particle Physics, Fudan University,
Shanghai 200433, China}
\affiliation[c]{School of Aeronautics and Astronautics, Shanghai Jiao Tong University, Shanghai 200240, China}

\emailAdd{mahdisg@yzu.edu.cn}
\emailAdd{xmeikuang@yzu.edu.cn}
\emailAdd{wangb@yzu.edu.cn}
\emailAdd{zhangchengyong@fudan.edu.cn}
\emailAdd{zhouyuting@yzu.edu.cn}

\abstract{In this work we study how entanglement of purification (EoP) and the new quantity of ``complexity of purification" are related to each other using the $E_P=E_W$ conjecture. First, we consider two strips in the same side of a boundary and study the relationships between the entanglement of purification of this mixed state and the parameters of the system such as dimension, temperature, length of the strips and the distance between them. Next, using the same setup, we introduce two definitions for the complexity of mixed states, complexity of purification (CoP) and the interval volume (VI). We study their connections to other parameters similar to the EoP case. Then, we extend our study to more general examples of BTZ black holes solution in massive gravity, charged black holes and multipartite systems. Finally, we give various interpretations of our results using resource theories such as LOCC and also bit thread picture.}

\begin{document} 
\maketitle
\flushbottom

\section{Introduction}

In holography, each surface, volume or any particular region of the bulk could have a specific quantum information meaning. One of the main examples is the duality between the entanglement entropy of a region $A$ in the boundary and the minimum area of the extremal codimension two hypersurface in the bulk that is homologous to the boundary region $A$ \cite{Ryu:2006bv,Ryu:2006ef}. Another example is the duality between the subregion volumes and the corresponding quantum computational complexities \cite{Alishahiha:2015rta, Brown:2015bva, Ghodrati:2017roz,Ghodrati:2018hss}.

The idea of searching for dualities between the geometry in the bulk and the quantum information quantities in the boundary inspired recent works such as \cite{Takayanagi:2017knl} to search for dualities for other entanglement measures (specifically for mixed states). Some examples are entanglement cost, entanglement distillation, entanglement of formation, squashed entanglement, and specifically the entanglement of purification (EoP).  For each of these quantities, one could search for their dual specific surfaces in the bulk through various minimization processes. Then, by studying their distinct properties, one could establish concrete links between geometry and  quantum information quantities and also find algebraic relations between them, such as sub or superadditivities, monogamy, etc. One then could even demonstrate those relations using only the geometrical intuitions from holography. 
 
In \cite{Takayanagi:2017knl}, Takayanagi and Umemoto proposed that the minimal cross section of the entanglement wedge, separating two regions of $A$ and $B$, called $E_W(\rho_{AB})$ is dual to the entanglement of purification between the two regions and this has been called the $E_P=E_W$ conjecture. Using this idea and also by using our intuitions from complexity of subsystems, we introduce two new quantities for the mixed states, which will be called ``complexity of purification (CoP)" and ``volume interval (VI)". Then, we will study their relations to the minimal entanglement wedge cross section and their behaviors under changing the parameters of the system.
  
Note that for defining such quantities for mixed states, one could use various proposals of holographic complexity such as ``complexity=action (CA)" \cite{Brown:2015lvg} conjecture, ``complexity=volume (CV)" \cite{Alishahiha:2015rta} conjecture or path integral optimization process \cite{Caputa:2017yrh}. Recently, there have been few studies where using CA or CV conjectures, the holographic  complexity of purification has been introduced, see for instance \cite{Du:2018uua,Agon:2018zso,Chen:2018mcc, Ling:2018xpc,Yang:2018gfq}. Note that the definition of CoP in field theory corresponds to the minimum number of gates needed to prepare a purified state out of a mixed state. The holographic dual for CoP has also been discussed in several recent works such as \cite{Agon:2018zso} and \cite{Caceres:2018blh}. 

To indirectly study various aspects of CV or CA and their relations to purification, various holographic tools such as differential geometry and kinematic space \cite{Czech:2015qta}, tensor network \cite{Swingle:2009bg}, or path integral optimization \cite{Caputa:2017yrh}, could also be implemented to propose and further study this new definition. For instance, in \cite{Abt:2018ywl}, using kinematic space, the direct relation between subregion complexity and entanglement entropy of intervals has been worked out. In their work,  the volume of a general region in the spatial slice has been written as an integral over the corresponding region in the kinematic space and in the dual CFT it has been written purely in terms of \textit{entanglement entropies}.  So similar studies, using the kinematic space, could then be implemented to find the connection between entanglement and complexity of purification.

To study this connection, we use however, a direct  \textit{geometrical approach} here. First in section \ref{EoPsec}, we review the definition of entanglement of purification for mixed states and its relation to mutual information. In section \ref{EoPtworegions}, we study the entanglement of purification for two strips in the background of Schwarzschild AdS black brane, which have the same size and are on the same side of the boundary, similar to the work of \cite{Yang:2018gfq}. In that work, the properties of holographic entanglement of purification (EoP) and also its relation to holographic mutual information (HMI) for two infinite, disjoint strips with the width $l$ and separation $D$ have been studied. 

In this work, for the first step in the calculations, we determine the critical $D_c$, for each length $l$ and for any dimension $d$, where the mutual information and as the result the EoP would become zero. We therefore, specify all the regions of the parameters where EoP could be non-zero in our model and re-derive the results of \cite{Yang:2018gfq}. We also study the behavior of EoP versus temperature in various dimensions. From our results we explain the physical behavior of EoP in different limits and circumstances.    
 
Then, based on what we have learnt about EoP, its connection to minimal wedge cross section and its behavior in different dimensions, we move to propose new definitions for the \textit{purification of complexity} based on $E_P=E_W$ conjecture and then we study their behaviors.

 In \cite{Agon:2018zso}, the authors suggested that the complexity of purification is the summation of two quantities of \textit{spectrum complexity} and \textit{base complexity}. Based on the definitions of these two complexities and their expectations from tensor network, they have suggested that ``holographic action" would match their definition for CoP and this would not be the case for the holographic volume.
 
On the other hand, in \cite{Caceres:2018blh}, based on some examples such as black holes with a large genus behind the horizon \cite{Fu:2018kcp}, it has been proposed that subregion complexity based on ``complexity=volume" conjecture could match with the expectational behavior of complexity of purification better.   
 
 In this work, we search for different \textit{volumes} which could be dual to complexity of purification (CoP) which match with our expectations from the intuition we got from EoP. We consider the interplay between the bulk geometry and the minimal wedge cross section. Specifically, we use the CV conjecture to search for new dualities and to define the suitable definition for the complexity of purification. We then study various properties of our proposed quantities.
 
Note that in \cite{Hubeny:2018ijt}, different arrangement of holographic entanglement entropy has been proposed to be dual to different entanglement measures. Similarly, arrangements of holographic complexities could be proposed and their structures could be analyzed which is the main idea of our work here. 
 
In section \ref{MixedStates} of this work, we consider the criteria that a new definition for the complexity of purification should satisfy. We use the idea of writing linear combinations of subregion complexities, similar to the work of \cite{Hubeny:2018ijt} which has been done for entanglement entropies and then we move to propose our definitions.
 
In section \ref{cop31} we extend the studies of \cite{Yang:2018gfq} for the EoP and thermofield double states to ``holographic complexity of purification" (CoP) and study the effect of distance between different subregions and their widths on the evolution of their corresponding subregion complexities and their purification complexities.  We also study the effect of temperature on each volume and its effect on CoP. 

Then, in section \ref{newBV}, we proposed another holographic measure for the complexity of correlations between two mixed state, where we have dubbed it ``volume of interval (VI)" which is the whole volume corresponding to each strip. This new definition shows various interesting behaviors which match with our intuition of quantum correlations. For instance, after a critical $l$, it linearly increases. For small values of $l$, it also first increases with $l$, however for some specific small $l$ it shows a decreasing behavior which could corresponds to the phenomenon of \textit{quantum locking} \cite{Takayanagi:2017knl,Horodecki:2005hnl}. This effect would occur when a correlation measure decreases by a large amount by tracing out only a few qubits. As the entanglement entropy $E_P$ has this property \cite{Christandl:2017knl}, one would expect that the purification complexity also shows this property as well, where actually could only be observed here through our new holographic measure, VI.  
 
 In order to learn more about the physical characteristics of EoP and CoP and how they would be related to other physical parameters of the system and to get further intuitions, in section \ref{purificationgeneral} we calculate EoP and CoP numerically in some more general cases. In section \ref{puremassive}, we calculate them for the case of BTZ black hole in massive gravity theory and study the effect of graviton mass parameter, which is dual to momentum dissipation effect in the boundary. In section \ref{ChargedBTZpure}, we study charged BTZ black holes and study how EoP and CoP would behave by changing charges, and then in section \ref{pureMultipartite} we consider the purification for multipartite systems. Note that all of these studies could be repeated for the case of thermal quenches, similar to \cite{Yang:2018gfq} which is the idea of our on-going work \cite{MG:2019KM}.

In section \ref{pureOperational}, we present some operational interpretations and then some more intuitions from the ``bit thread" picture for the behaviors of EoP and CoP to give explanations of what we have observed in our various examples.

Finally, we conclude with a discussion in section \ref{purediscussion}.

\section{Entanglement of mixed states}\label{EoPsec}

When a quantum system is pure, the only way to characterize the quantum entanglement of a bipartite system  would be the von Neumann entropy of the reduced density matrix,
\begin{gather}
S_A:= -tr \rho_A \log \rho_A.
\end{gather}

However, when the system is in a mixed state, there would be several different quantities which could describe the classical or quantum correlations between the two systems $A$ and $B$.

One of these quantities is the mutual information (MI) which is defined as follows
\begin{gather}
I(A:B) = S(\rho_A)+S(\rho_B)-S(\rho_{AB}),
\end{gather}
where $AB=A\cup B$. Another quantity which could be a measure of the correlation between mixed states, is the entanglement of purification $E_P(A:B)$, which is defined by the minimum entanglement entropy for all possible purifications of the mixed state.

 This quantity is defined as follows
 \begin{gather}
E_P(A:B)= \text{min}_{\rho_{AB}= Tr_{A' B' \ket{ \psi} \bra{\psi} } } S(\rho_{AA'}),
\end{gather}
where $\ket{\psi}$ is a pure state on the enlarged Hilbert space $\mathcal{H_A}\otimes \mathcal{H_B} \otimes \mathcal{H_A'} \otimes \mathcal{H_B'}$. 

Note that $\mathcal{H_A}\otimes \mathcal{H_B}$ is the initial Hilbert space, where the mixed state $\rho_{AB}$ lives. One could enlarge these states by adding $\mathcal{H_{A'}}$ (or $\mathcal{H_B'}$).

The relationship between entanglement of purification $E_P$ and the mutual information, $I(A:B)$ is
\begin{gather}\label{eq:inequality}
\frac{1}{2} I(A:B)\le E_P(A:B) \le \text{min} \{S(\rho_A),S(\rho_B)\}.
\end{gather}

When the mutual information (MI) is zero, the \textit{``classical"} entanglement of purification (EoP) is zero as well. Also, if $AB$ is a pure state, this inequality would be saturated in both sides. It worths mentioning that the entanglement of purification satisfies the strong superadditivity \cite{Takayanagi:2017knl}, while entanglement entropy satisfies strong subadditivity, and another important property of mutual information is monogamy \cite{Hayden:2011ag}.

As first introduced in \cite{Takayanagi:2017knl}, the holographic dual of this quantity is the minimal cross section of the entanglement wedge $E_W(\rho_{AB})$ which corresponds to the correlation between the two disconnected subsystems $A$ and $B$ on the boundary. When the system is pure, $E_W$ becomes equal to the entanglement entropy, which interestingly is also the same scenario in the field theory side. 

Therefore,  there would be three quantities which characterize the correlations between mixed states where all of them have holographic duals, namely, entanglement entropy, mutual information and entanglement of purification. As it has been shown in \cite{Hayden:2011ag}, the mutual information among arbitrary disjoint spatial regions $A$, $B$, $C$ obeys the inequality and the monogamy relation $I(A:B \cup C) \ge I(A:B)+ I(A:C)$. This means that the correlations in holographic theories arise primarily from ``entanglement" rather than the classical correlations. This fact would be very important when one considers the relationships between complexity and its growth rates with the correlations between the subregions.

Now after this introduction, we move to study EoP for two strips and get more physics on how the entangled pairs behave and get further intuitions to be prepared for defining CoP.

\subsection{Entanglement of purification (EoP) for two subregions}\label{EoPtworegions}
Similar to \cite{Yang:2018gfq}, in the following setup,  two subregions $A$ and $B$ which are infinite strips separated by $D$ and are on the same side of the boundary could be considered as
\begin{gather}
A:= \{l+D/2 >x_1>D/2, -\infty <x_i<\infty, i=2,3,...,d-1\} \nonumber\\
B:= \{ -l-D/2 < x_1 < -D/2, -\infty < x_i < \infty, i=2,3,...,d-1\}.
\end{gather}

Then, using the inequality \ref{eq:inequality}, one could find the critical distance between them $(D_c)$, which the EoP drops to zero.

Note that here $S_A=S_B=S(l)$ and $S_{AB}= S(2l+D)+S(D)$. So the mutual information of $AB$ would be
\begin{gather}\label{eq:lD}
I(D,l)=S_A + S_B - S_{AB}=2S(l)-S(D)-S(2l+D).
\end{gather}

\begin{figure}[ht!]
 \centering
  \includegraphics[width=6.5cm] {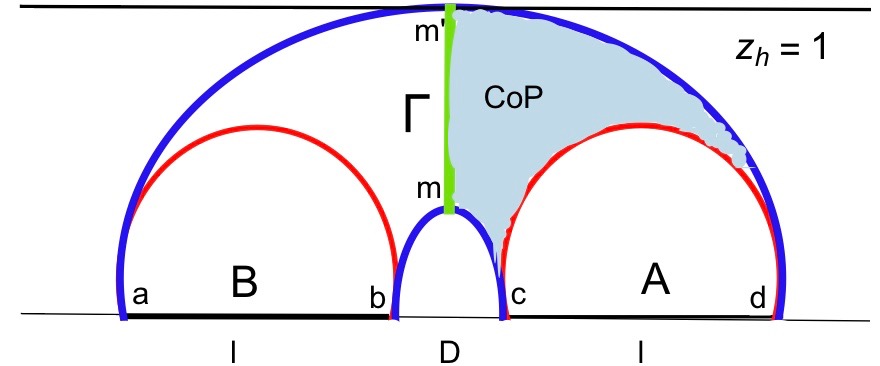}
  \caption{Two strips of $A$ and $B$ with length $l$ and with the distance $D$ between them. The two turning points corresponding to region $ad$ and $bc$ are $m$ and $m'$ and $\Gamma$ is the cross section of ``connected" entanglement wedge.}
 \label{fig:strips}
\end{figure}

The setup of the strips is shown in figure \ref{fig:strips}.
\vspace*{6px}

The critical $D_c$ for each dimension could be found by setting $I(D,l)=0$.

Now, considering the Schwarzchild AdS black brane in $(d+1)$-dimensions as, 
\begin{gather}\label{metric}
ds^2=\frac{1}{z^2} \left[ -f(z)dt^2+\frac{dz^2}{f(z)}+d\vec{x}^2_{d-1} \right],  \ \ \ \ \ f(z):=1-z^d/z_h^d,
\end{gather}
for the case of $d=2$, corresponding to a planar BTZ black hole, the authors of \cite{Takayanagi:2017knl} have found the entanglement wedge as
\begin{gather}
E_W=\frac{c}{3} \text{min} \left[ A^{(1)}, A^{(2)} \right],
\end{gather}
where
\begin{gather}\label{eq:area}
A^{(1)}= \log \frac{\beta}{\pi \epsilon},\ \ \ \ \ \ \ 
A^{(2)} = \log \frac{\beta \sinh\left(\frac{\pi l}{\beta} \right)  }{ \pi \epsilon}.
\end{gather}

From equations \ref{eq:lD} and \ref{eq:area}, one then could write
\begin{gather}
\sinh \left(\frac{l}{2}\right)^2= \sinh \left( \frac{D_c}{2} \right) \sinh \left( \frac{2l+D_c}{2} \right),
\end{gather}
and then from that, the critical $D_c (2,l)$ could be found as \cite{Chen:2018mcc}
\begin{gather}
\cosh \frac{D_c (2,l) }{2} =\sqrt{1+2 \sqrt{2 \cosh l} \cosh \frac{l}{2} +2 \cosh l  } \left [ \cosh \frac{3l}{2} - \sqrt{2} (\cosh l)^{3/2} \right].
\end{gather}

For the geometry \ref{metric}, assuming only $x_1(z)$ is a function of $z$ coordinate, the induced metric could be found, and then taking the minimum of area could give us the functional dependence of $x'(z)$,
\begin{gather}
\sqrt{-g}=\sqrt{{x'_1}^2+\frac{1}{f(z)}} \left( \frac{1}{z} \right)^{d-1}, \ \ \ \ \ 
{x'_1}=\frac{1}{ \sqrt{ (1-\frac{z^d}{z_h^d}) \left ( \frac{z_0^{2d-2}}{z^{2d-2}} -1\right) } }.
\end{gather}

So the width of the strip and the holographic entanglement entropy could be written as
\begin{gather}
w=2 \int_\delta^{z_0} dz \frac{1}{\sqrt{(1-\frac{z^d}{z_d^d}) \left( \frac{z_0^{2d-2} }{z^{2d-2} }-1  \right)}}, \nonumber\\
S(w)=\frac{2 V_{d-2} }{4G_N} \int_\delta^{z_0} \frac{dz}{z^{d-1}} \frac{1}{\sqrt{(1-\frac{z^d}{z_h^d})\left(1-\frac{z^{2d-2}}{z_0^{2d-2}} \right)  }}.
\end{gather}

Note that $V_{d-2}=\int dx^{d-2}$, and also $z_0$ is the turning point of the minimal surface.

In the following parts, we fix temperature by assuming $z_h=1$. Later we also study the effect of temperature on EoP and CoP by varying $z_h$.

\begin{figure}[ht!]
 \centering
  \includegraphics[width=7cm] {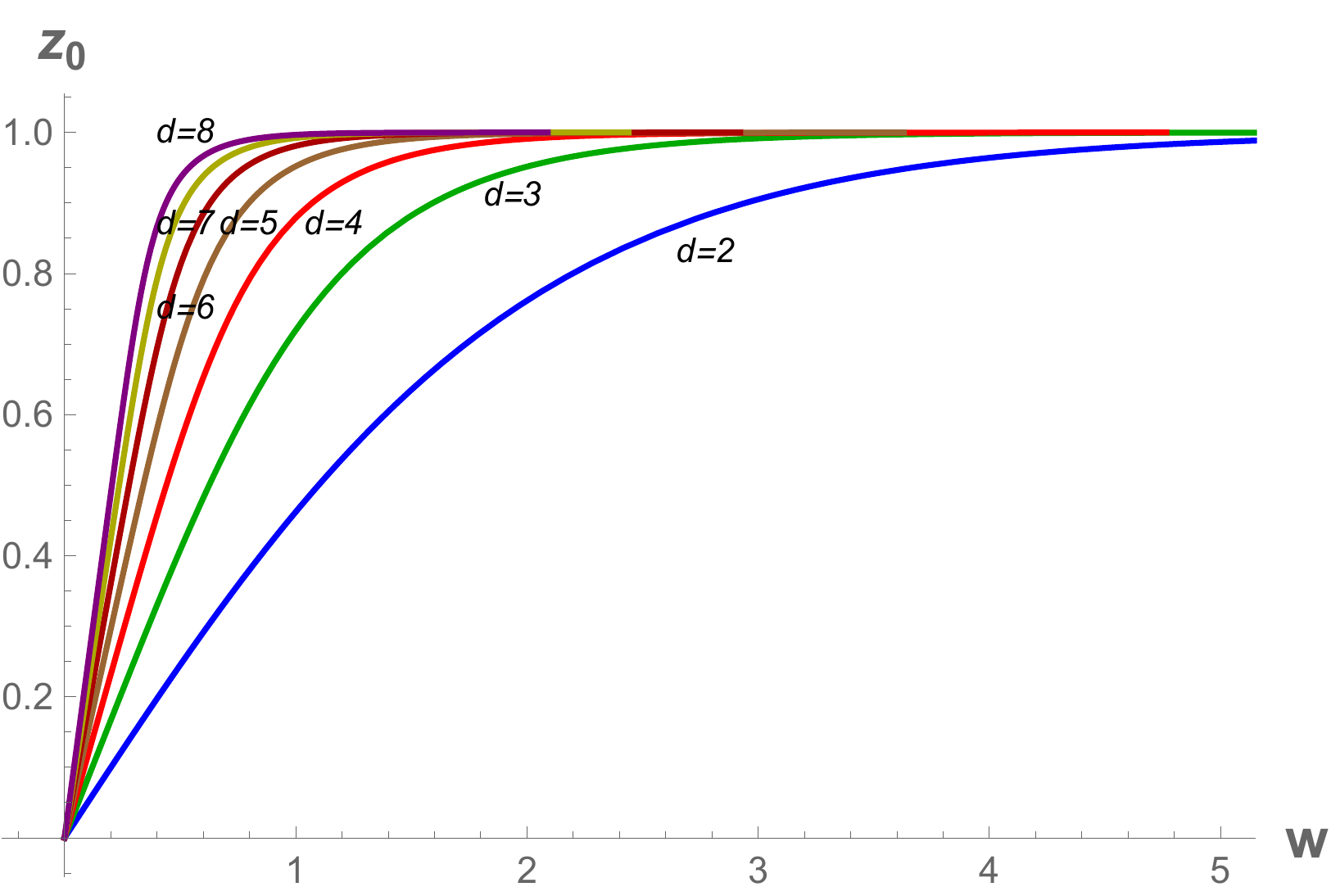}
  \caption{The relationship between turning point and width of ``one" strip. }
 \label{fig:z0w}
\end{figure}

The plot of turning point $z_0$ in the bulk, versus the width of any strip $w$ is shown in figure \ref{fig:z0w}.

It could be seen that in higher dimensions, the turning point reaches to its maximum value at lower $w$ while the maximum value of $z_0$ is one which is the horizon temperature. Note also that for any specific width $w$, for higher dimensions, the turning point is deeper inside the bulk, as $z_0$ is bigger.

\begin{figure}[ht!]
 \centering
  \includegraphics[width=5cm] {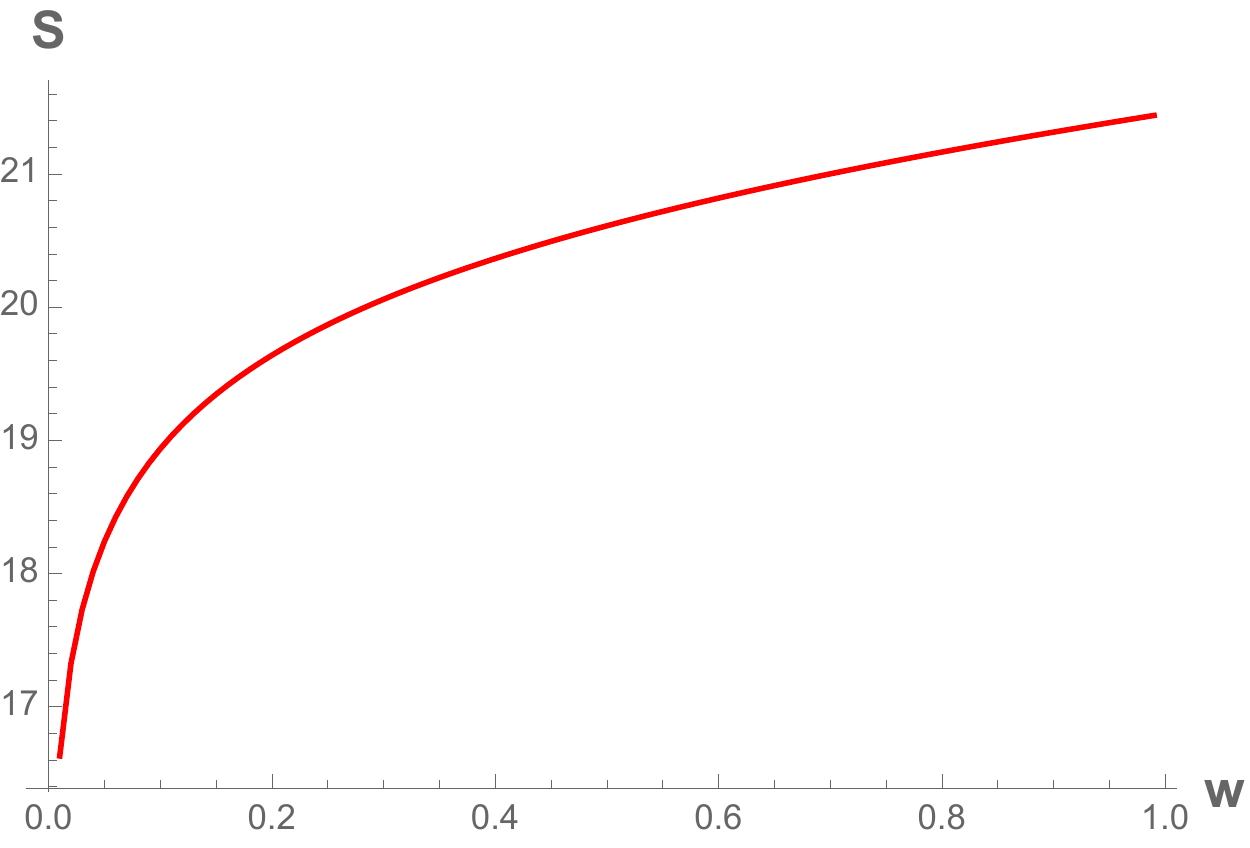} \ \ \ \ \ \ \ 
    \includegraphics[width=5.6cm] {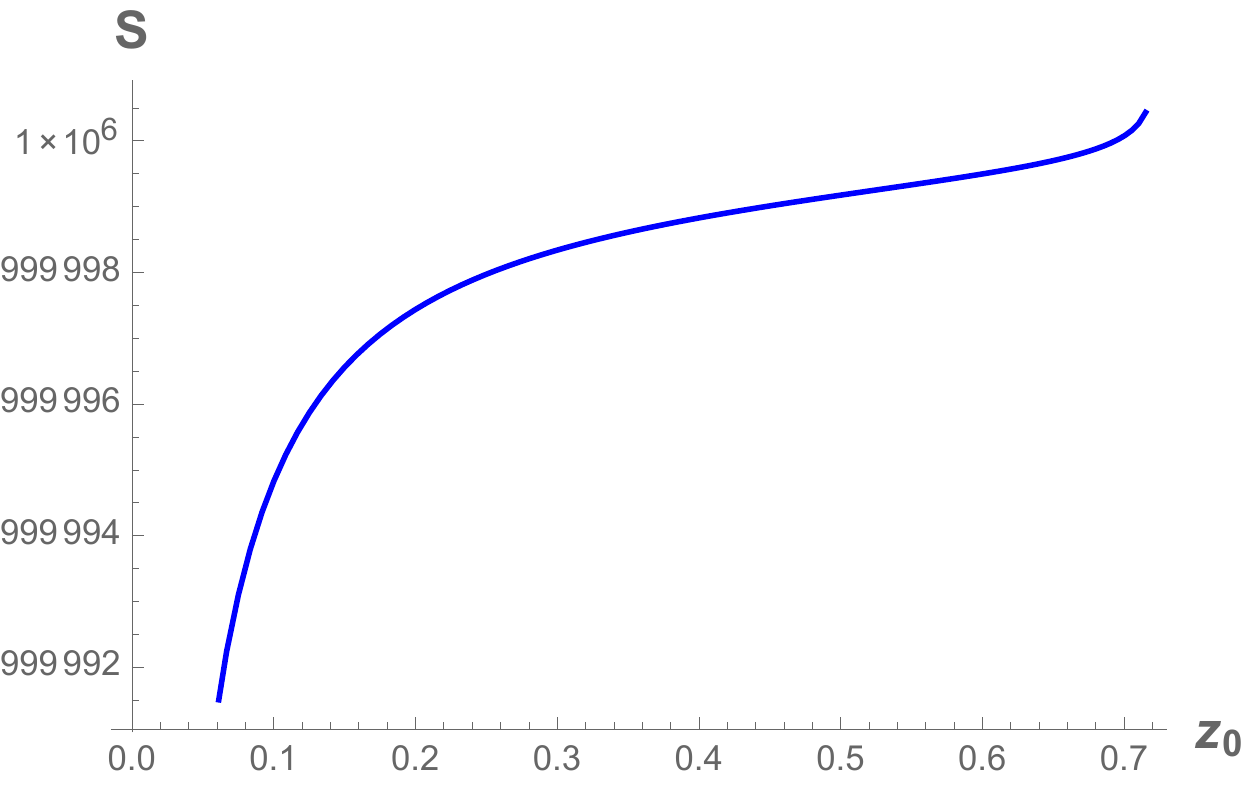}
  \caption{The relationship between entanglement entropy and the width of one strip $w$ and turning point $z_0$ for $d=3$.}
 \label{fig:SW22}
\end{figure}

The relationship between $S$ and $w$ is also shown in figure \ref{fig:SW22}.

So as one would expect, increasing the width of the strip which is proportional to the number of quantum gates in the system would increase the entanglement entropy. The increasing rate is higher for lower number of gates and it slows down for bigger $w$ which corresponds to higher number of gates. Also, note that increasing dimension, $d$, causes that the rate of growth of entanglement entropy becomes much bigger for smaller number of gates and as we will see this would also be the case for the subregion complexity and also CoP.

\begin{figure}[ht!]
 \centering
  \includegraphics[width=5.8cm] {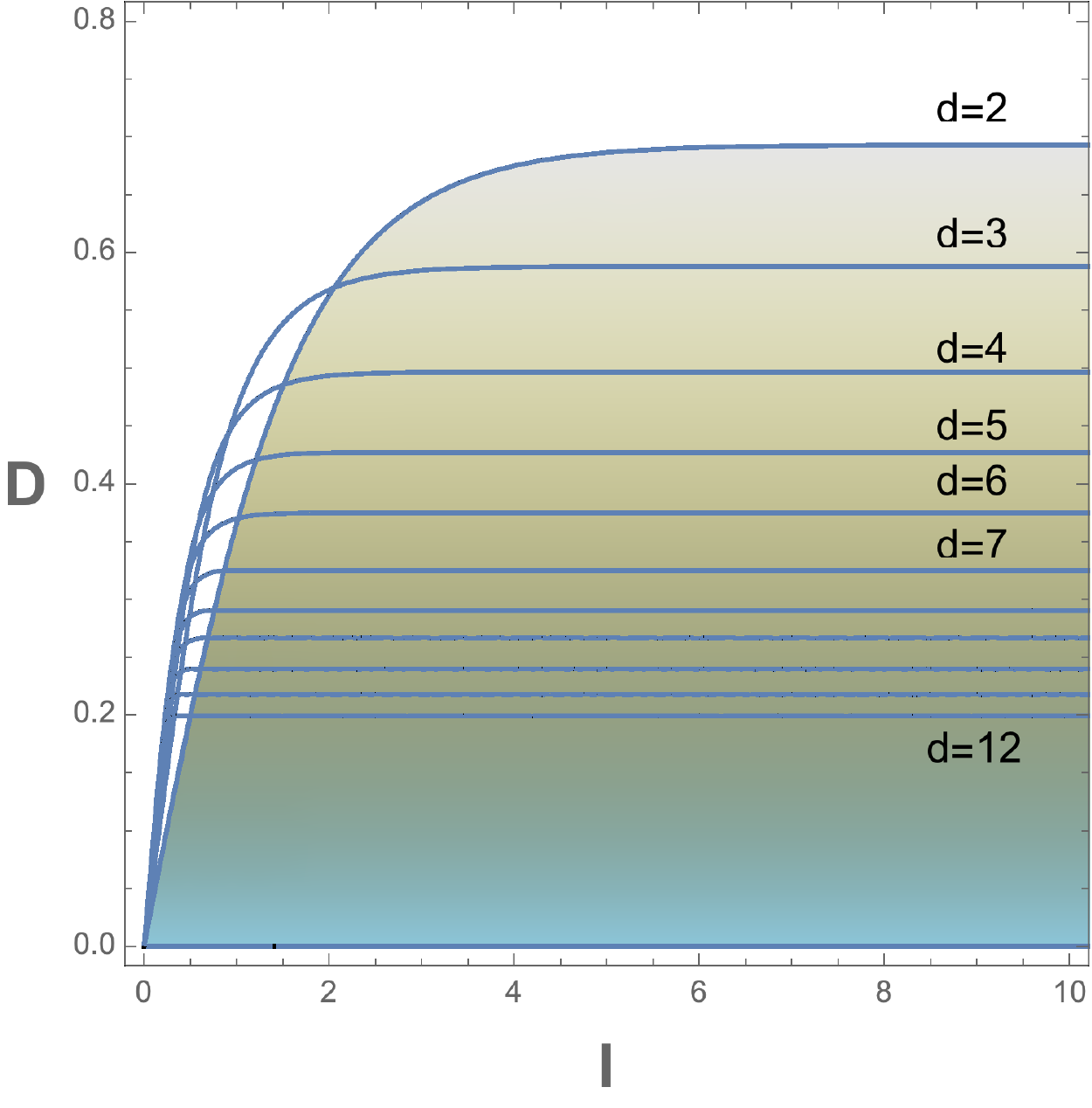}
  \caption{For each dimension, the region below the lines have non-vanishing EoP.}
 \label{fig:region11}
\end{figure}

\begin{figure}[ht!]
 \centering
  \includegraphics[width=6cm] {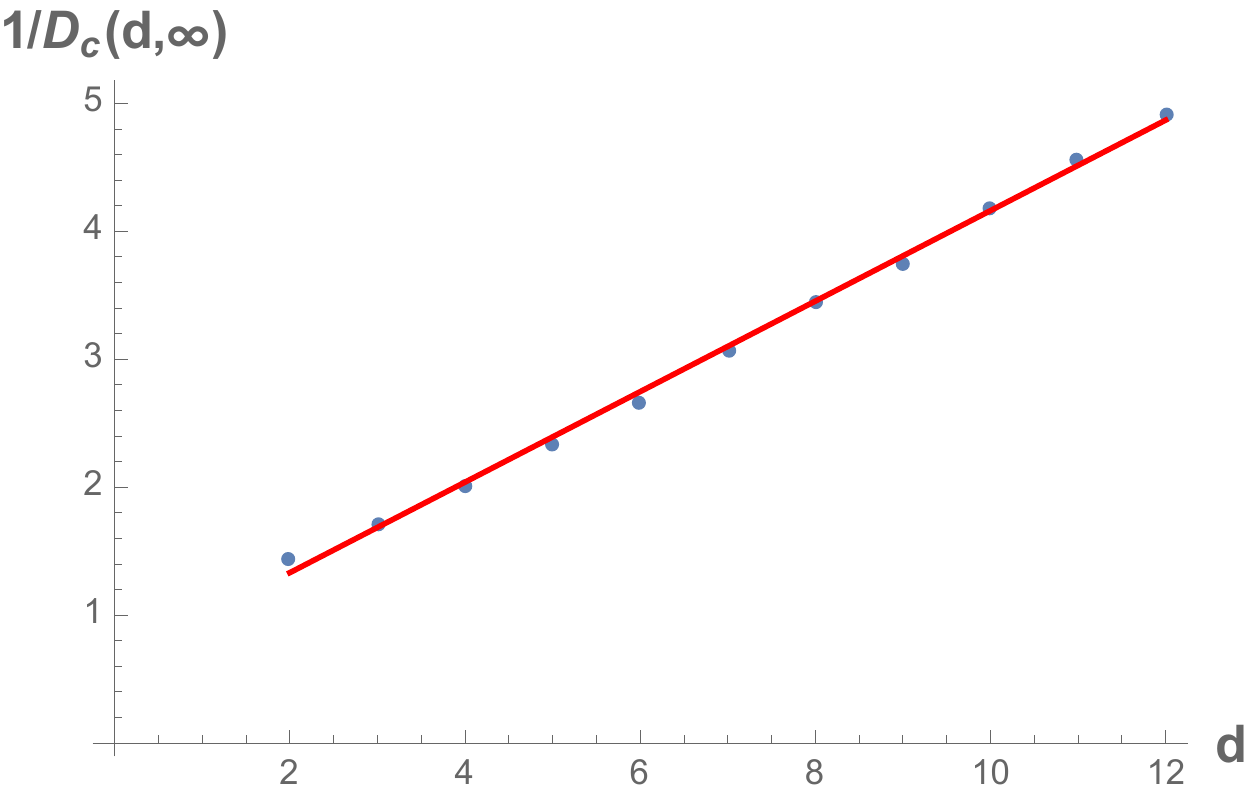}\ \ \ \ \ \ \ 
    \includegraphics[width=6cm] {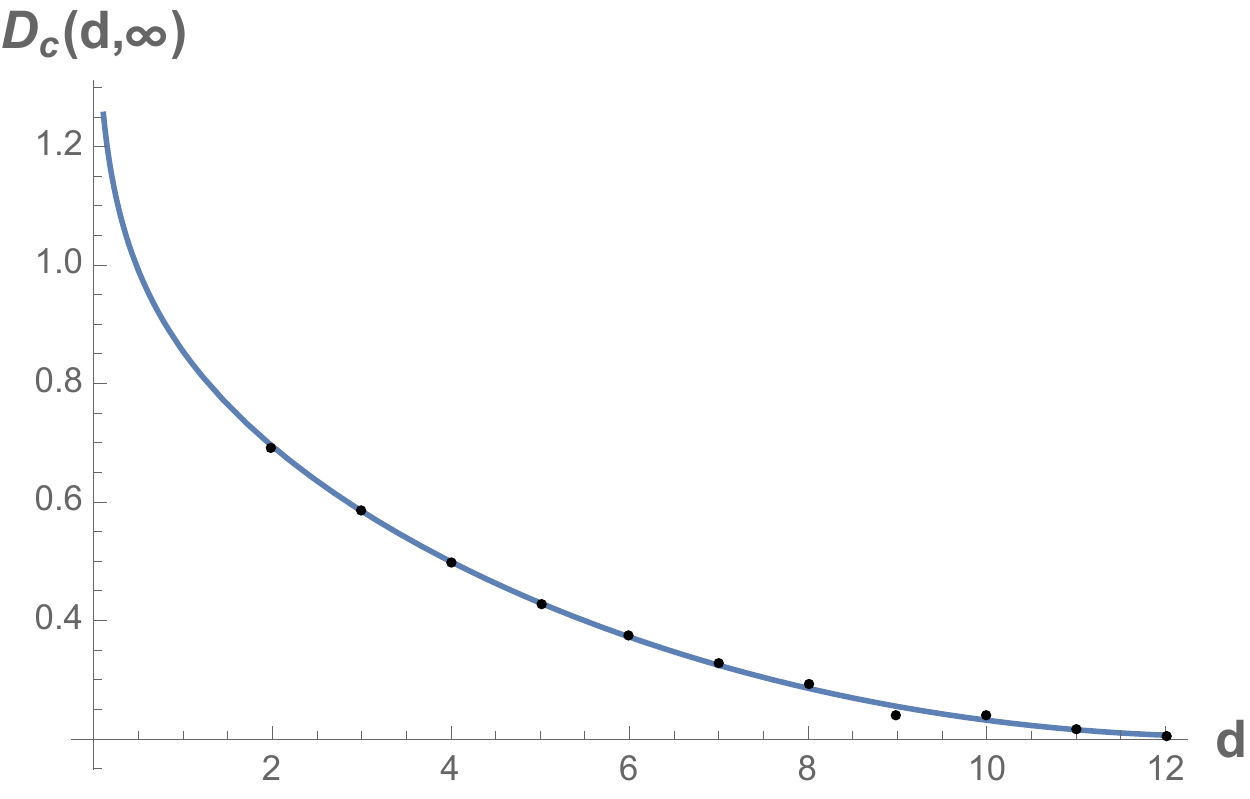}
  \caption{The relationship between $D_c(d,\infty)$ and dimension $d$.}
 \label{fig:regionfunction}
\end{figure}

Now, using the relation \ref{eq:lD}, one could find the critical distance between the two strips, $D_c$, for any length of strips $l$ that the mutual information and therefore EoP becomes zero. Therefore, in figure \ref{fig:region11}, we could specify the non-zero regions. The relationship between $D_c$  and $d$ is also shown in figure \ref{fig:regionfunction}. 

For the linear function between $1/D_c(d, \infty)$ and $d$ one finds
\begin{gather}
D_c(d,\infty)^{-1} \simeq 0.62+0.35d,
\end{gather}

and for a curved function between $D_c(d, \infty)$ and $d$ one could also write,
\begin{gather}
D_c(d,\infty) \simeq 0.888- 0.284 \log (d).
\end{gather}

 So by increasing dimension, the critical distance between the two regions would decrease by  approximately a logarithmic function. This could be explained by considering how increasing the dimension of the system, could let the classical and quantum correlations to spread in more space dimensions \cite{Cevolani:2016uua,PhysRevA.86.032110}, and as the result this critical distance would decrease, so to keep the necessary correlation in the $x$-axis strong enough to get a non-zero EoP. This would decrease logarithmically with the dimension, which later could be examined further with intuitions from the spreading of information in higher dimensional systems and other factors.

Now for the case that $D$ and $l$ would be smaller than $D_c(d,l)$, and they would be in the suitable regions for each dimension, and therefore the EoP would be non-zero, one could find the entanglement of purification by finding the area of the minimal cross section, $\Gamma$, as follows
\begin{gather}
\Gamma=\int_{z_D}^{z_{2l+D}}  \frac{dz}{z^{d-1} \sqrt{1-\frac{z^d}{z_h^d}}},
\end{gather}
which would result in
\begin{equation}
\frac{4}{V_{d-2}} E(l,D) =\begin{cases}
     \ln \frac{\tanh (\frac{D+2l}{2} )}{ \tanh (\frac{D}{2} )}+\ln \frac{1+\sqrt{1- \frac{\tanh^2 \left( \frac{D}{2}\right )} {z_h^2} }}{1+\sqrt{1- \frac{\tanh^2  \left( \frac{D+2l}{2} \right) }{z_h^2}} }, & d=2,\\ \\
    \frac{-4z^{-(d-2)}\sqrt{1-\frac{z^d}{z_h^d}}+(d-4) \frac{z^2}{z_h^d} {}_2F_1(\frac{1}{2}, \frac{2}{d}, \frac{d+2}{d}, \frac{z^d}{z_h^d} ) }{4(d-2)} \Big |^{z_{2l+D}}_{z_D}, & d>2.
  \end{cases}
\end{equation}

Note that for the case of $d=2$ one can get an analytic function. Also, by taking $z_h=1$, the equation (2.7) of \cite{Yang:2018gfq} could be re-derived. 

\begin{figure}[ht!]
 \centering
  \includegraphics[width=7cm] {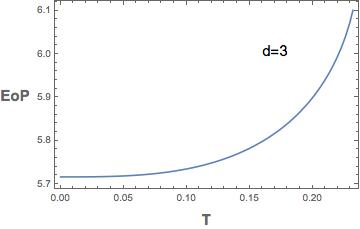}
    \includegraphics[width=7cm] {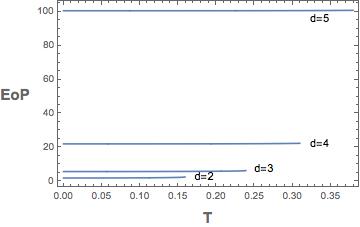}
  \caption{In the left figure, EoP versus $T$ is shown for $d=3$. In the right figure, EoP curves for different dimensions are shown. For both cases we take $l=20$ and $D=0.3$.}
 \label{fig:Eopp1122}
\end{figure}

From this equation, one could gain several results. First, regarding the functional dependence to $z_h$ or temperature, one could see from figure \ref{fig:Eopp1122} that increasing temperature at each dimension would increase EoP.  Increasing dimension would also cause that EoP jumps suddenly. In fact, dimension would have a bigger effect on increasing EoP relative to temperature. Comparing EoP in different dimensions, it actually seems that EoP does not change much with temperature and it just looks like it is relatively constant.  However, at a specific temperature, at each dimension, EoP suddenly diverges. Note that at any $d$, we chose a specific $l$ and $D$ to make sure the mutual information and therefore EoP is non-zero.

\begin{figure}[ht!]
 \centering
  \includegraphics[width=6cm] {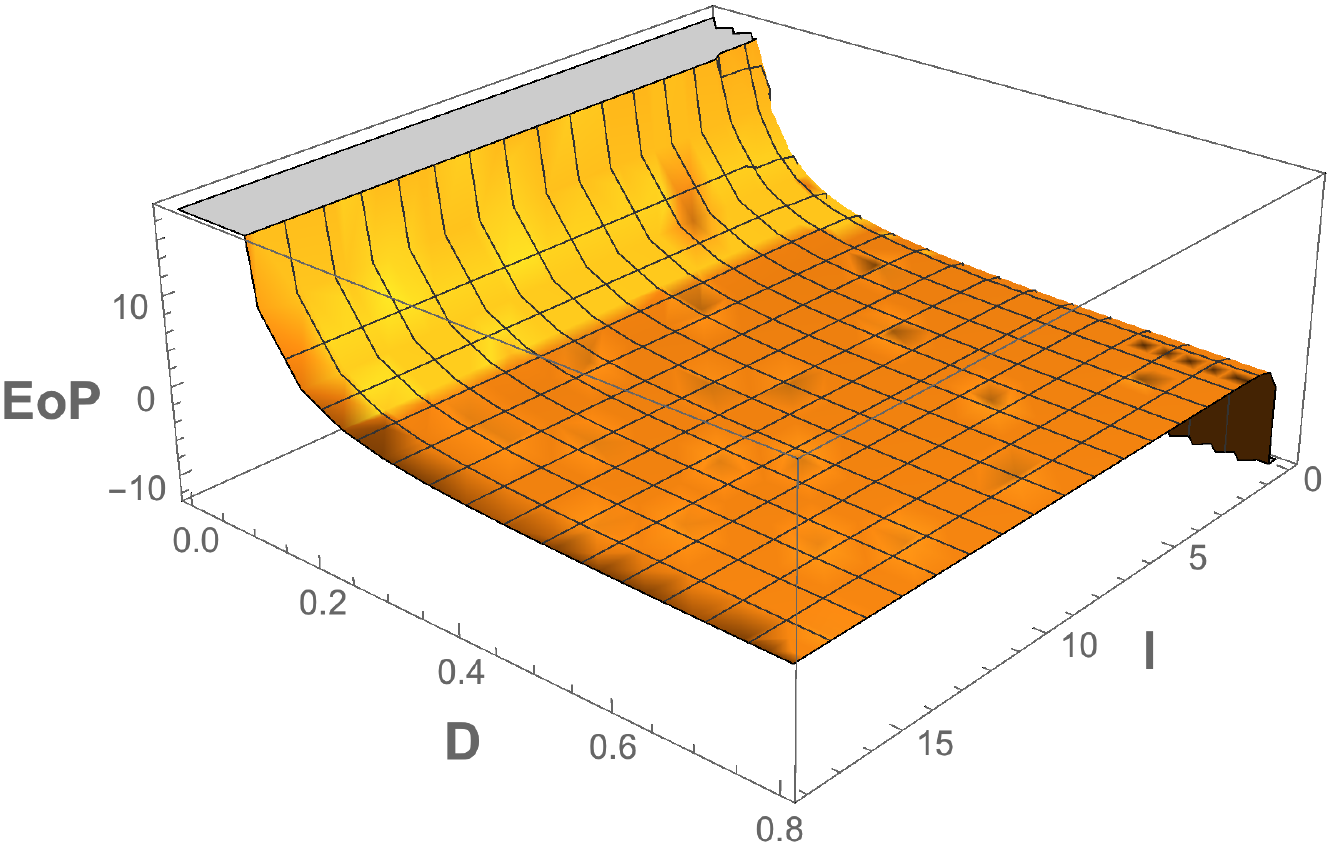}
  \caption{ The plot of EoP in three dimensions for different $l$ and $D$ for $d=4$.}
 \label{fig:EoP3D}
\end{figure}

Now we consider the dependence of EoP with respect to different length scales, $l$ and $D$. Without considering the parts where mutual information and therefore EoP drop to zero, and for the case of $z_h=1$, the three dimensional plot of EoP versus $D$ and $l$ is shown in figure \ref{fig:EoP3D}.

\begin{figure}[ht!]
 \centering
  \includegraphics[width=5cm] {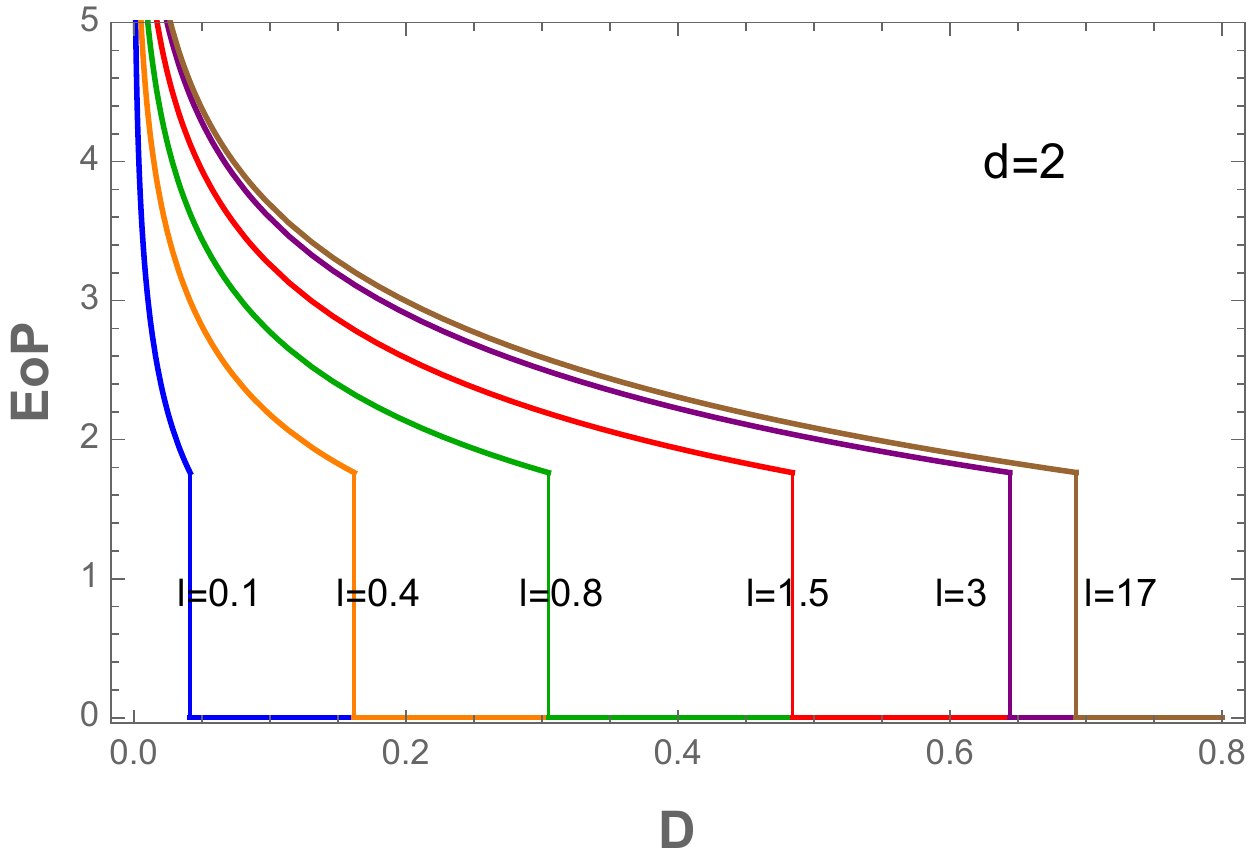}
    \includegraphics[width=4.9cm] {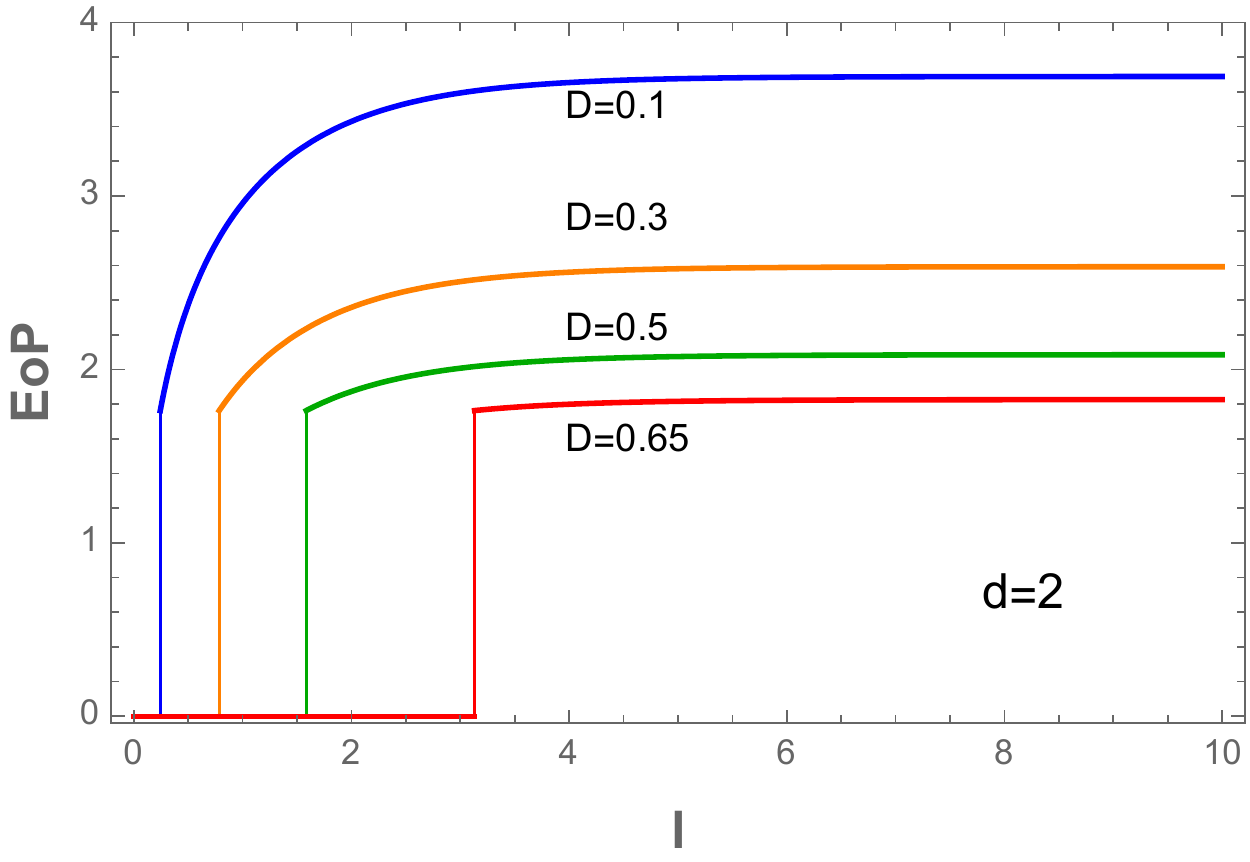} \label{fig:EoPplots2}
    \includegraphics[width=4.9cm] {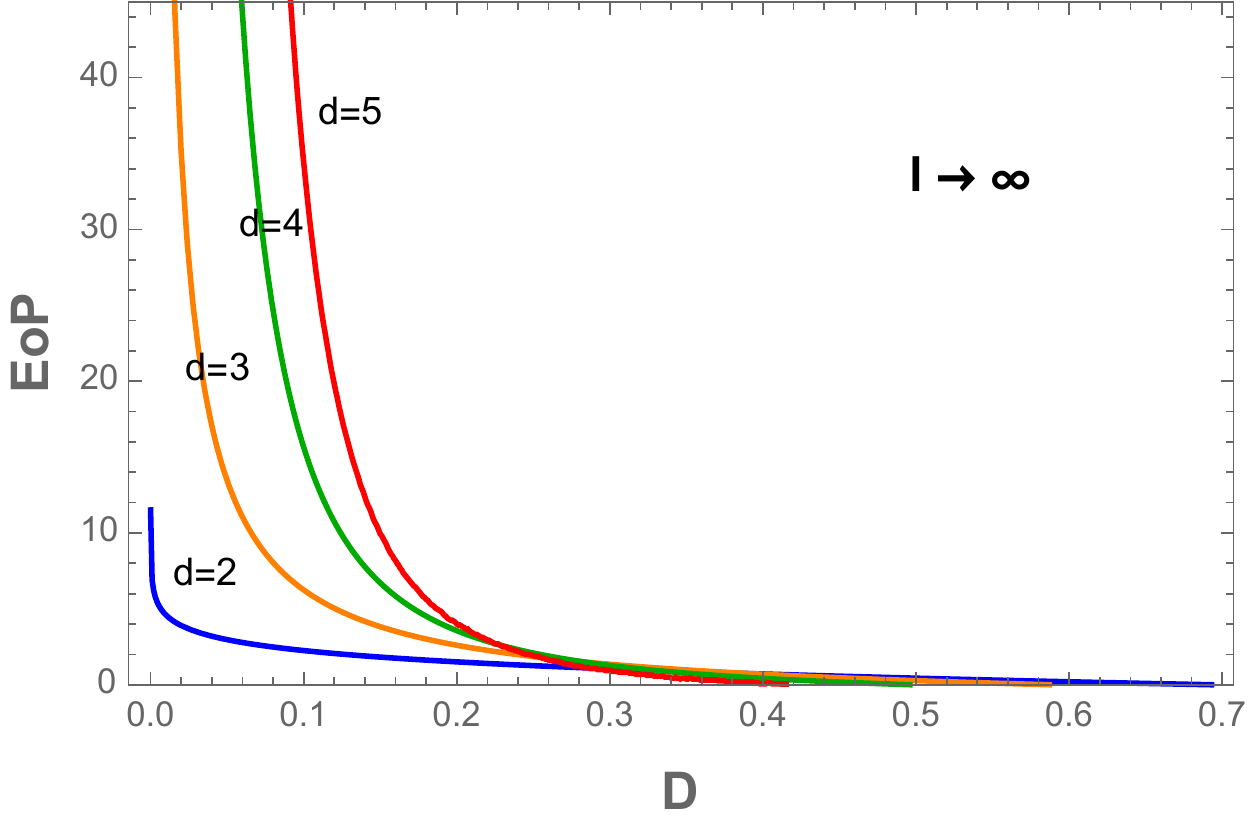}
  \caption{EoP in the unit of $4/V_{d-2}$ for different $l$ and $D$ when $d=2$, for Schwarzchild AdS black brane.}
 \label{fig:EoPplots}
\end{figure}

From figure \ref{fig:EoP3D}, one could see that generally by increasing $D$, in any dimension, after $D_c$, EoP decreases until it becomes relatively constant or drops to zero. Also, in any dimension, for higher $l$, EoP becomes constant.
 We checked this plots for different dimensions numerically and found that for any $d$, it behaves relatively the same way. In any case, with increasing dimensions, for any particular $D$, EoP would be higher which also could be seen from the third, right plot of figure \ref{fig:EoPplots}.

The more precise plots between EoP and various parameters are shown in figure \ref{fig:EoPplots}. Note that these plots have already been shown in \cite{Yang:2018gfq} and we brought them here to later compare with the corresponding plots of CoP.
 
From figure \ref{fig:EoPplots}, one could notice that increasing the width of strip $l$ would increase the minimum critical distance, $D_c$, where there are still non-zero EoP. Increasing $D$ above this critical $D_c$ makes EoP zero. This is because for bigger $l$, there are more degrees of freedom which could correlate to each other and therefore even at bigger $D$, there could still be some correlations if $l$ would be big enough. 

On the other hand if $l$ be small, by increasing $D$, EoP would fall off much faster. However, when $l$ becomes big enough, then EoP would not depend much on $l$ anymore and increasing $l$ cannot change the falling down behavior of EoP. This is because when $l$ increases, the furthest parts of the strips are less correlated and only the qubits which still are close to each other could become correlated to each other and participate in EoP. Therefore, increasing $l$, after certain points, cannot change the behavior of EoP, and this is specifically obvious from the middle plots of figure \ref{fig:EoPplots}.

From the last figure, for the case of $l \to \infty$, one can notice that for a specific distance between the two strips, and for higher dimensions, EoP would be bigger and this could be because more degrees of freedom through different dimensions could correlate to each other and therefore EoP would be bigger.  Also, in the tensor network intuition, each tensor in the network would actually represent a volume of space of order $\ell_{AdS}^{d-1}$. So by increasing the dimension $d$, the corresponding bulk volume of each tensor in the network of strips would increase, which would lead to the bigger EoP.

Note that quantum entanglement is non-increasing under LOCC which is the basis for the definition for EoP. So increasing $l$ which is proportional to the number of local operators, would not change EoP much. However, as we will see, this would not be the case for the complexity of purification which we will study in the next part.

One point that worths to mention is that the sharp drops in the plots of EoP which could be considered as a second order phase transition could be removed by considering the quantum effects. Another point is that due to the phase transition which could be observed as the Ryu-Takayanagi surfaces become disconnected, one could infer that the qubits are actually behave in a non-local way. Then the entanglement of purification and the complexity of purification could characterize how many qubits are actually purely in the system $A$, and how many qubits are actually shared with the other system $B$. The effect of non-locality on both quantities could therefore be examined.

Also, comparing these plots with those of \cite{Bhattacharyya:2018sbw}, which have been derived numerically from the field theory sides, one could see that the general behaviors from both holography and direct calculations are similar.

\section{Complexity of mixed states}\label{MixedStates}

In this section we study various holographic measures for complexity of mixed states. We first calculate the complexity of purification (CoP) using subregion complexity and ``complexity=volume" proposal, and then we introduce another measure which we call \textit{``volume of interval" } (VI) and then we study their properties.

By introducing CoP, for the subregion complexity, we want to see how the rate of complexification in one region would affect the rate of complexity growth in the other region and how the classical and quantum correlations between the two subsystems would play a role on the relationship between their complexity growth rates. For instance, one would like to check if one can increase the rate of growth of complexification by adding another system which could get correlated with the first one. Then one wants to  study how by changing the correlations, or the complexity growth rate in one system, the rate of growth in the other subsystem would change. So for doing that we need to study the complexity of purification holographically.

Now for defining complexity of purification one could choose several methods which we will first give a brief sketch here. In the boundary side, for defining mixed state complexity qualitatively, similar to the pure state, one should choose an appropriate reference state and a set of gates which scale with the purifying Hilbert spaces and then the complexity of purification would be proportional to the minimum number of gates needed to prepare an arbitrary purification of the mixed state.

Similar to entanglement of purification \cite{doi:10.1063/1.1498001}, one could also define the (regularized) complexity of purification which could be dual to the computational cost of creating the state $\rho$ using negligible communication from maximally entangled states and then one could find some lower bounds using the mutual information.

Also,  in \cite{Agon:2018zso}, to any mixed state $\rho$, the authors have associated two basic measures of complexity. One is the \textit{spectrum complexity} which measures how much it would be difficult to construct a mixed state $\rho_{spec}$ with the same spectrum as $\rho$. Then, there is \textit{basis complexity}, which measures the difficulty of constructing $\rho$ from $\rho_{spec}$. Then the complexity of purification is the sum of these two complexities.

Furthermore,  the mixed state information metric for the case of $\text{AdS}_3 /\text{CFT}_2$  has been studied recently in \cite{Lan:2018yxm}. Using such metric, one could define new measures for the complexity of purification.

Recently, also in \cite{Guo:2019azy}, by using the Reeh-Schlieder theorem and the surface/state correspondence, the authors provided a proof for the holographic EoP.  They have used some unitary transformation that act on a subregion which in the bulk would be dual to the deformation of curves in the AdS space, while the boundary is invariant.  Using their picture and the state/surface correspondence, one could explain how the volume of a specific subregion is dual to the complexity of purification. Specifically the number of these unitary transformations which is compatible with all the conditions of surface/state correspondence could lead to the proof for the holographic CoP and even the upper Lloyd's bound. 

Also, in \cite{Guo:2019azy}, the authors have studied the final difference between the EoP and entanglement entropy after a projective measurement which leads one to a better understanding of the sources for each one. This calculation could be repeated for the CoP as well which could lead to a better definition of it. 

Another way to gain further information about the nature of quantum correlations between various patches of the system and its dynamical behaviors could be defining new quantum information measures by combining the previously defined ones. For instance mutual information has been defined by linear combinations of entropies. Similarly, using the linear combination of complexities, one could also define new quantum computational measures.

So one might think that similar to \cite{Hubeny:2018ijt}, these definitions should have two properties of being \textit{primitive} and \textit{faithful}. In fact in \cite{Hubeny:2018ijt}, the general form of the information quantities has been proposed to be like
\begin{gather}
Q(\vec{S})=q_A S(A)+q_B S(B)+q_C S(C)+\nonumber\\
q_{AB} S(AB)+q_{AC} S(AC)+q_{BC} S(BC)+q_{ABC} S(ABC),
\end{gather}
where $\vec{S}$ is the entropy vector defined as
\begin{gather}
\vec{S}= \{ S(A), S(B), S(C), S(AB), S(AC), S(BC), S(ABC)\},
\end{gather}
and $q_i$s are some rational coefficients.
Obviously this definition of entropy space, which consist of all the \textit{``linear"} combinations of entanglement entropies could be generalized to n-partite systems as well.  

For the complexity measures then, one could write
\begin{gather}
Q(\vec{\mathcal{C}})=q_A \mathcal{C}(A)+q_B \mathcal{C}(B)+q_C \mathcal{C}(C)+\nonumber\\
q_{AB} \mathcal{C}(AB)+q_{AC} \mathcal{C}(AC)+q_{BC} \mathcal{C}(BC)+q_{ABC} \mathcal{C}(ABC),
\end{gather}
where $\vec{\mathcal{C}}$ is the entropy vector defined as
\begin{gather}
\vec{\mathcal{C}}= \{ \mathcal{C}(A), \mathcal{C}(B), \mathcal{C}(C), \mathcal{C}(AB), \mathcal{C}(AC), \mathcal{C}(BC), \mathcal{C}(ABC)\},
\end{gather}
and $q_i$s are some rational coefficients. Some combinations of these complexities would be complexity of purification which is dual to the minimum number of quantum gates which prepare the purification of the mixed states. This way one could similarly avoid the UV divergences in the definition of CoP which would be of utmost interest for us.

 Moreover, one could generate equalities and inequalities  similar to the ones written for entropy. Note that for the entropy case we have the strong subadditivity (SSA) which means the amount of correlation is monotonic under inclusion. We numerically see that this is also the case for the complexity of purification. 

In fact, for the entropy case one has the monogamy of mutual information (MMI) \cite{Hayden:2011ag, Hubeny:2018bri}, which actually is the superadditivity of mutual information, or the negativity of tripartite information $I_3 (A:BC)$ in the following form
\begin{gather}
S(AB)+S(BC)+S(AC) \ge S(A) +S(B)+S(C)+S(ABC).
\end{gather} 
One then should check how significant this inequality would be for the case of linear combinations of complexities (volumes) such as complexity of purification.

Note that a qualitative definition for CoP is the minimum number of gates which would be required to prepare an arbitrary purification of the given mixed state. Then, for the case of CV, and for a region $A$ in a boundary Cauchy slice $\sigma$ where its complement is $B:= \sigma \setminus A$ one would find superadditivity property for $\mathcal{C}^V$,
\begin{gather}
\mathcal{C}^V(A)+\mathcal{C}^V(B) \le \mathcal{C}^V (\sigma),
\end{gather}

 while for the CA case one finds
\begin{gather}
\mathcal{C}^A(A)+\mathcal{C}^A(B) \ge \mathcal{C}^A (\sigma),
\end{gather}
and this difference would be a problem for the holographic complexity conjecture. One needs to choose which of CV or CA should be used for defining CoP. Here we take the CV proposal to define our measures.

Note also that complexity shows non-local behaviors \cite{Fu:2018kcp}, which would have a significant role on the behavior of complexity of purification and one should consider this point in defining complexity. Therefore, some proposals such as bit thread picture \cite{Agon:2018lwq, Freedman:2016zud,Cui:2018dyq} could be used in defining and further studying CoP. We return to this point in section \ref{pureOperational}.

Now here, first, similar to \cite{Hubeny:2018ijt} we propose the \textit{complexity vector} and the complexity space which could be defined as the linear combinations of various volumes of the bulk (where actually here the coefficients are either $1$ or $-1$), specifically the sections of the bulk which are inside the Ryu-Takayanagi surfaces and are homologous to various regions of the boundary.  Then those combinations which are UV finite would be of considerable interest.

\subsection{Complexity of purification (CoP) for two subregions}\label{cop31}
Similar to the terms for mutual information and based on studies in \cite{Alishahiha:2018lfv}, one could define a new quantity associated with two subregions $A$ and $B$ as follows
\begin{gather}
\mathcal{C} (A |B)=\mathcal{C} (A)+\mathcal{C} (B)-\mathcal{C} (A\cup  B),
\end{gather}
where for our case, it would be
\begin{gather}
C(A|B)= 2C(l)+C(D)-C(2l+D).
\end{gather}

This is our main definition for the complexity of purification. Note that this quantity could be thought of as mutual complexity which is always non-negative and also symmetric under the exchange of $A$ and $B$, where all $\mathcal{C}$'s here are evaluated using CV proposal. We would like to study the properties of this quantity which could also be thought of as a measure of correlations between the two subsystems. 

 \begin{figure}[ht!]
 \centering
  \includegraphics[width=6.5cm] {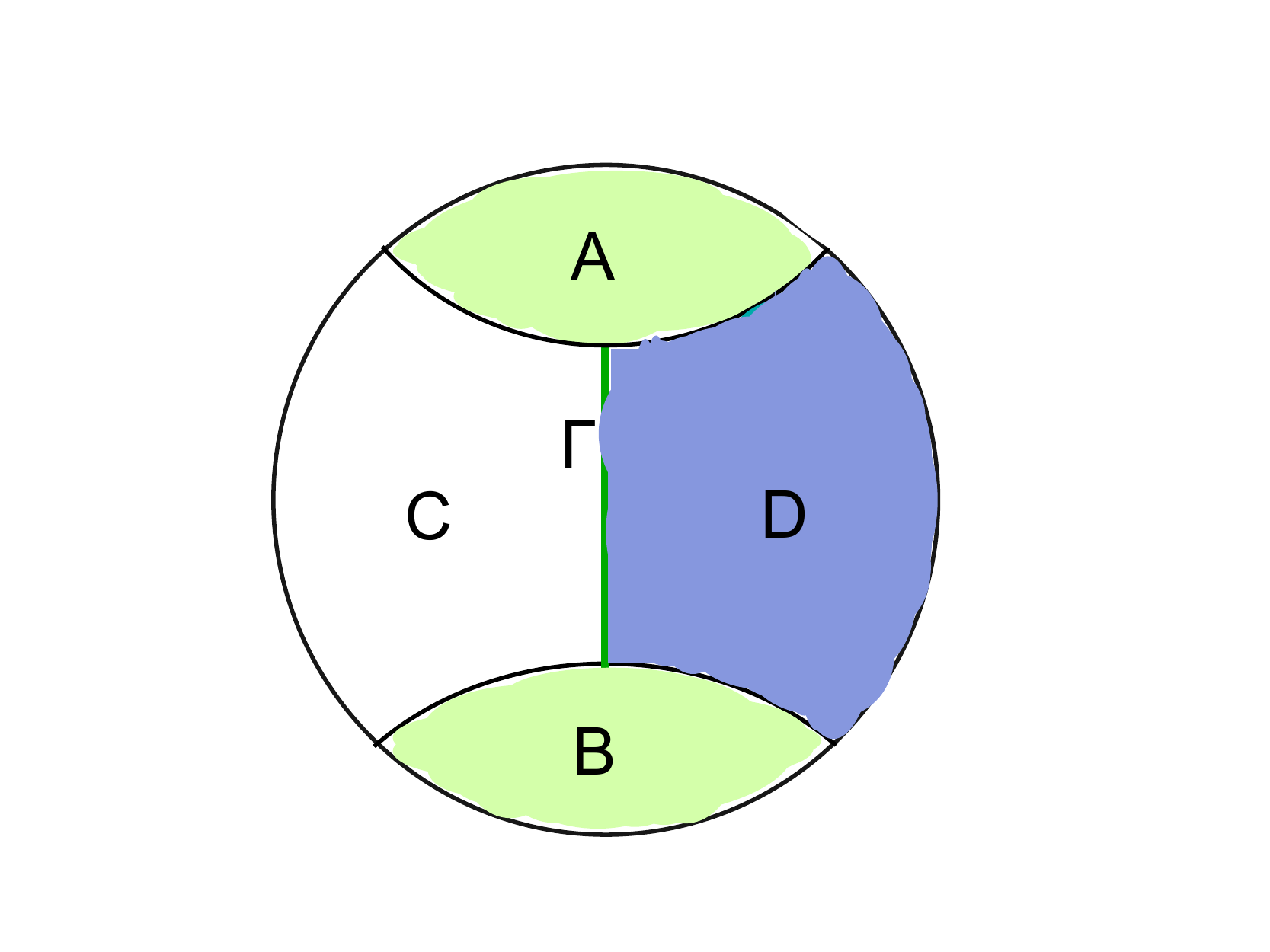}
  \caption{As our main definition, the volume $D$ is proposed to be the ``complexity of purification" between $A$ and $B$.}
 \label{fig:CoPregion2}
\end{figure}

We take the ``complexity of purification" (CoP) as the volume between the boundary and the surface $\Gamma$. This region shown in figure \ref{fig:CoPregion2}, is the volume of region $D$ and it would be be calculated as
\[
\boxed{CoP (A,B)= \frac{V_D}{8\pi  G}= \frac{1}{8 \pi  G} \left (\frac{V_{ABCD} -V_A-V_B}{2} \right).}
\]\label{eq:mainDef}

\begin{figure}[ht!]
 \centering
  \includegraphics[width=7cm] {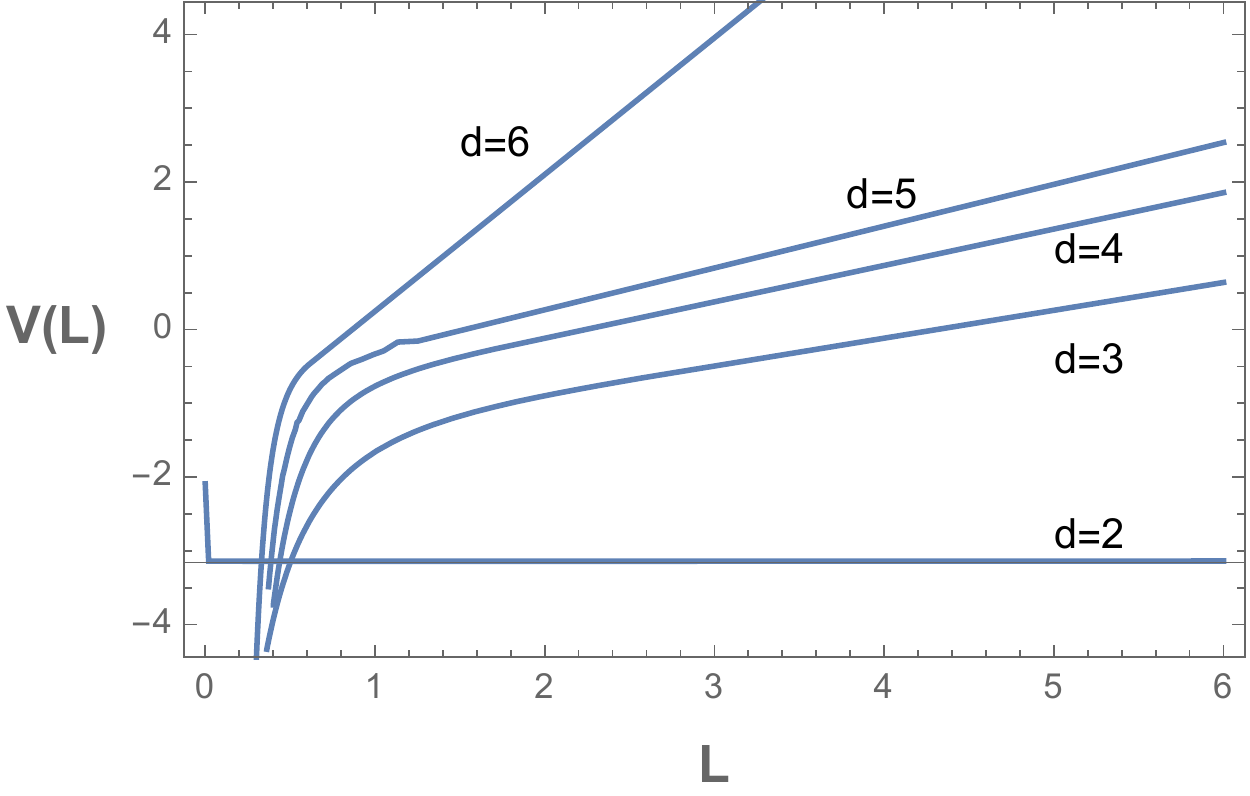}
    \includegraphics[width=6.55cm] {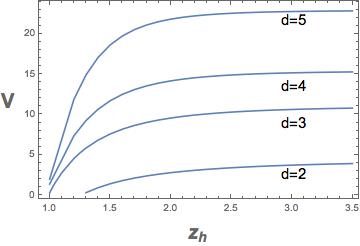}
  \caption{The volume $V(L)$ corresponding to each length of strip $L$ for various $d$.}
 \label{fig:CoPvolume}
\end{figure}

This quantity consists of linear combination of three volumes. The behavior of each volume versus the length of the boundary $L$ is shown in figure \ref{fig:CoPvolume}. Note that in this figure we set different cutoffs for each dimension to make the curve smooth for that specific $d$, and this way we could compare the well behaved curves. From this figure one could see that by increasing the dimension $d$, the volume increases while for the case of $d=2$, it is a constant as we have also found analytically. 

For our setup shown in figure \ref{fig:strips}, which consists of two strips with width $l$ and distance $D$ between them, the complexity of purification which is associated to the volume of the region shown in blue, would be
\begin{gather}
CoP \sim \frac{1}{2} \big (C(2l+D)-2C(l)-C(D)  \big)=-\frac{1}{2} C(A|B).
\end{gather} 

From the volume of subregion $D$, which actually is the subregion complexity \cite{Ben-Ami:2016qex, Alishahiha:2015rta}, the complexity of purification for the two strips, could be found as
\begin{gather}\label{EQQ}
V_D= 2L^{d-2} \Bigg ( \int_\delta^{z_{2l+D}} \frac{dz}{z^d \sqrt{1-z^d}} \int_z^{z_{2l+D}} \frac{dZ}{\sqrt{(1-Z^d)(\frac{z_{2l+D}^{2d-2}    }{Z^{2d-2}  }-1)}} \nonumber\\ -  \int_\delta^{z_D} \frac{dz}{z^d \sqrt{1-z^d} } \int_z^{z_D} \frac{dZ}{\sqrt{(1-Z^d)(\frac{z_D^{2d-2}}{Z^{2d-2}}-1)} }  \nonumber\\
 -2 \int_\delta^{z_l} \frac{dz}{z^d \sqrt{1-z^d} }  \int_z^{z_l} \frac{dZ}{\sqrt{ (1-Z^d)(\frac{z_l^{2d-2}}{Z^{2d-2}}-1) }}\Bigg).
\end{gather}

\vspace{15px}

For the case of $d=2$, the solution would be as follows
\begin{gather}
V_D=\left(-\pi-\frac{1}{\delta}  \arctanh \left(\frac{1}{z_{2l+D}}\right)\right )-\left(-\pi-\frac{1}{\delta} \arctanh \left(\frac{1}{z_D} \right)\right)-2\left(-\pi-\frac{1}{\delta} \arctanh\left(\frac{1}{z_l} \right)\right )\nonumber\\=
 2\pi+ \frac{1}{\delta} \bigg[ 2 \arctanh \left(\coth\left(\frac{l}{2}\right)\right)+\arctanh \left(\coth\left(\frac{D}{2}\right)\right)-\arctanh \left(\coth\left(\frac{2l+D}{2}\right)\right) \bigg]\nonumber\\
=2\pi-\frac{i \pi}{\delta},
\end{gather}
 where one can notice that the universal and real part is just a constant $2\pi$ which matches with the results of \cite{Ben-Ami:2016qex,Abt:2017pmf}. The factor of $i$ could be removed by considering the real  part of squares in each term and the whole divergent term should be removed as well.

\begin{figure}[ht!]
 \centering
  \includegraphics[width=6.5cm] {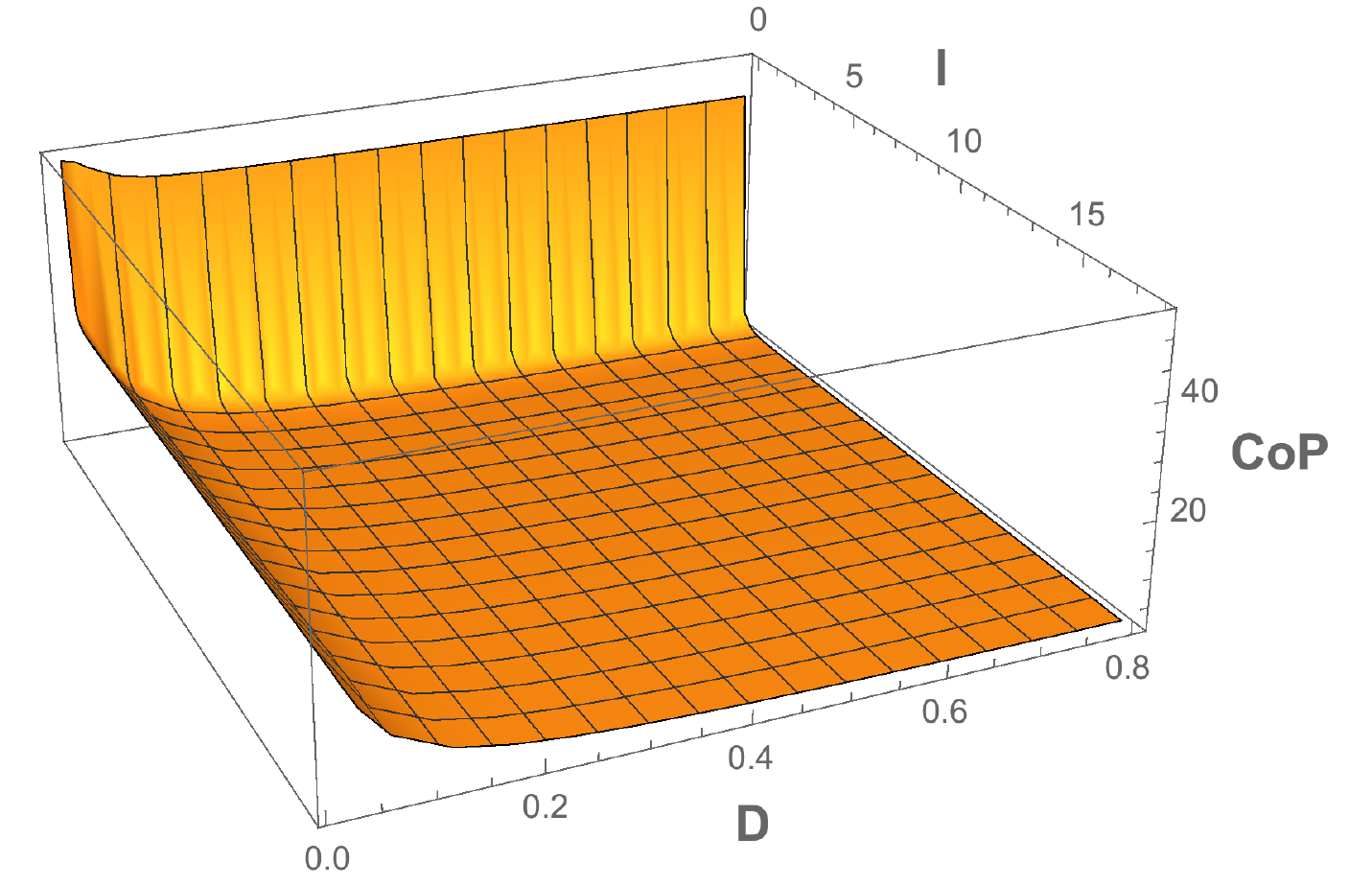}
  \caption{The relationship between complexity of purification and $D$, $l$ for $d=3$.}
 \label{fig:COPnew2}
\end{figure}

Now for higher dimensions which are bigger than $d=2$, we can solve \ref{EQQ} numerically and find the behavior of complexity of purification versus $D$ and $l$. Note that similar to \cite{Ben-Ami:2016qex}, the divergent term of pure $\text{AdS}_3$ is in the form of $\frac{L(z_0)}{2 (\text{d}-1) \delta ^{\text{d}-1}}$ and it should be subtracted to get the desired result. The plot is shown in figure \ref{fig:COPnew2}.

One can notice that it is non-zero only for small $D$ and with increasing the distance between the strips it decreases. It also does not change much with increasing the width of intervals, and in these respects, the behavior is actually very similar to the EoP. Also, increasing dimension $d$, would decrease CoP greatly.

\subsection{CoP for non-symmetrical systems}

\begin{figure}[ht!]
 \centering
  \includegraphics[width=6cm] {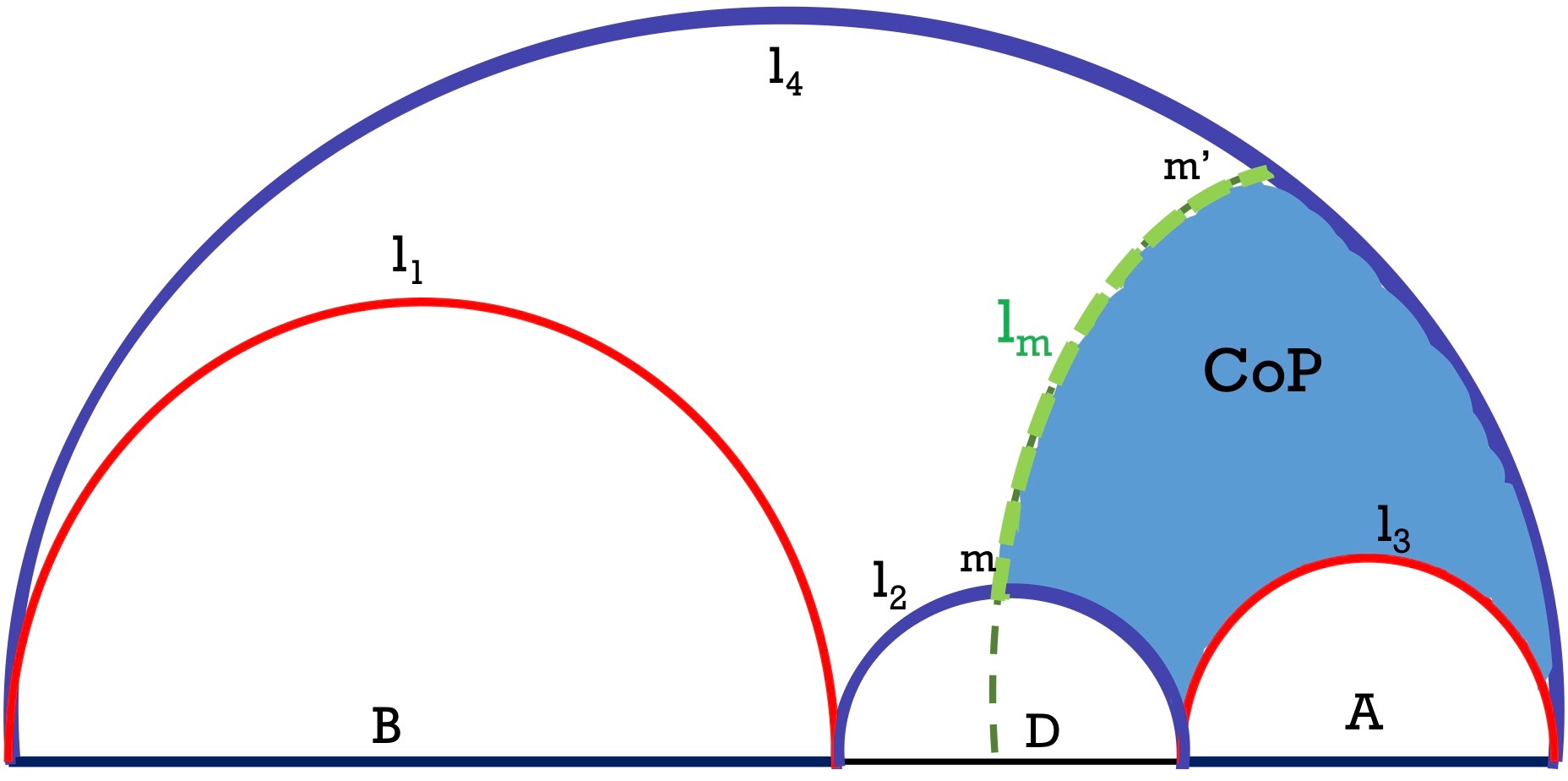}
  \caption{The definition of complexity of purification for non-symmetrical case.}
 \label{fig:nonsymmetric}
\end{figure}

For the non-symmetric case, the definition of CoP would be a bit more complicated. The corresponding volume is shown by the blue region in figure \ref{fig:nonsymmetric}. Of course, to calculate this region one could not simply use a factor of $\frac{1}{2}$ as in the relation of \ref{eq:mainDef}.

For this case, using the algorithm presented in \cite{Liu:2019qje}, first one should find the length of the minimal wedge cross section for such a non-symmetrical configuration. For doing that, the corresponding turning points, $m$ and $m'$ on the the HRT surfaces $l_2$ and $l_4$ should be found. Then, by a direct integration, one could find the volume behind the dark green line and then by removing the region below $l_3$ and also the part which is below surface $l_2$ and in the right part of the green line, one could find the volume corresponding to CoP which has been shown in figure \ref{fig:nonsymmetric}.

 We leave finding the general relation of this volume and the calculation of its various examples to future works.

\subsection{The new measure: The Interval Volume (VI) }\label{newBV}
\begin{figure}[ht!]
 \centering
  \includegraphics[width=7cm] {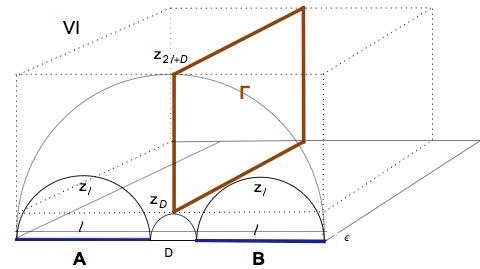}
  \caption{The corresponding region for the new measure of correlation that we called $VI(A,B)$ and its relation with the minimal surface $\Gamma$ is shown, where the width of strip is $l$, the length of strip is infinite and also a cutoff $\epsilon$ is needed. We show that this volume functional has interesting features and it could be a measure of correlation between $A$ and $B$.}
 \label{fig:CoPvolumeinterval}
\end{figure}

Considering the surface of the minimal wedge cross section, $\Gamma$, and its arrangement with the boundary, as shown in figure \ref{fig:CoPvolumeinterval}, one could define another functional as 
\begin{gather}
VI= \frac{1}{2} \left ( \int_\epsilon^{z_{2l+D}} \frac{dz}{z^d \sqrt{f(z)}} -\int_\epsilon^{z_D} \frac{dz}{z^d \sqrt{f(z)} } -2 \int_\epsilon^{z_l} \frac{dz}{z^d \sqrt{f(z)} } \right ). 
\end{gather}

Taking
\begin{gather}
G(z) \equiv \int^z_0 \frac{dz}{z^d \sqrt{f(z)} } =\frac{ -2 z^{1-d} \sqrt{1-z^d} +z(d-2)  _2F_1 \left( \frac{1}{2}, \frac{1}{d}, \frac{d+1}{d},z^d  \right)    }{2(d-1)},
\end{gather}
this new definition which we call the volume of interval (VI) could be written as
\begin{gather}
VI=\frac{1}{2} \left( G(z_{2l+D})-G(z_D) \right) -G(z_l) +G(\epsilon).
\end{gather}

The first two terms are the finite parts and the last term is the divergent term which could be removed by a cutoff or counter terms. 

The finite part which is independent of the cut off and therefore is universal for each case would be
\begin{equation}
\frac{4}{V_{d-1}} C_E(l,D) =\begin{cases}
 \frac{1}{2}  \Big( \text{csch}(\frac{D}{2})+2\text{csch}(\frac{l}{2})-\text{csch}(\frac{2l+D}{2}) \Big) , & d=2,\\ \\
\frac{1}{2}G(z_{2l+D})- \frac{1}{2}G(z_D) -G(z_l)  , & d>2.
  \end{cases}
\end{equation}

 \begin{figure}[ht!]
 \centering
  \includegraphics[width=5cm] {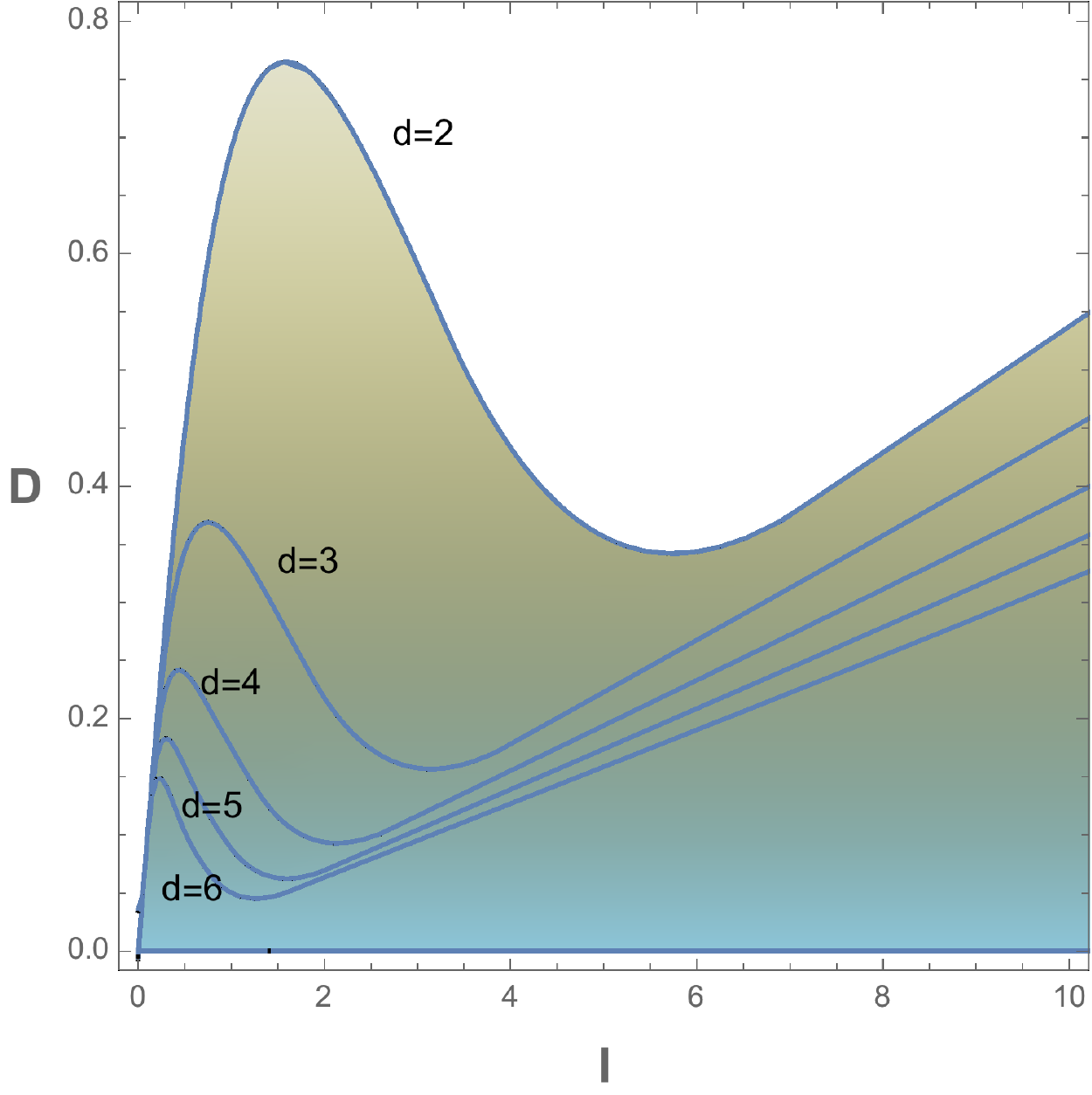}
    \includegraphics[width=6cm] {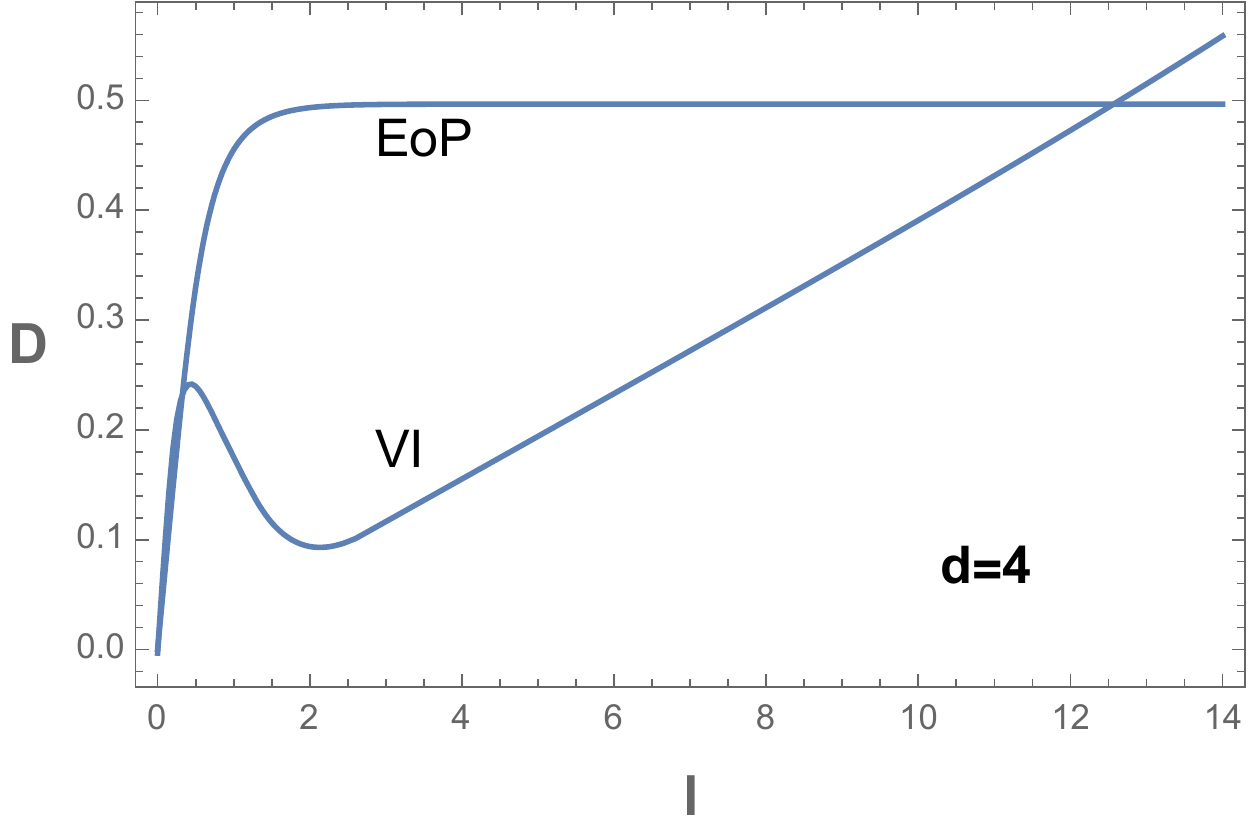}
  \caption{Left: The region of $D$ versus $l$ for different dimensions where VI is positive. The lines are where VI is zero. Right: Comparison between entanglement of purification and VI.}
 \label{fig:COP2new2222}
\end{figure}

In the left section of figure \ref{fig:COP2new2222}, for each dimension and for $D$ versus $l$, the region where VI is positive is shown. One could see that after a specific value of $l$, the relationship is linear, but for smaller $l$, there would be a maximum at any dimension $d$. 

Also note that, for a specific width of strips $l$, by increasing dimension, the distance $D$ between the strips should be reduced in order to get a positive or non-zero VI. 

In the right figure of \ref{fig:COP2new2222}, the curve where EoP is zero is compared with the corresponding curve for VI. Note that the region below each curve is the region where each quantity is positive.   Also, note that for each dimension, the specific $l$ where EoP becomes constant is approximately where the minimum of VI is located

 \begin{figure}[ht!]
 \centering
  \includegraphics[width=5.75cm] {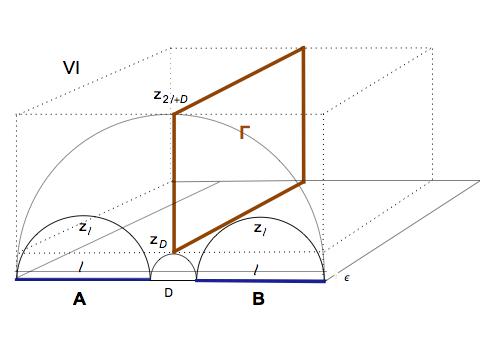}
    \includegraphics[width=5.75cm] {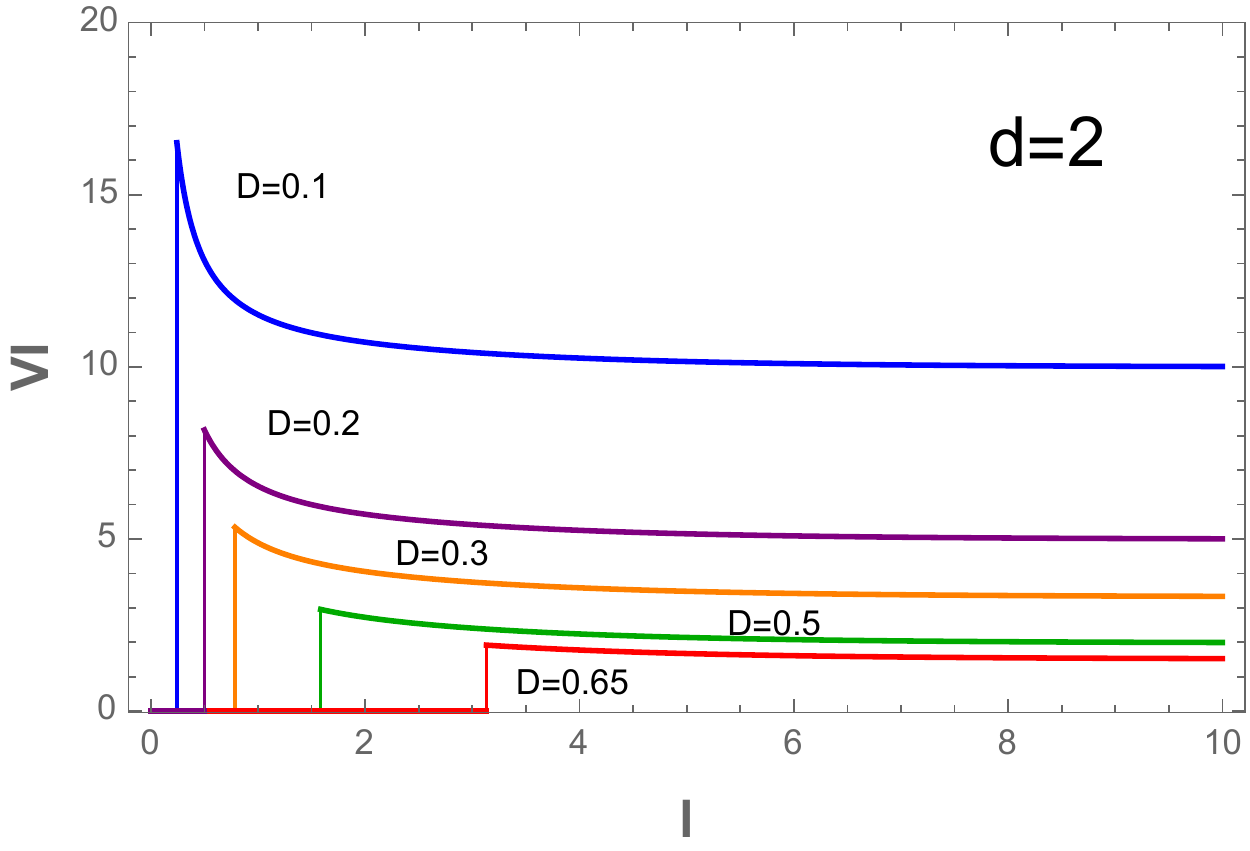}
  \caption{Volume of Interval (VI) for different $l$ and $D$ when $d=2$ for Schwarzchild AdS black brane.}
 \label{fig:COPnew}
\end{figure}

Note that in order to find the critical $D_c$ or the minimum $l$ where VI is non-zero, we used the same regions where we have found in the previous section for EoP and mutual information, because if we assume that there would be no mutual information and therefore no entanglement of purification, then the complexity of purification would be zero as well.

Now, from the left figure of \ref{fig:COPnew}, one can see that VI would decrease by increasing the distance between the two strips, $D$.  Also, one could see that the minimum non-zero value of VI would decrease by increasing the width of strips $l$. In the right sector of figure \ref{fig:COPnew}, the relationship with the length of strips $l$ is shown.

\begin{figure}[ht!]
 \centering
  \includegraphics[width=6.5cm] {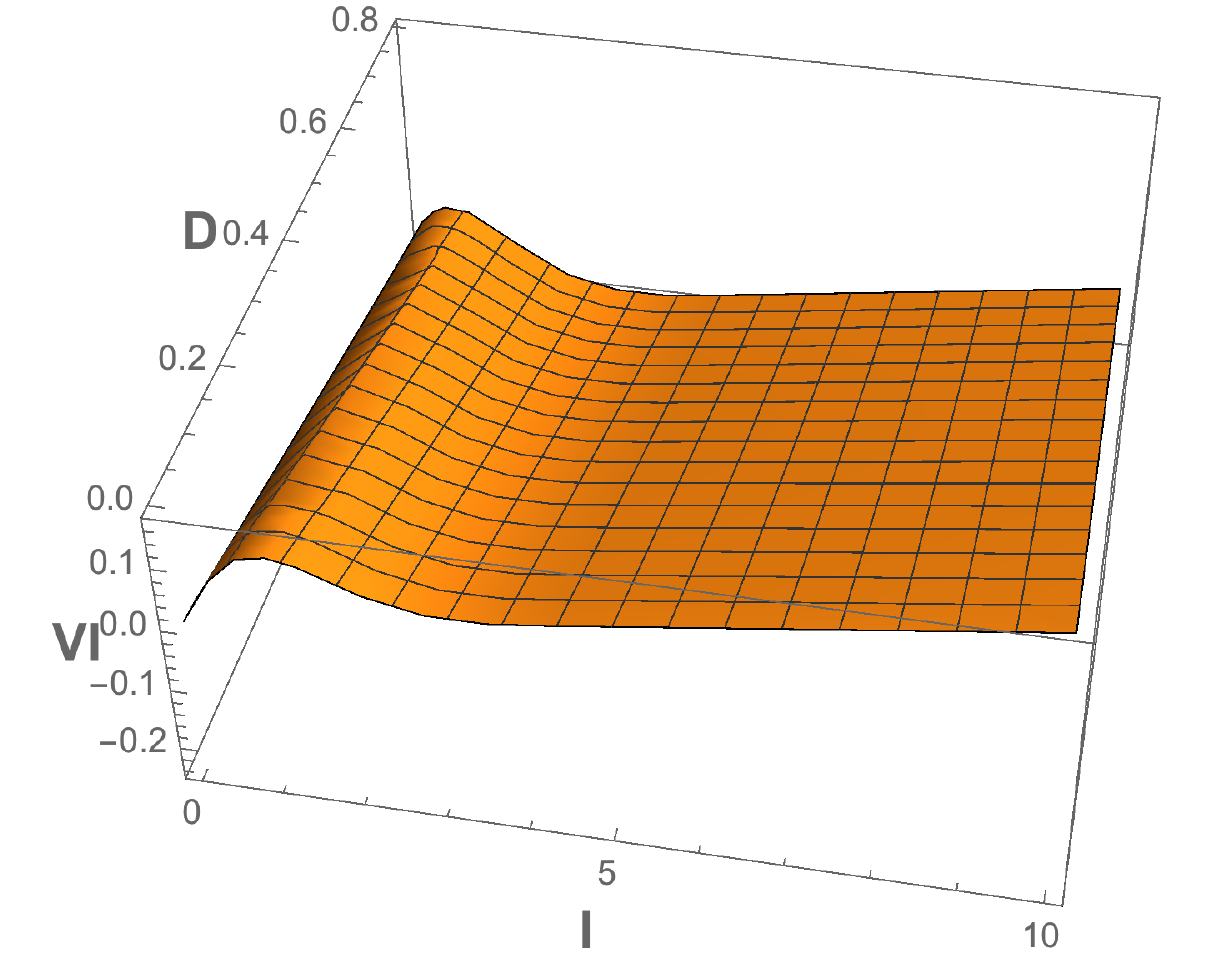}
  \caption{The relationship between the new defined correlation measure VI and the parameters $D$, $l$ for $d=3$.}
 \label{fig:COPnew22}
\end{figure}

 From the three dimensional figure \ref{fig:COPnew22}, one could see that by increasing $D$, CoP monotonically and linearly decreases which is intuitionally correct.
 
Note also that for small strip widths $l$, and for just a small range, CoP decreases. In fact, increasing $l$ by a small value $\delta l$ corresponds to addying a few qubits to the system. After the critical width $l_c$, this quantity increases linearly which matches with the expectations. This peculiar behavior for small $l$, which corresponds to small number of qubits, could be explained by \textit{quantum locking effect} which also has been observed in the behavior of EoP \cite{Takayanagi:2017knl}. Note also that with bigger $D$, this critical $l_c$ becomes slightly smaller, as with increasing $D$ the classical correlation between the gates in the two systems decreases and the quantum locking effect could occur with these smaller number of gates.

 Note that we propose this behavior is connected to locking, because in several works such as \cite{1499048} arbitrary drops (locking effect) have been observed in ``all" different correlation measures, including squashed entanglement, entanglement of formation, entanglement cost, intrinsic information, accessible information, logarithmic negativity and entanglement of purification. 
As by several ways we showed that EoP and CoP are connected, therefore we conjecture that quantum locking should also be observed by CoP, which we detected its trace here.

Note also that locking is closely related to the irreversibility of information theoretical task, therefore, definitely one would expect that it could be observed through complexity as well. In fact, this drop could be observed by increasing some parameters such as mass or separation D through the new correlation measures such as VI.

So $VI$ could be a good measure for  quantum correlations of mixed states, or the computation rates from the quantum mechanical sources. However, the functional we have defined and studied here is \textit{not} CoP as it violates the second and probably the third property of CoP defined in \cite{Caceres:2018blh} which are
\begin{gather}
\text{Positivity:} \ \ \ \ \ \ \ C_A^P>0, \nonumber\\
\text{Monotonicity:} \ \ \ \ \ \ \ C_{A+\delta A}^P > C_A^P, \nonumber\\
\text{Weak Superadditivity:} \ \ \ \ \ \ C_A^P+C_{\delta A}^P < 2C_{A+\delta A}^P.
\end{gather}

One though could study the relationship between this measure and the CoP as in figure \ref{fig:COP2new2222}. 

As our next project \cite{MG:2019KM}, we would like to calculate both of these quantities in a dynamical setup such as a quenched system, then study their behaviors with time and then  compare the results with the experimental data in the spin-spin correlations in various models  \cite{PhysRevLett.120.070501,PhysRevB.98.024302}.

 By calculating butterfly velocities and scrambling times \cite{Hayden:2007cs, Sekino:2008he, Giddings:2017cpl}, one could then study holographically how the information and with what ``speed" could flow between the regions $A$ and $B$ along the surface $\Gamma$, or rather far from it into the bulk and in the region $D$. Therefore, a new characteristic time similar to the scrambling time, would be defined using this measure. Also both EoP and CoP would be implemented to probe various phase important transitions such as phases interpolating between Mott insulator and superfluid phase \cite{PhysRevB.98.024302}. We leave these explorations to our upcoming works.

\section{Purification of other more general cases}\label{purificationgeneral}
To gain further intuitions about the quantum correlations between the two regions, the same study could be done for massive backgrounds such as BTZ black hole solutions in the massive gravity theories, or charged black holes, or even rotating solutions, and then the effect of each physical parameter on the correlation and entanglement and complexity of purifications could be studied. We consider these cases in the next parts and then explain our results using various pictures. We also generalize our definition to n-partite systems.

\subsection{Purification of BTZ black hole solution in massive gravity theory}\label{puremassive}
To study the effect of momentum dissipation in the boundary field theory, one could study EoP and CoP for massive BTZ black holes.

\begin{figure}[ht!] 
\centering
\includegraphics[width=6cm]{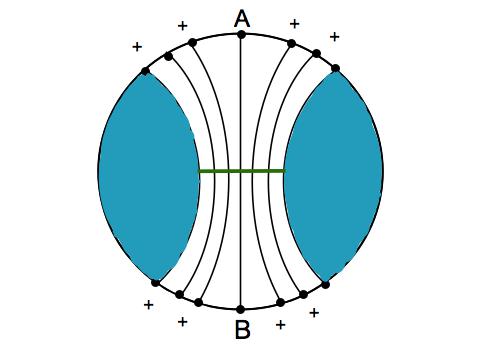}
  \caption{Entanglement wedge cross section between two regions of $A$ and $B$. We will study the effects of mass parameter $m$ and charge $q$ on EoP and CoP and explain the results using bit thread picture.}
 \label{fig:massiveBTZ2}
\end{figure}

Before any calculation, one could expect that in the background of BTZ black hole solution of massive gravity theory, both EoP and CoP would be lower than the massless case. This is because in these theories, the graviton gain mass and therefore the diffeomorphism invariance is broken. 
Considering bit threads in the construction of figure \ref{fig:massiveBTZ2}, one expects that dissipations would lower the correlations between the two regions. This mathematically could be examined by associating ``momentum" to bit threads and determining their behaviors in such systems. Note that similar ideas of generalizing bit thread picture by varying the thickness, density or orientation of bit threads in various backgrounds such as higher-curvature gravity has already been done in works such as \cite{Harper:2018sdd}.

So for checking our expectations, we first write the metric of massive BTZ black hole in the following form \cite{Hendi:2016pvx}
\begin{equation}
ds^2=\frac{1}{z^2}[-f(z)dt^2+\frac{dz^2}{f(z)}+dx^ { 2 }]~~~\mathrm{with} ~~~f(z)=-\Lambda-m_0 z^2-2q^2 z^2 \ln (\frac{1}{z \ell})+m^2 c c_1 z.
\label{Metric}
\end{equation}%

The above geometry is a solution to Einstein equations for the three dimensional Einstein-massive gravity with the action \cite{Hendi:2016pvx,Ghodrati:2016ggy,Ghodrati:2016vvf,Ghodrati:2016tdy}
\begin{equation}
\mathcal{I}=-\frac{1}{16\pi }\int d^{3}x\sqrt{-g}\left[ \mathcal{R}-2\Lambda+L(\mathcal{F})+m^{2}\sum_{i}^{4}c_{i}\mathcal{U}_{i}(g,h)\right],
\label{Action}
\end{equation}%
where $\mathcal{R}$ is the scalar curvature, $L(\mathcal{F})$ is an arbitrary Lagrangian of electrodynamics and $\Lambda$ is the cosmological constant.

Also the fixed symmetric tensor satisfies the relation $h_{\mu \nu }=diag(0,0,c^{2}h_{ij})$ and the corresponding symmetric polynomials  $\mathcal{U}_{i}$ could be
evaluated as $\mathcal{U}_{1}=c/r$ and $\mathcal{U}_{2}=\mathcal{U}_{3}=\mathcal{U}_{4}=0$\footnote{For any symmetric tensor, the symmetric polynomials of the eigenvalues of the $d\times d$ matrix $\mathcal{K}_{\nu }^{\mu }=\sqrt{%
g^{\mu \alpha }h_{\alpha \nu }}$ are written as
\begin{eqnarray}\label{eq-Ui}
\mathcal{U}_{1} &=&\left[ \mathcal{K}\right] ,\;\;\;\;\;\mathcal{U}_{2}=%
\left[ \mathcal{K}\right] ^{2}-\left[ \mathcal{K}^{2}\right] ,\;\;\;\;\;%
\mathcal{U}_{3}=\left[ \mathcal{K}\right] ^{3}-3\left[ \mathcal{K}\right] %
\left[ \mathcal{K}^{2}\right] +2\left[ \mathcal{K}^{3}\right] ,  \notag \\
&&\mathcal{U}_{4}=\left[ \mathcal{K}\right] ^{4}-6\left[ \mathcal{K}^{2}%
\right] \left[ \mathcal{K}\right] ^{2}+8\left[ \mathcal{K}^{3}\right] \left[
\mathcal{K}\right] +3\left[ \mathcal{K}^{2}\right] ^{2}-6\left[ \mathcal{K}%
^{4}\right].
\end{eqnarray}}.

In \ref{Metric}, $m_0$ is an integration constant which is related to the total mass of black hole.  In the following part we set $m_0=1$. So our solution is a  ``massive BTZ black hole" in a ``massive gravity theory" where the graviton has the finite mass $m$. Here we are interested on the effect of this parameter $m$.

  Also we set cosmological constant $\Lambda=-1$. In addition, since here we are not interested on the effect of charge, we drop $L(\mathcal{F})$ from the action and consider $q=0$ in the metric.

Note that in the action, $m$ is the mass of graviton in the theory. In the holographic framework, the massive terms in the gravitational action break the diffeomorphism symmetry in the bulk, which as mentioned correspond to momentum dissipations in the dual boundary field theory \cite{Blake:2013bqa}.

In addition, in the following study, we set $c=c_1=1$ without loss of generality.

So, for the geometry \eqref{Metric}, the induced metric would become
\begin{equation}
\sqrt{-g}=\sqrt{x^{\prime}_{1}+\frac{1}{f(z)}}\left(\frac{1}{z}\right).
\end{equation}
Minimizing the above equation gives us
\begin{equation}
\frac{x^{\prime}}{z\sqrt{x^{\prime^{2}}+\frac{1}{f}}}=\frac{1}{z_{0}},  \ \ \ \ \ \ \ 
x^ { \prime } = \frac { 1 } { \sqrt { f \left(  \frac { z _ { 0 }^2 } { z^2 }  - 1 \right) } }.
\end{equation}
Thus, the width of the strip and its holographic entanglement entropy could be evaluated as
\begin{gather}
w = 2 \int _ { \delta } ^ { z _ { 0 } } d z \frac { 1 } { \sqrt { f  \left( \frac { z _ { 0 } ^ { 2 } } { z ^ {  2 } } - 1 \right) } }, \nonumber\\
 S(\omega)=\frac{1}{2}\int _ { \delta } ^ { z _ { 0 } } \frac { d z } { z  } \frac { 1 } { \sqrt {f (z)\left( 1 - \frac { z ^ { 2 } } { z _ { 0 } ^ { 2 } } \right) } }.
 \end{gather}

\begin{figure}[ht!]
\centering
\includegraphics[width=0.4\textwidth]{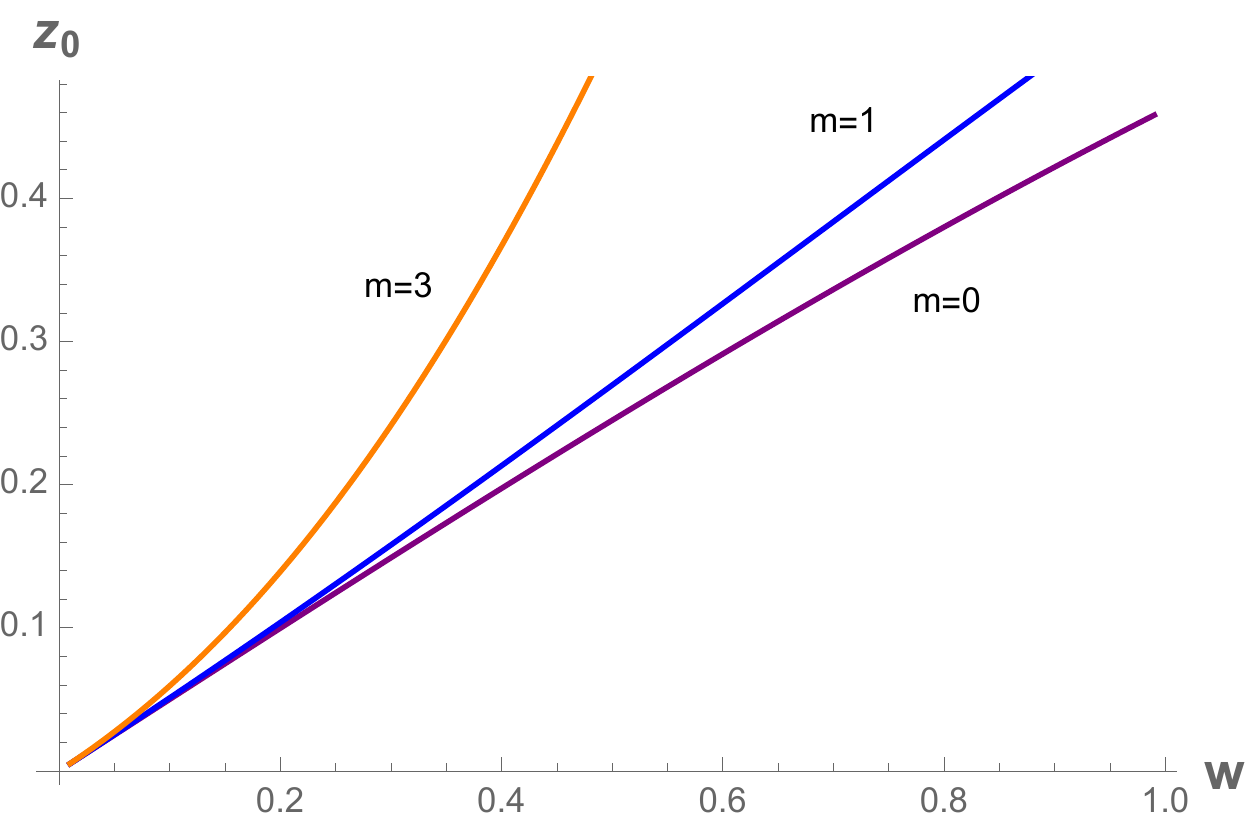}\hspace{1cm}
\includegraphics[width=0.4\textwidth]{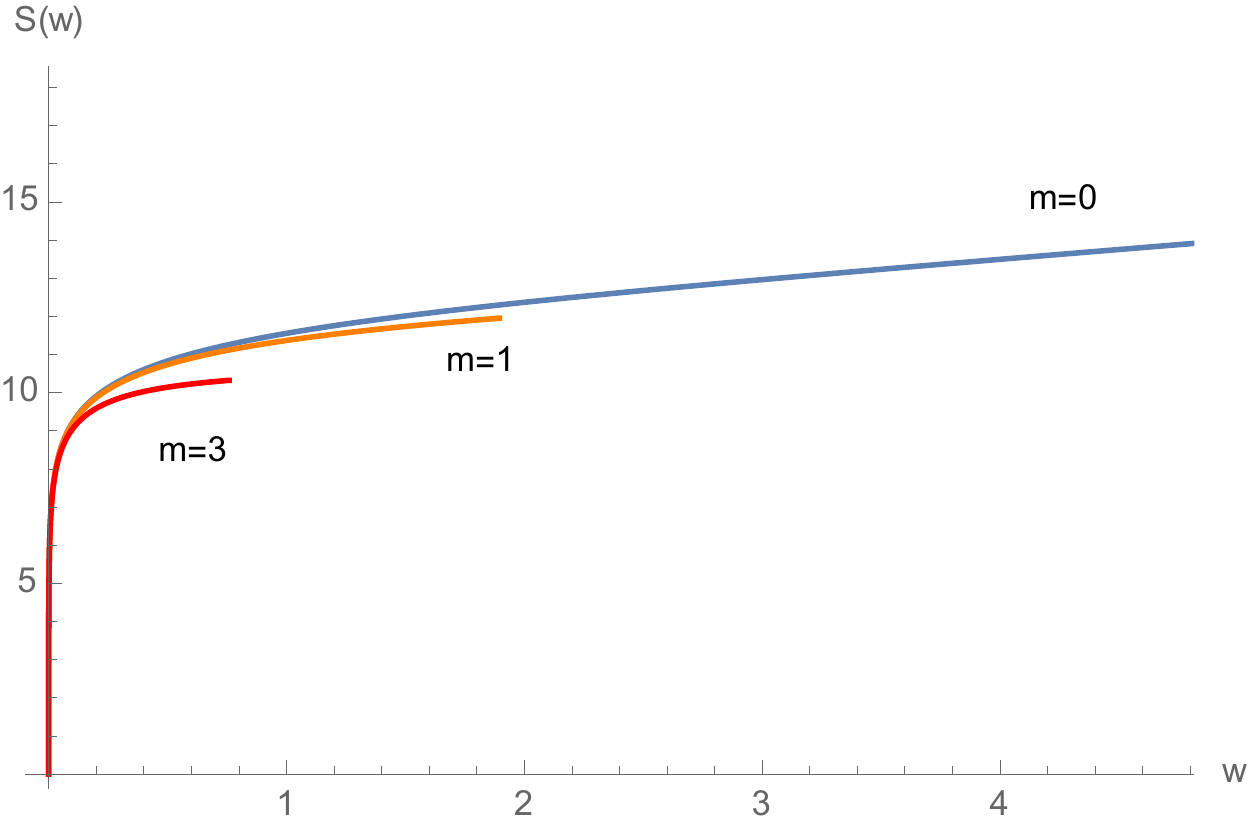}
\caption{The turning point (left) and the related holographic entanglement entropy (right) versus $w$ for different $m$.}\label{fig-z0-w-S}
\end{figure}

Then, in figure \ref{fig-z0-w-S}, we show the position of turning points as a function of width of strip $w$ for different $m$. We also show the holographic entanglement entropy $S(w)$ versus $w$ for different mass parameter.

One could notice that for any specific width $w$, with increasing mass $m$, the turning point would go deeper into the bulk and becomes bigger, while increasing $m$ for any $w$ could decrease the entanglement entropy $S(w)$. This is because when the backgrounds become massive, some entanglement between pairs would be broken down.  

\subsubsection{EoP in massive BTZ}

\begin{figure}[ht!]
\centering
\includegraphics[width=0.35\textwidth]{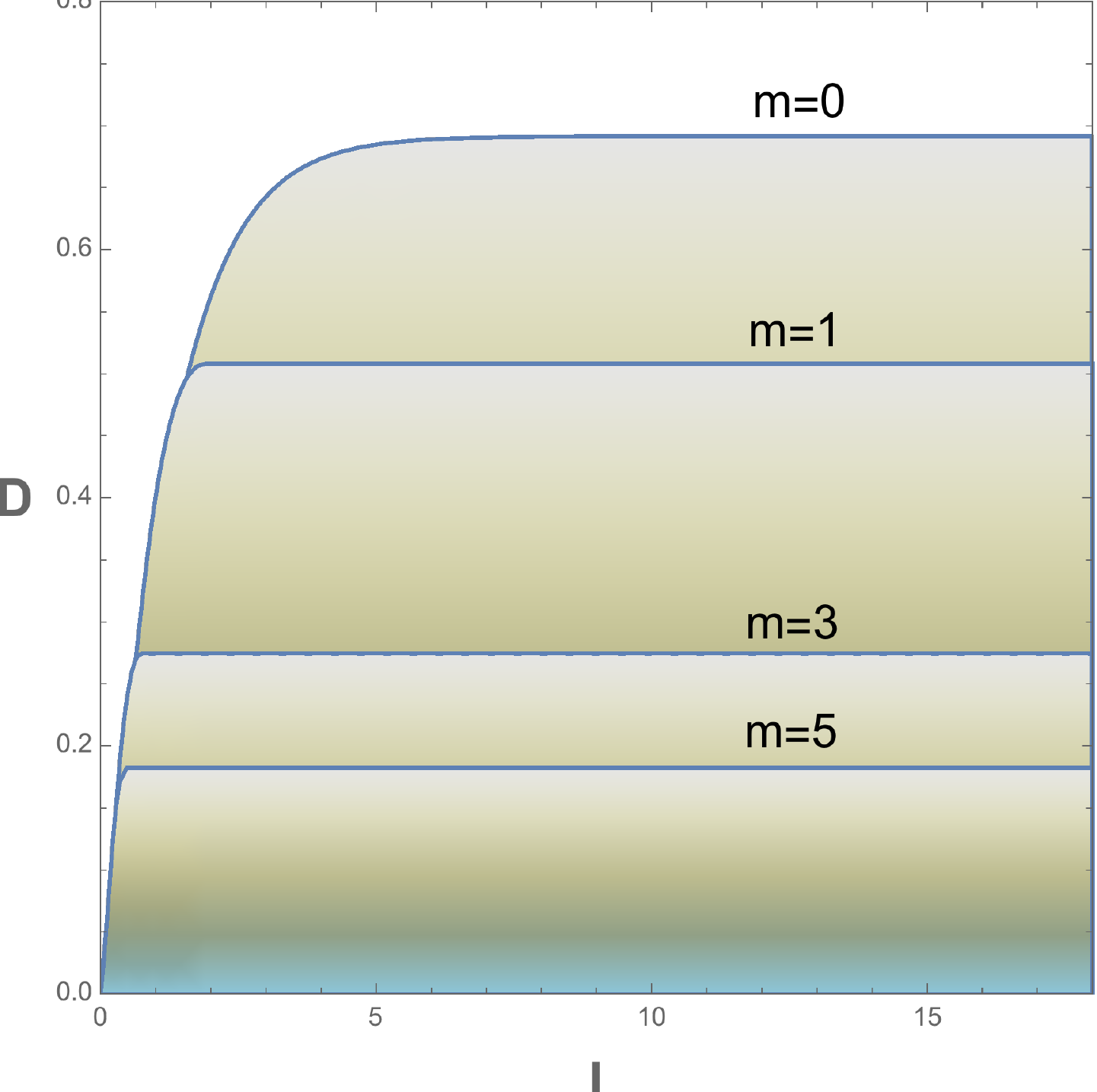}
\caption{For each $m$, the region below the lines have non-vanishing EoP.}\label{fig-LDeop0}
\end{figure}

In figure \ref{fig-LDeop0}, we show the regions with non-vanishing EoP. One could notice that as $m$ increases, the critical distance $D_{c}(m, l\to\infty)$ which makes the mutual information and EoP zero, would decrease. So by increasing $m$ the strips need to be closer to each other in order to keep the correlations between them constant. Again, one could see that the mass parameter actually could break the correlations and lower the entanglement among the quibits of the two regions, as we have indeed expected from the bit thread picture.

We then could fit the critical $D_{c}(m,l\to\infty)$ as a function of $m$ as
\begin{equation}
D_{c}^{-1}(l\to\infty)\simeq1.28+0.805 m,
\end{equation}
or
\begin{equation}
D_{c}(l\to\infty)\simeq0.7-0.2919\log m.
\end{equation}

Therefore, one could associate a logarithmic function between the decreasing strength of correlations and the increasing of graviton's mass parameter. 

\begin{figure}[ht!]
\centering
\includegraphics[width=0.4\textwidth]{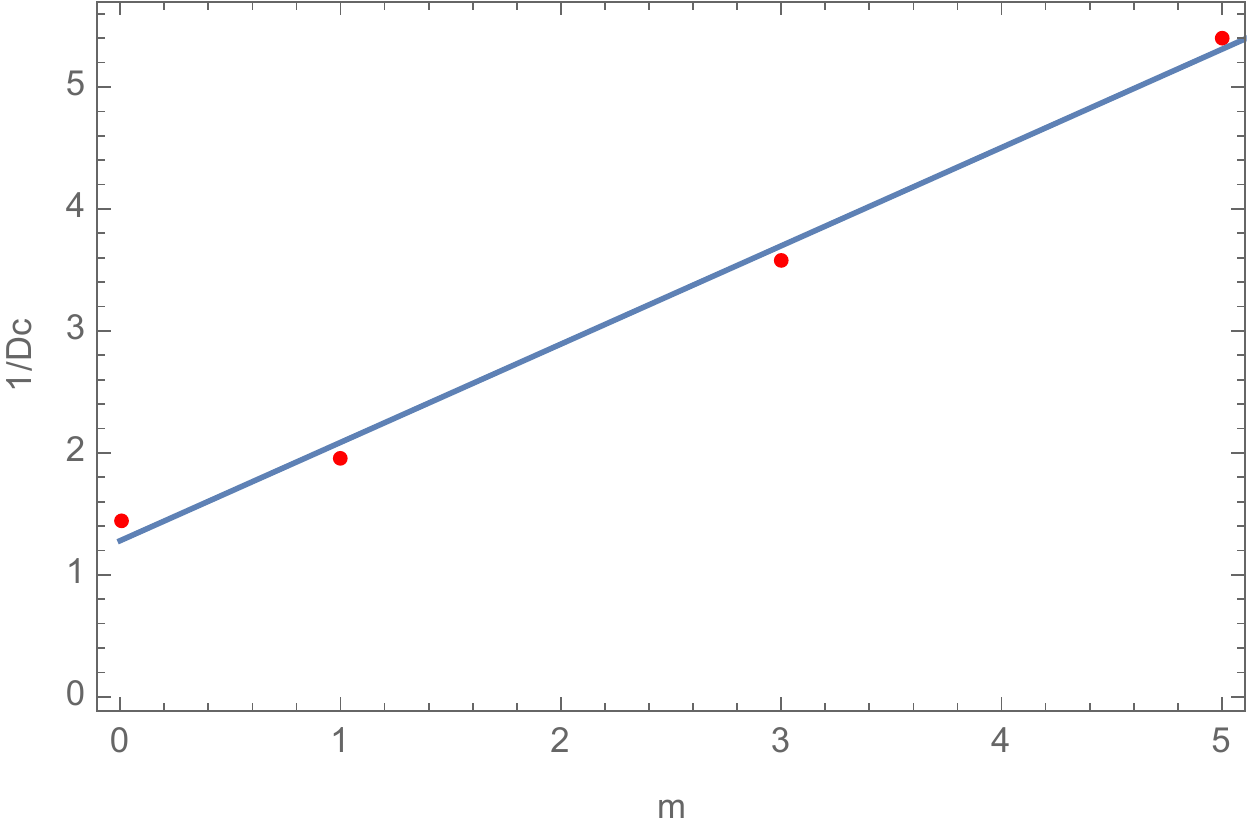}\hspace{1cm}
\includegraphics[width=0.4\textwidth]{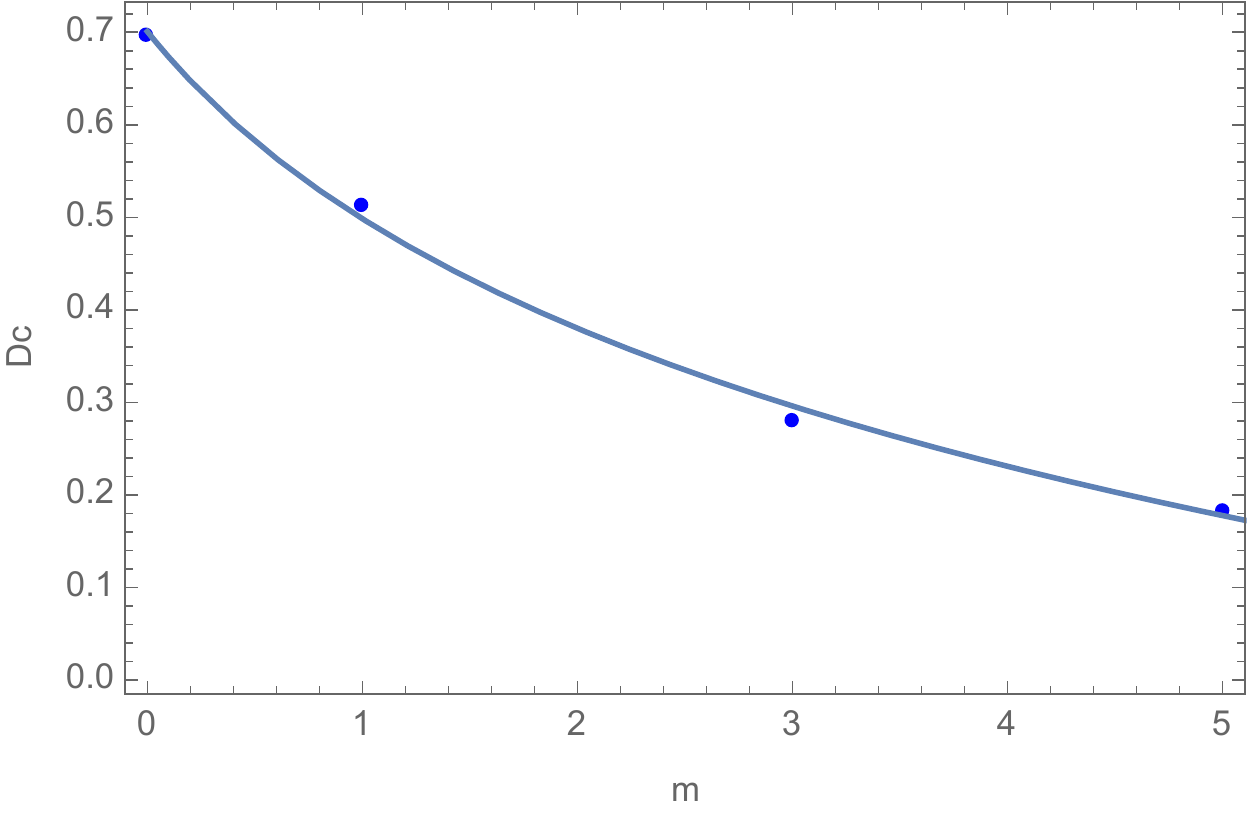}
\caption{Left: The relationship between $1/D_{c}$ and $m$. Right: The relationship between $D_{c}$ and $m$.}\label{fig-Dcfit}
\end{figure}

The fitting lines are shown in figure \ref{fig-Dcfit}.

For the cases where $D$ and $l$ are smaller than $D_{c}(d, l)$ and $l_c$, and the MI and EoP are non-zero, the area of surface $\Gamma$ in this model could be written as
\begin{equation}
\Gamma = \int _ { z _ { D } } ^ { z _ { 2 l + D } } \frac { d z } { z \sqrt { 1 - z ^ { 2 }+m^{2}cc_{1}z } },
\end{equation}
and therefore the entanglement of purification would be calculated as
\begin{equation}
\left. 2 E ( l , D )=\frac{\log z}{\log(2+m^{2}z+2\sqrt{1+(m^{2}-z)z})} \right| _ { z _ { D } } ^ { z _ { 2 l + D } }.
\end{equation}

\begin{figure}[ht!]
\centering
\includegraphics[width=0.3\textwidth]{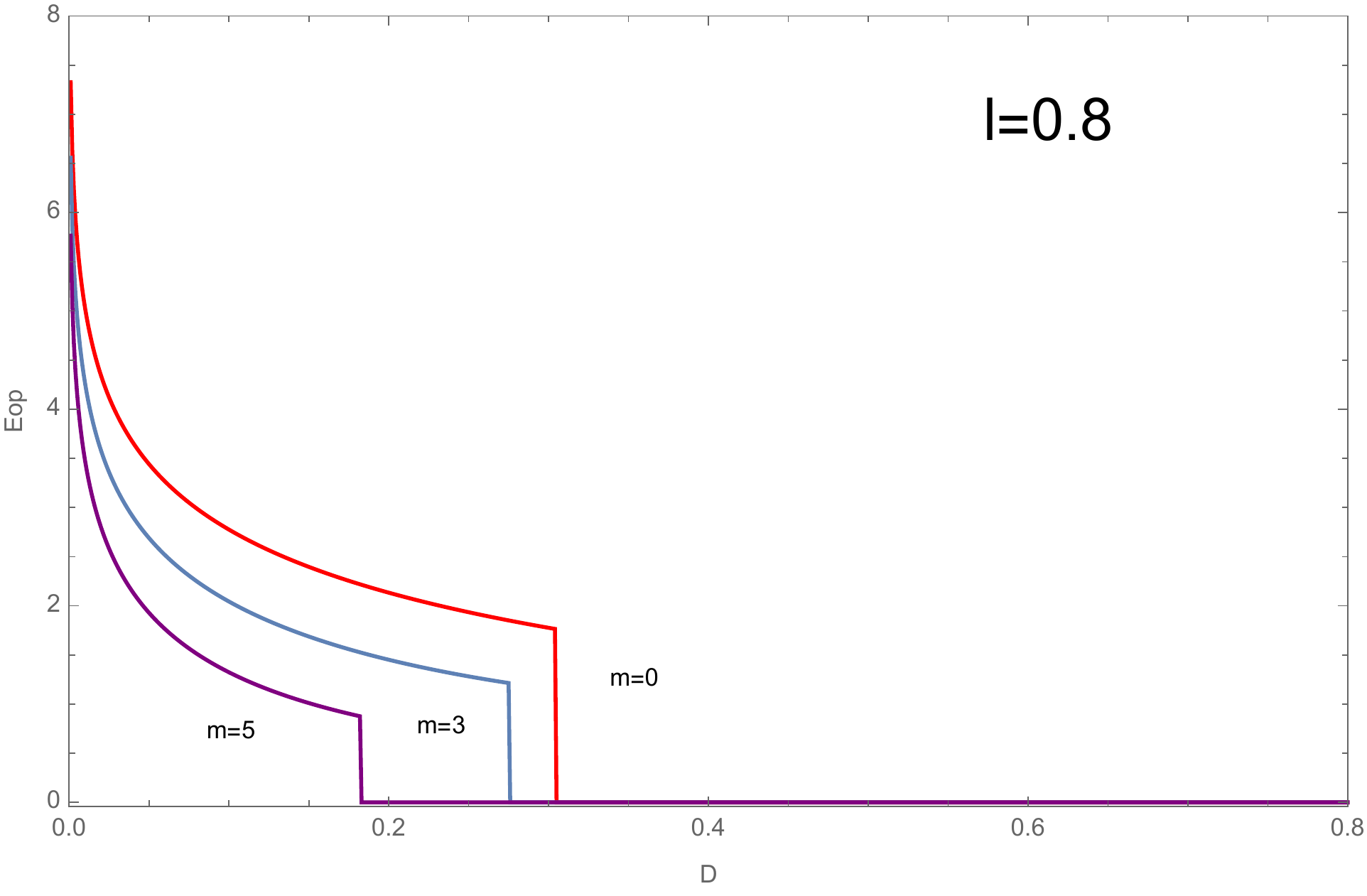}\hspace{0.5cm}
\includegraphics[width=0.3\textwidth]{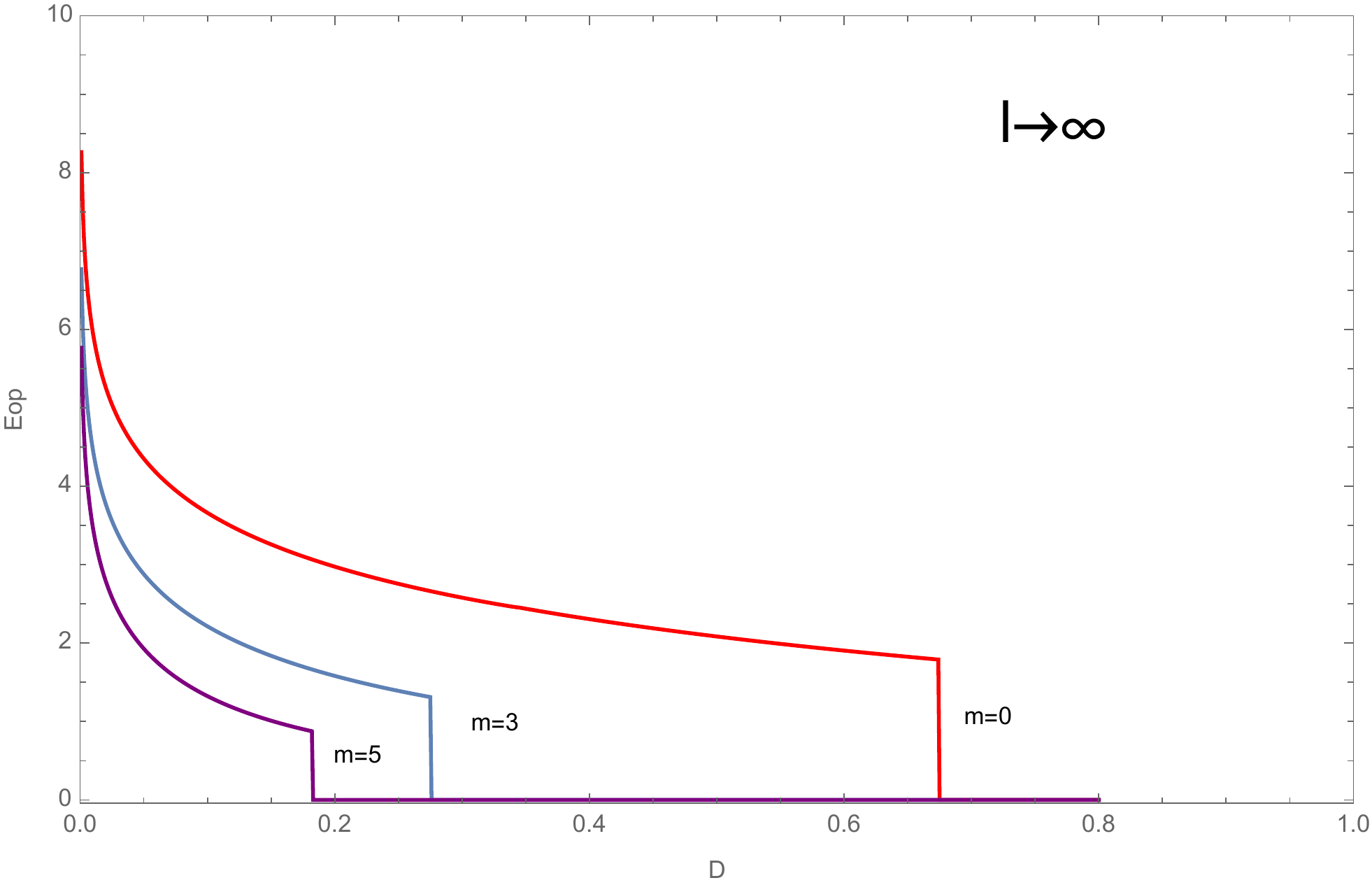}\hspace{0.5cm}
\includegraphics[width=0.3\textwidth]{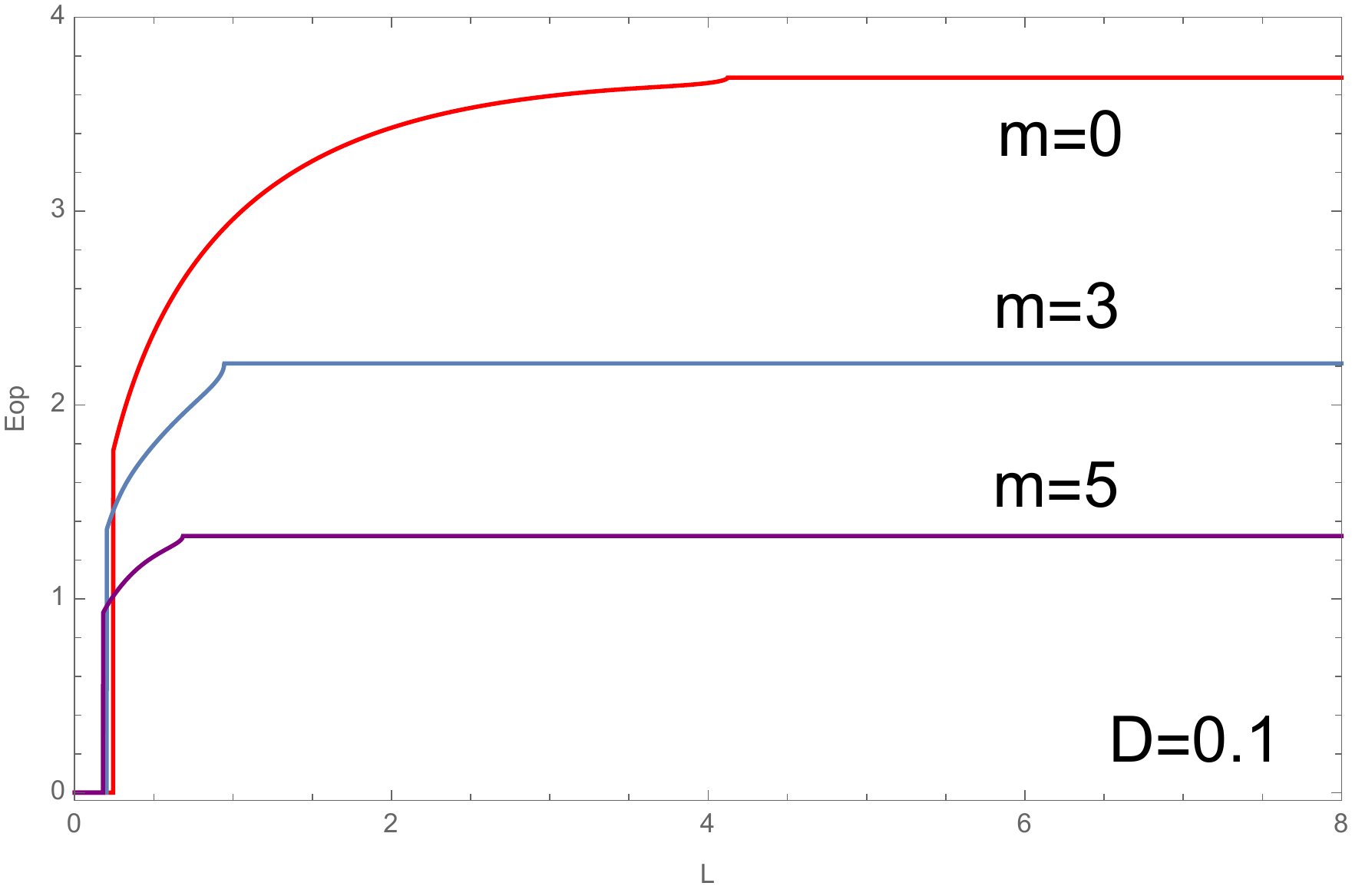}
\caption{EoP as function of  $D$ with $l=0.8$ (left)  and $l=\infty$ (middle), and EoP as a function of $l$ with $D=0.1$ (right).}\label{fig-D-EOP}
\end{figure}

Then, this equation could be studied numerically. In figure \ref{fig-D-EOP}, we show EoP as a function of $D$ for fixed finite $l=0.8$ and also infinite $l$ case. We also present EoP as a function of $l$ for fixed $D=0.1$. One could notice that as we have expected, in all cases, the EoP for larger momentum relaxations would be smaller as $m$ breaks the correlation and therefore decreases the EoP.

Then, we study EoP as a function of $m$ with fixed $D$ and $l$ in the left and middle plots of figure \ref{fig-m-EOP}, where it is obvious that again EoP would fall down as $m$ increases. In the right plot of figure \ref{fig-m-EOP}, one could  recognize a phase transition and a phase diagram in the $m-D_{c}$ plane, namely, above the line, one would have $\text{EoP}=0$ and below the line, the EoP is positive.

\begin{figure}[ht!]
\centering
\includegraphics[width=0.4\textwidth]{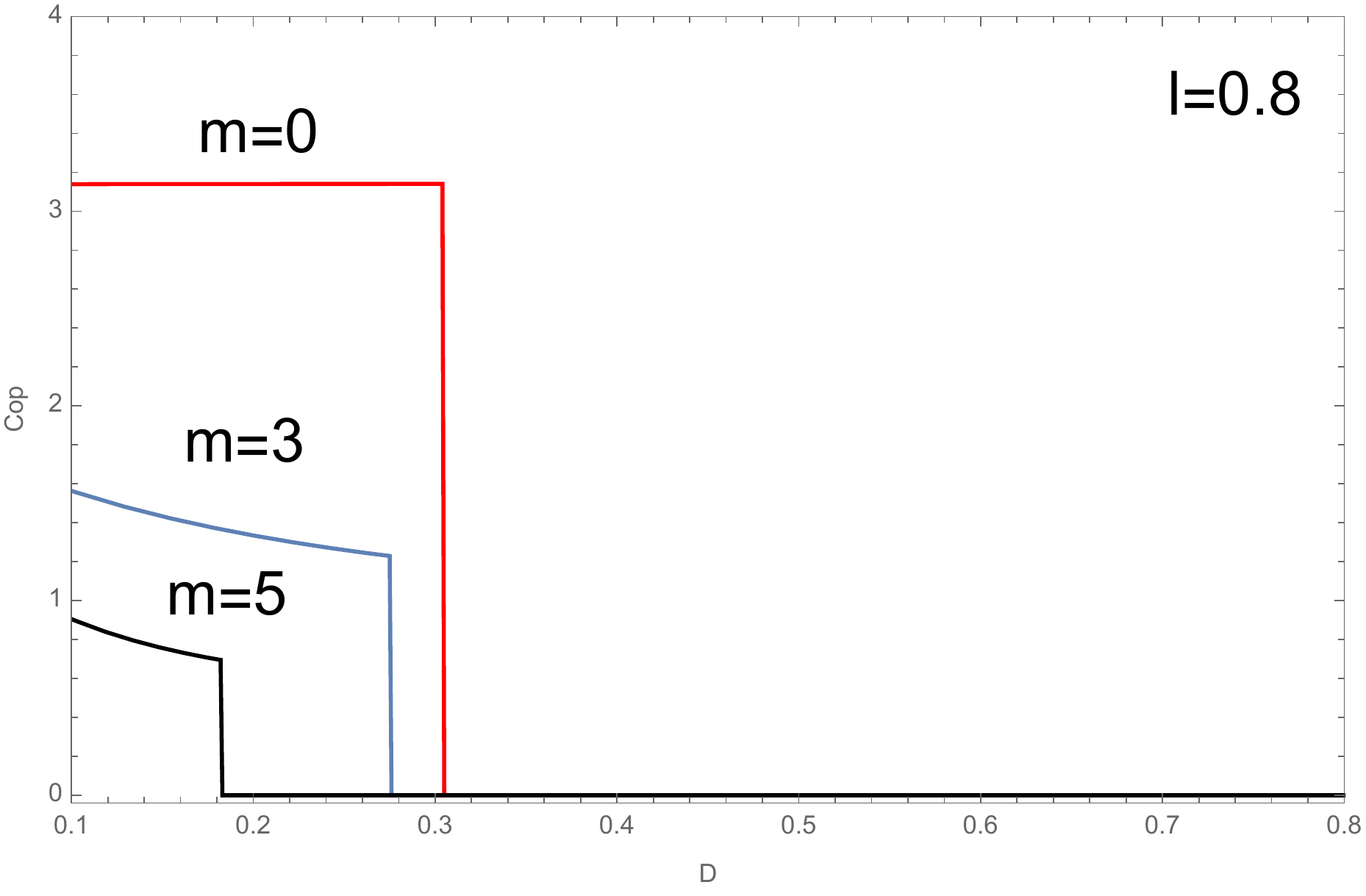}\hspace{0.5cm}
\includegraphics[width=0.4\textwidth]{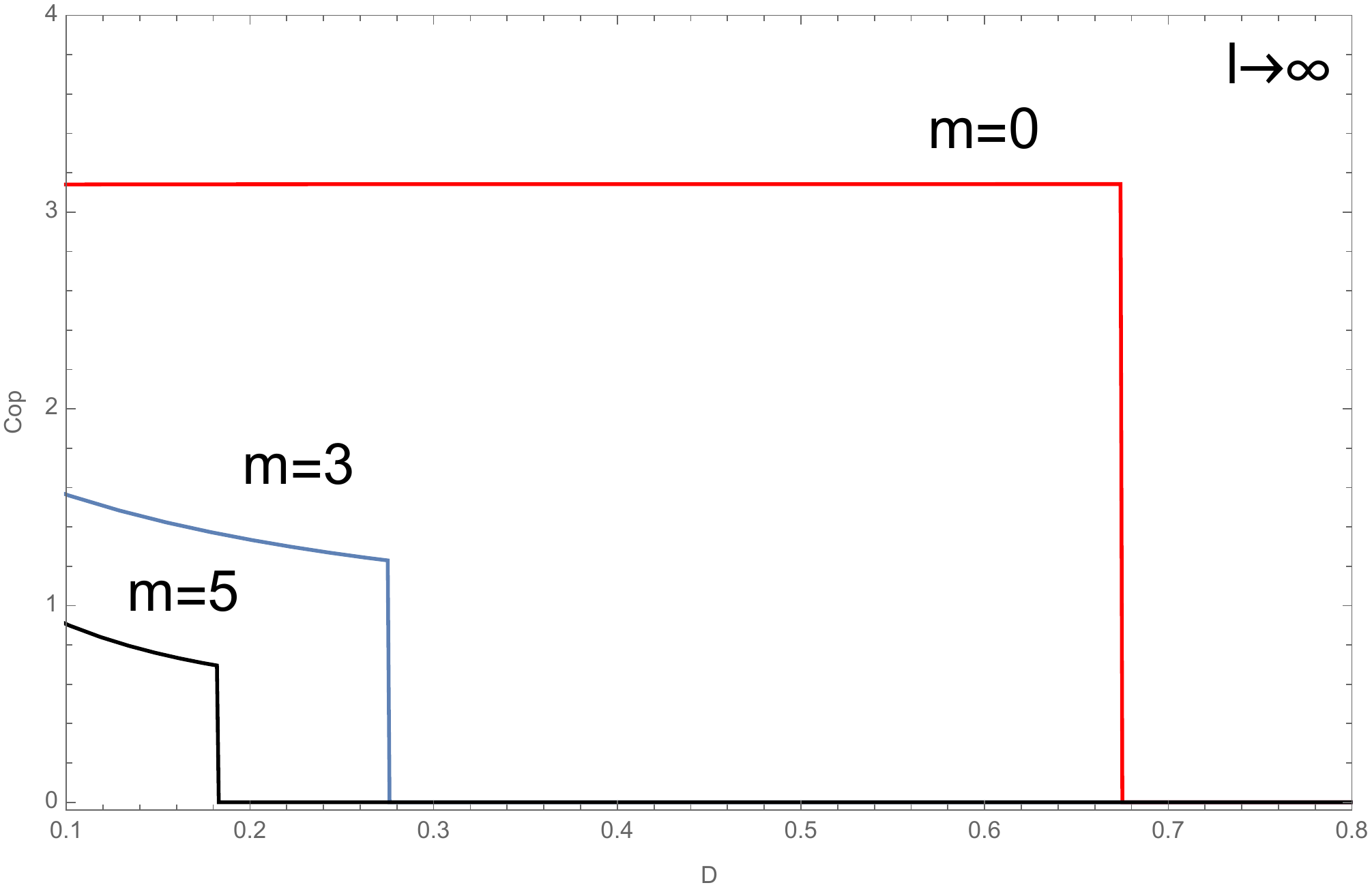}
\caption{Left: EoP as function of $m$ with fixed $D=0.1$ and $l=0.8$. Right: Phase diagram in the $m-D_{c}$ plane.}\label{fig-m-EOP}
\end{figure}

\begin{figure}[ht!]
\centering
\includegraphics[width=0.4\textwidth]{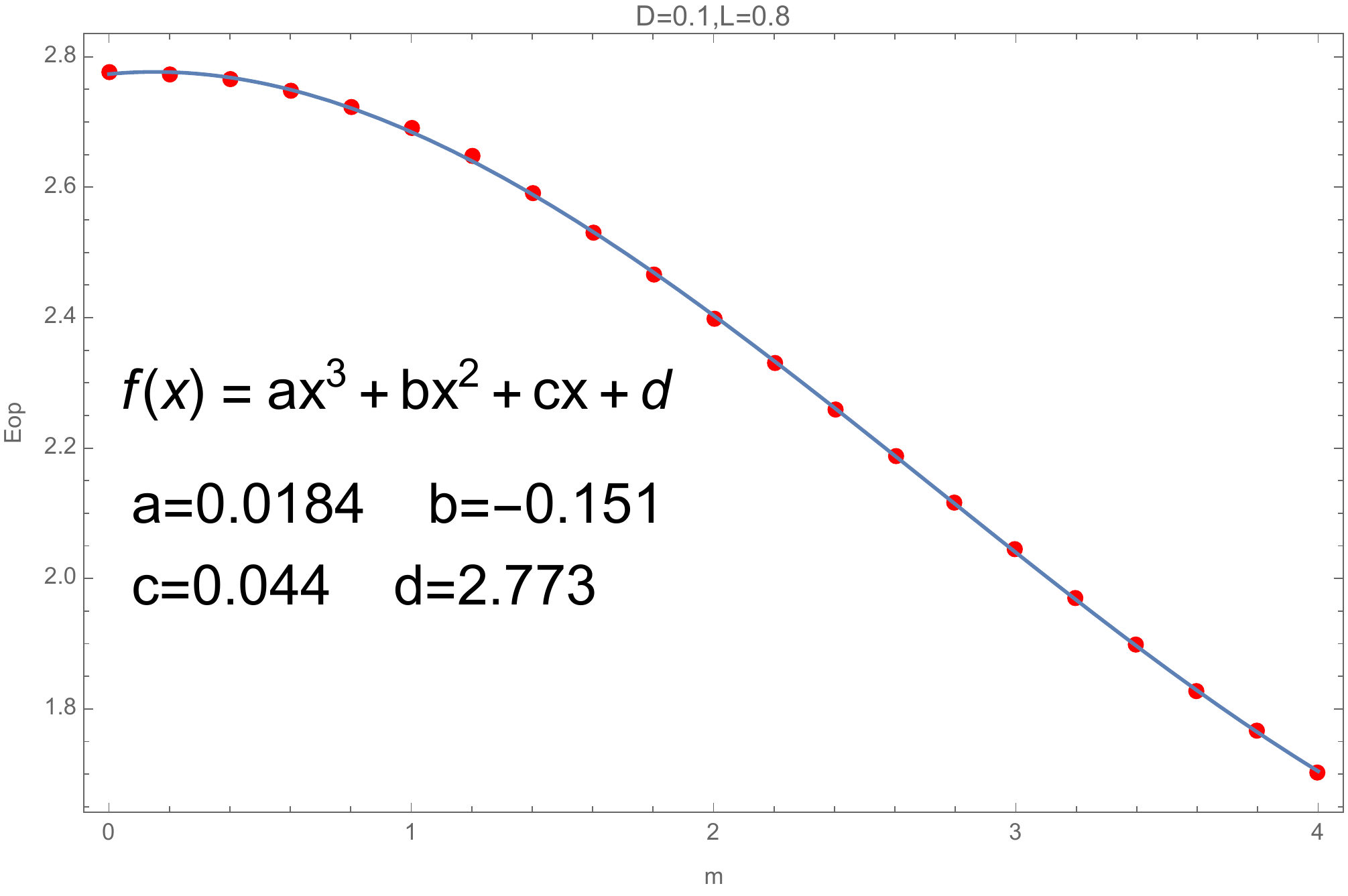}
\caption{EoP as function of $m$ with fixed $D=0.1$ and $l=0.8$.}\label{fig-m-EOP}
\end{figure}

\subsubsection{CoP in massive BTZ}

Now after studying EoP, we study the effect of the mass parameter on the complexity of purification. Again, we expect the mass parameter $m$ would decrease CoP and it even would have a higher effect on CoP than EoP. We check this expectation in what follows. 

With the definition that we have proposed in the previous section, CoP in massive gravity theories could be evaluated as
\begin{eqnarray}
CoP&=&\int_\delta^{z_{2l+D}}\frac{dz}{z^2\sqrt{1-z^{2}+m^{2}cc_{1}z}}\int_z^{z_{2l+D}}
\frac{dZ}{\sqrt{(1-z^{2}+m^{2}cc_{1}z)(z_{2l+D}^2/Z^2-1)}}\nonumber\\
&-&\int_\delta^{z_{D}}\frac{dz}{z^2\sqrt{1-z^{2}+m^{2}cc_{1}z}}\int_z^{z_{D}}
\frac{dZ}{\sqrt{(1-z^{2}+m^{2}cc_{1}z)(z_{D}^2/Z^2-1)}}\nonumber\\
&-&2\int_\delta^{z_{l}}\frac{dz}{z^2\sqrt{1-z^{2}+m^{2}cc_{1}z}}\int_z^{z_{l}}
\frac{dZ}{\sqrt{(1-z^{2}+m^{2}cc_{1}z)(z_{l}^2/Z^2-1)}}.
\end{eqnarray}

\begin{figure}[ht!]
\centering
\includegraphics[width=0.3\textwidth]{pic9.pdf}\hspace{0.5cm}
\includegraphics[width=0.3\textwidth]{pic10.pdf}\hspace{0.5cm}
\includegraphics[width=0.3\textwidth]{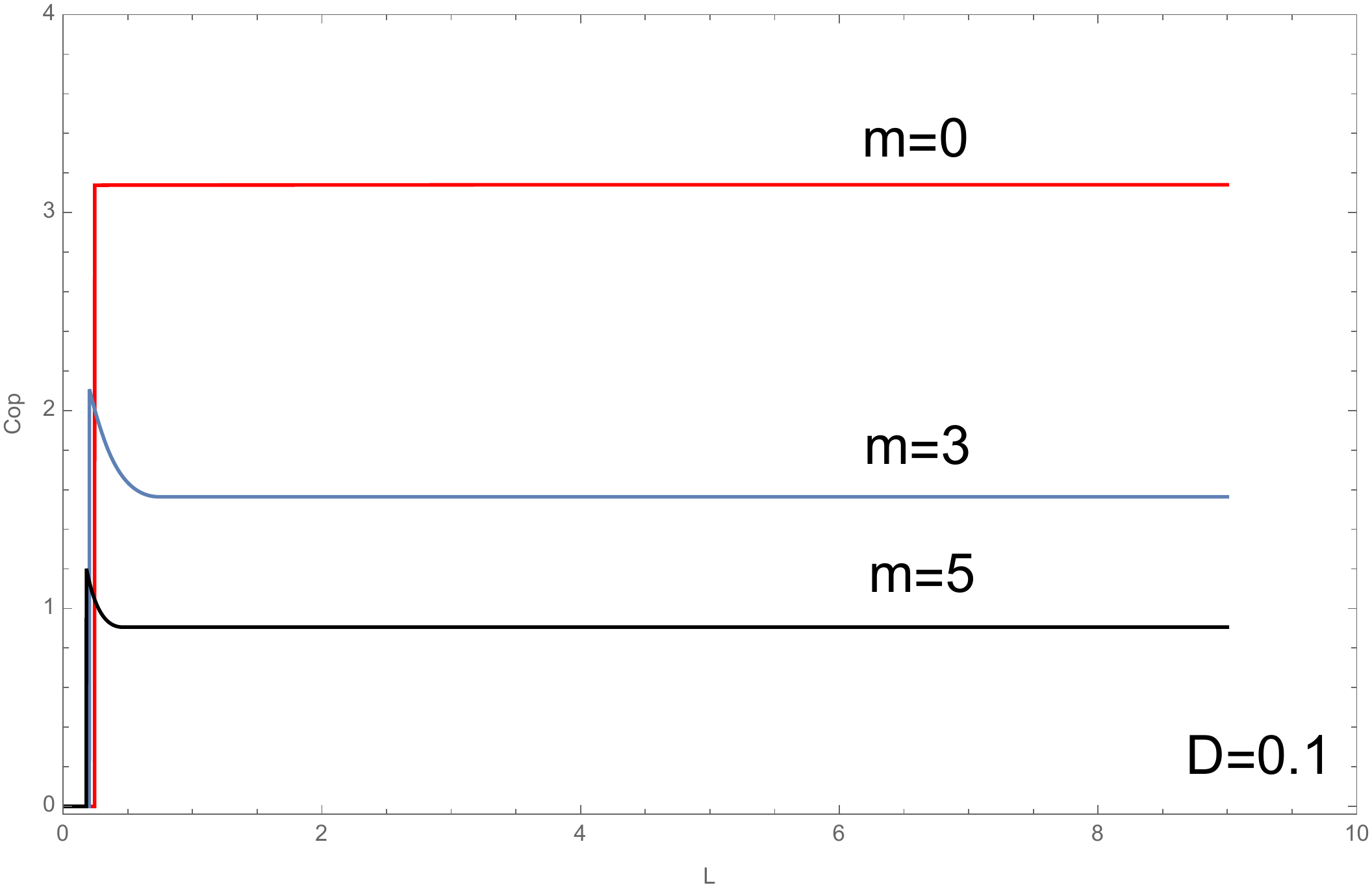}
\caption{CoP as function of  $D$ with $l=0.8$ (left)  and $l=\infty$ (middle), and CoP as a function of $l$ with $D=0.1$ (right).}\label{fig-D-COP}
\end{figure}

The numerical results of the CoP as a function of $D$ for fixed and finite $l$, and then as a function of $l$ for fixed $D$ are shown in figure \ref{fig-D-COP}.  One could see that increasing the parameter $m$ would decrease CoP. Also, one could see that $m$ has higher effect on CoP than EoP as the coefficient $a$ for the fitting function for CoP has been found to be bigger than the corresponding one for EoP.  So these results match with what we have expected from the bit thread picture.

However, note that unlike the case of Schwarzschild black hole, for the case of three dimensional bulk systems ($d=2$), the mass term makes the CoP to depend on both $D$ and $l$. So, unlike regular Schwarzschild black hole with $d=2$, CoP would not be a constant $2\pi$ for the regions where the mutual information is positive. This is because the mass term introduces some more degrees of freedom in the bulk and therefore all the degrees of freedom would not be only the \textit{topological} ones.

\begin{figure}[ht!]
\centering
\includegraphics[width=0.4\textwidth]{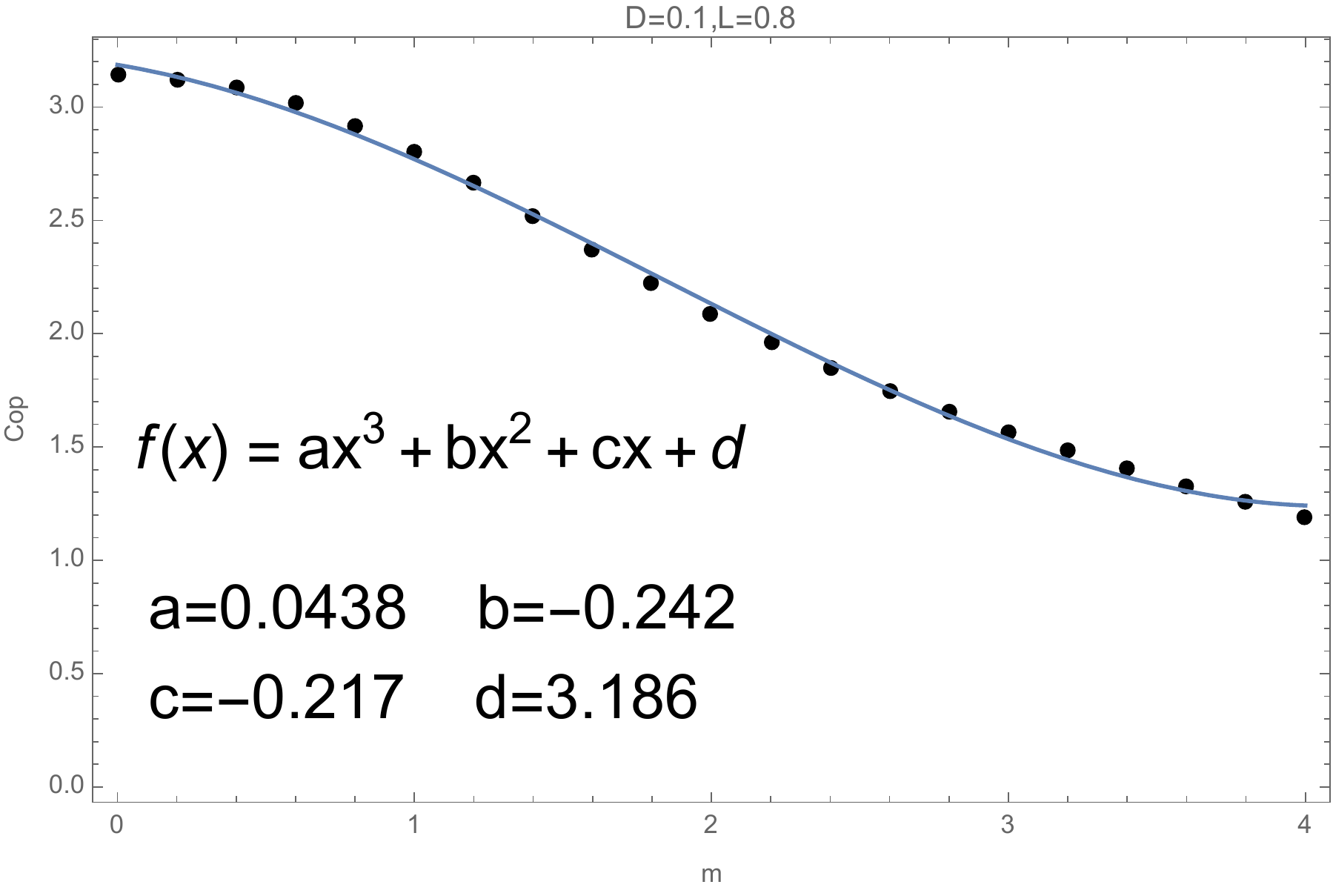}
\caption{CoP as function of $m$ with fixed $D=0.1$ and $l=0.8$.}\label{fig-m-COP}
\end{figure}

The functional dependence between $m$ and CoP for the fixed $D$ and $l$ is shown in figure \ref{fig-m-COP}. 

\subsection{Purification of charged black holes}\label{ChargedBTZpure}

To study the effect of charge on the correlation and therefore entanglement of purification, we consider the metric of Reissner Nordstr\"{o}m (RN) black hole with a planar horizon in the $\text{AdS}_{d+1}$ spacetime, \cite{Huang:2016zoz, Konoplya:2008rq} as in the following form
\begin{gather}
ds^2=\frac{1}{z^2}\left [ -f(z)dt^2+\frac{dz^2}{f(z)} + d \vec{x}^2_{d-1}      \right ] ,\nonumber\\
f(z) = 1+z^2 -2M z^d +Q^2 z^{2(d-1)},
\end{gather}
where the coordinate has been changed as $z=\frac{1}{r}$, and the AdS length scale has been set to one. Now the length of the strip and the area could be found as 
\begin{gather}
w=2 \int_\delta^{z_0} \frac{dz}{\sqrt{f}} \frac{1}{\sqrt {\frac{z_0^{2d-2}}{z^{2d-2}} -1  } } , \nonumber\\
S= \frac{2V_{d-2}}{4 G_N}   \int_\delta^{z_0}  \frac{dz}{z^{d-1} } \frac{1}{\sqrt{f}}  \frac{1} { \sqrt{1-\frac{z^{2d-2} }{z_0^{2d-2}}}}.
\end{gather}

One could observe that by increasing the charge of RN black hole, entanglement entropy decrease. From the bit thread picture we expect that both EoP and CoP would decrease as well.
\begin{figure}[ht!] 
\centering
\includegraphics[width=6.cm]{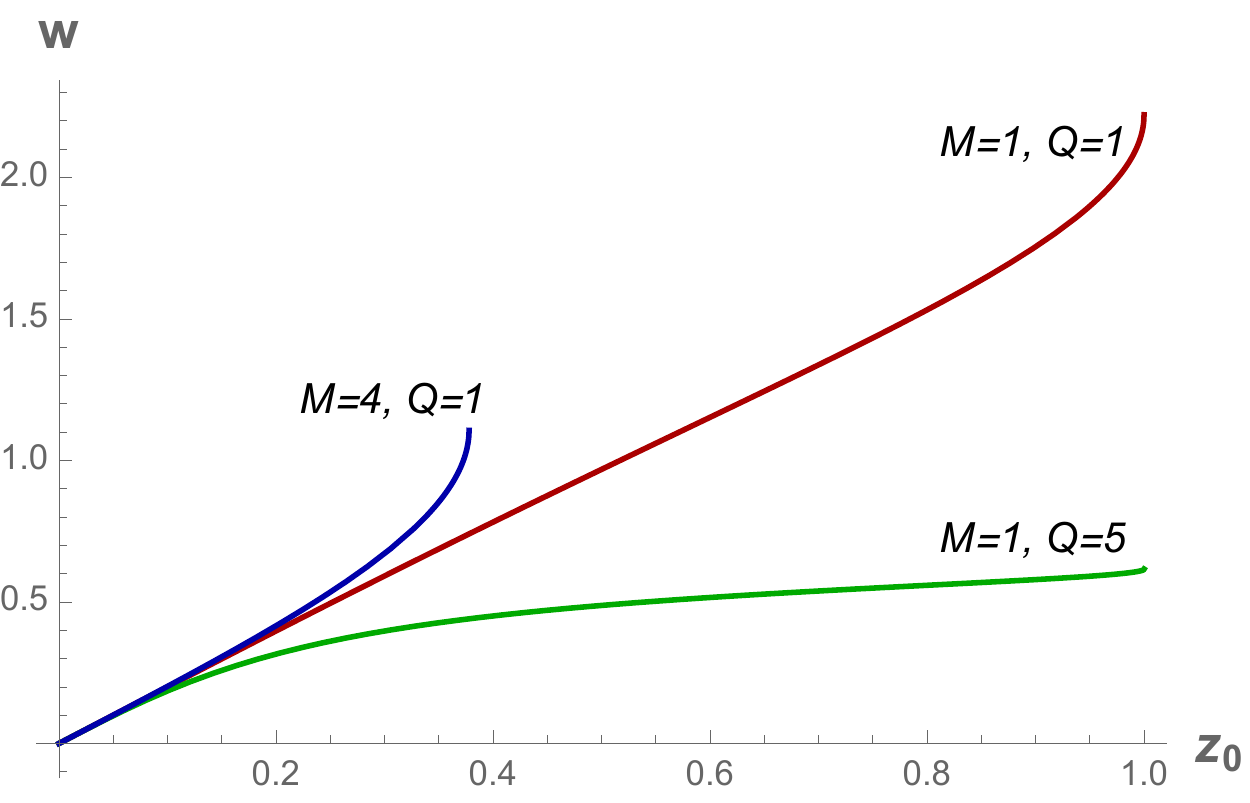} \ \ \ \ \ \  \ 
\includegraphics[width=6cm]{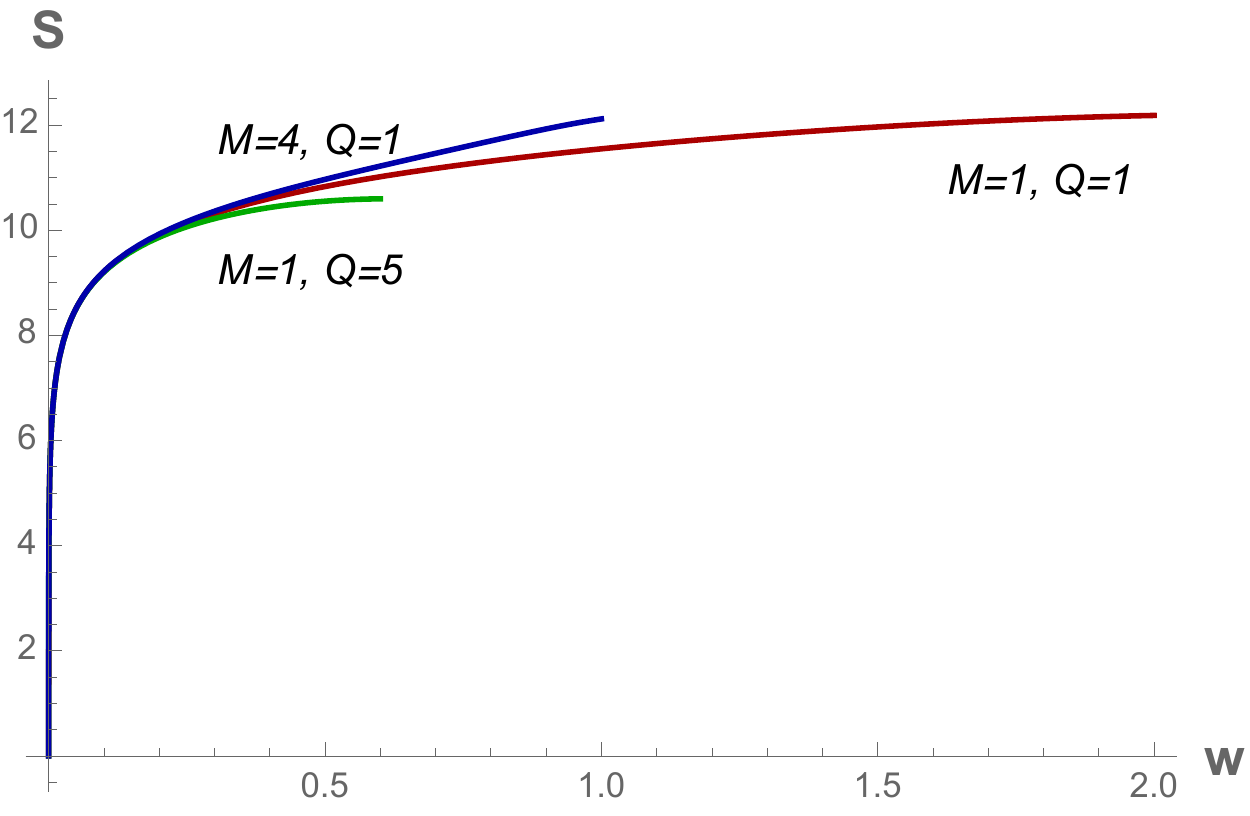}
  \caption{ $w$ versus $z_0$ and $S$ versus $w$ for various $M$ and $Q$. }
 \label{fig:RN}
\end{figure}

The plot of entropy versus width of the strip $w$ has been shown in figure \ref{fig:RN}.
One could notice that increasing $Q$ could provoke a phase transition, while increasing $M$ could prevent it and so they have an opposite effect on phase transitions, entropy and EoP.

For calculating the entanglement of purification, the area of minimum cross section $\Gamma=\int_{z_D}^{z_{2l+D}} \frac{dz}{z^{d-1} \sqrt{f(z)} }$ could then be found. As it could be difficult for the RN case, we turn to the simpler charged BTZ black hole.

\subsubsection{EoP in charged BTZ}

The metric of charged BTZ black hole is \cite{Carlip:1995qv}
\begin{equation}
ds^2=\frac{1}{z^2}[-f(z)dt^2+\frac{dz^2}{f(z)}+dx^ { 2 }]~~~\mathrm{where} ~~~f(z)=1 - z ^ { 2 } + \frac{q^{2}}{2}z^{2}\ln{z}.
\label{MetricChargedBTZ}
\end{equation}%

For this metric we have
\begin{gather}
w=2\int_0^{z_0} \frac{dz}{\sqrt{f(z) \left(f(z_0)\frac{z_0^2}{z^2}-1  \right)} },\nonumber\\
S=\frac{2V_0}{4G_N} \int_\delta^{z_0} dz \frac{1}{\sqrt{f(z)  \left(1-\frac{z^2}{z_0^2 f(z_0)} \right) } }.
\end{gather}

\begin{figure}[ht!]
\centering
\includegraphics[width=0.5\textwidth]{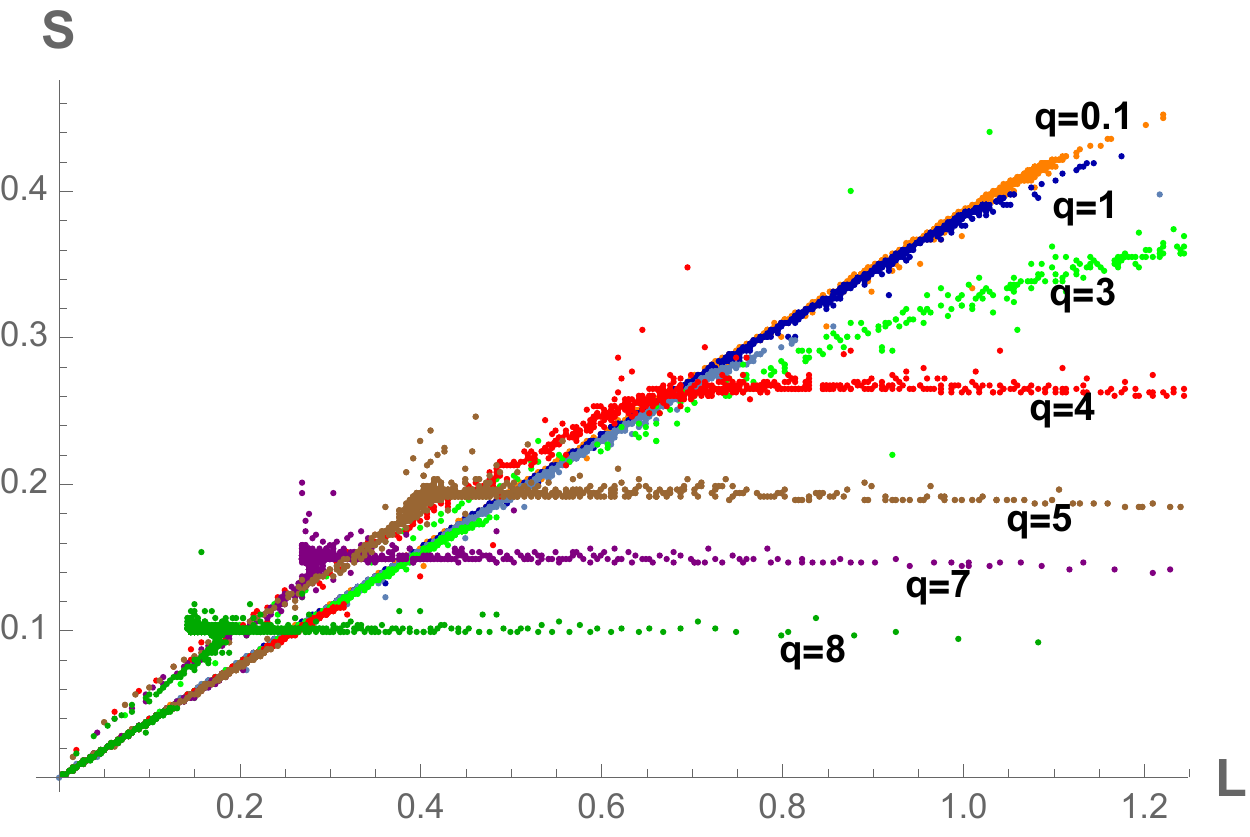}\hspace{1cm}
\caption{The relationship between $S(w)$ and $w$ for charged BTZ black hole.}\label{fig:MychargedPlot}
\end{figure}

The plot for the relationship between entanglement entropy and the width of strip for the charged BTZ is shown in \ref{fig:MychargedPlot}.  Similar to the RN case, one can see that increasing $q$, the charge of BTZ would decrease entanglement entropy and as we will see it would decrease mutual information, EoP and CoP. Therefore, we propose that decreasing of correlations among two regions by the effect of the same sign charge, would be a universal behavior.

\begin{figure}[ht!]
\centering
\includegraphics[width=0.4\textwidth]{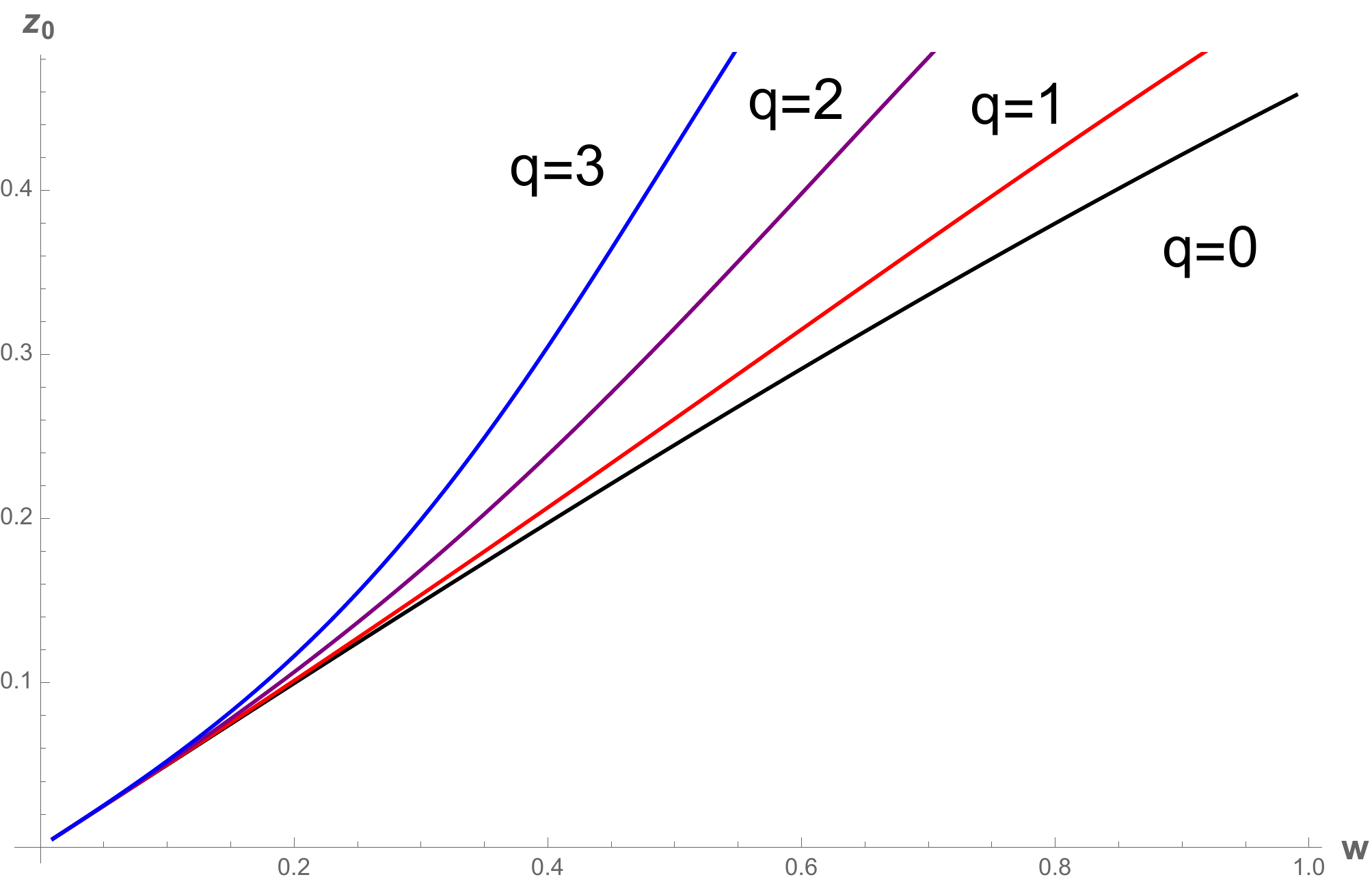}\hspace{1cm}
\includegraphics[width=0.4\textwidth]{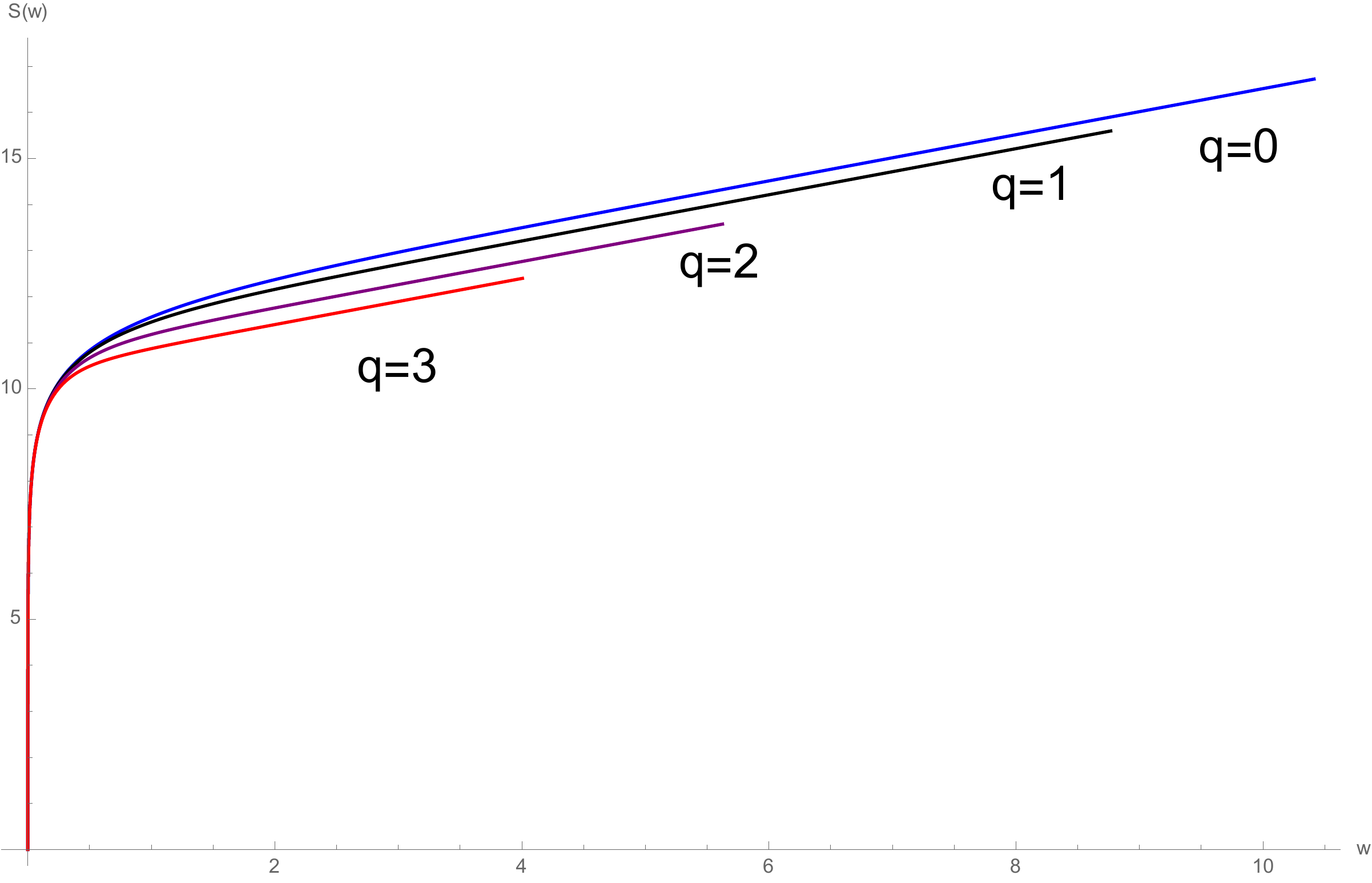}
\caption{The width of the strip (left) and the related holographic entanglement entropy (right) for different $q$.}\label{fig-w-z02}
\end{figure}

Also, as shown in figure \ref{fig:MychargedPlot}, for the case of charged BTZ black hole, we detected a first order butterfly shaped phase transition. More plots are shown in figure \ref{fig-w-z02}.

\begin{figure}[ht!]
\centering
\includegraphics[width=0.35\textwidth]{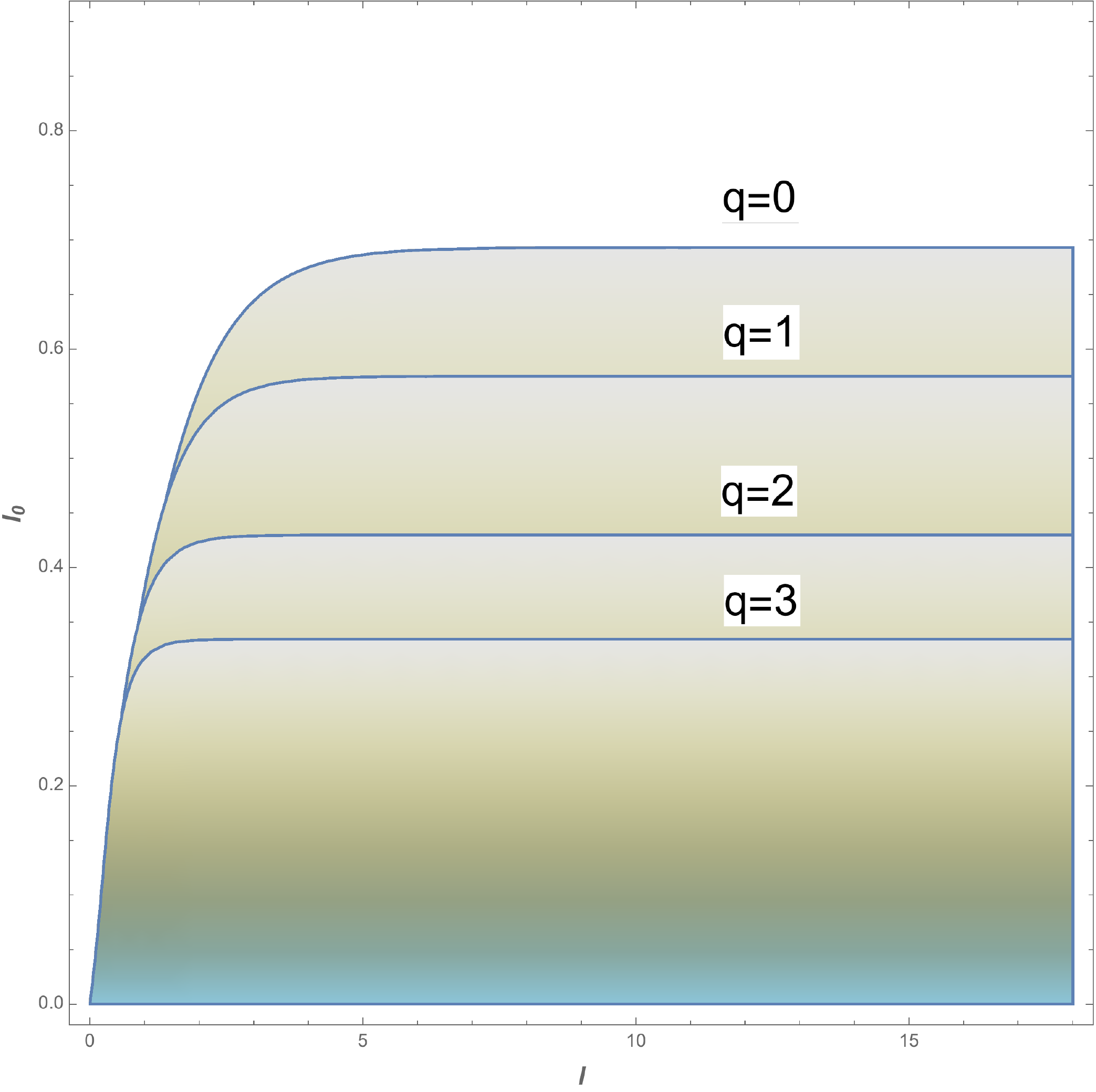}
\caption{For each $q$, the region below the lines have non-vanishing EoP.}\label{fig-LDeop00}
\end{figure}

\begin{figure}[ht!]
\centering
\includegraphics[width=0.4\textwidth]{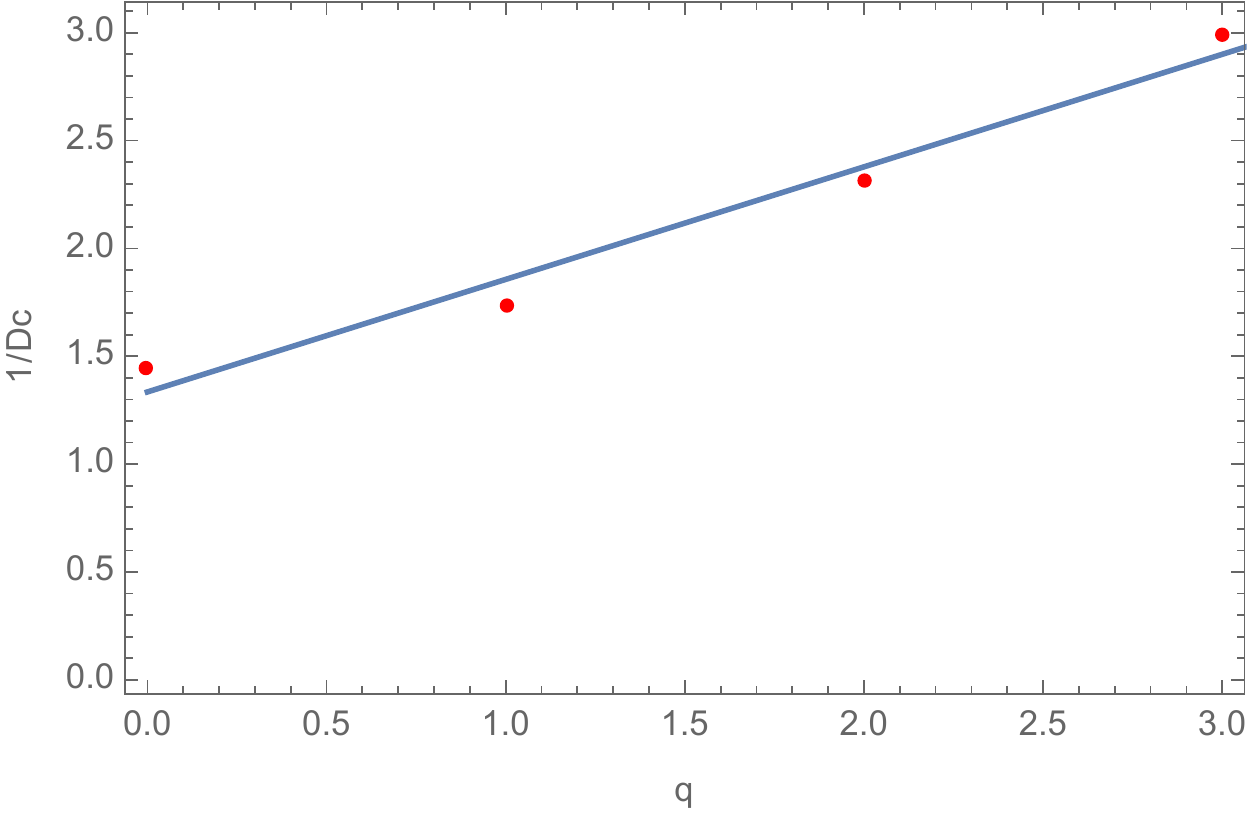}\hspace{1cm}
\includegraphics[width=0.4\textwidth]{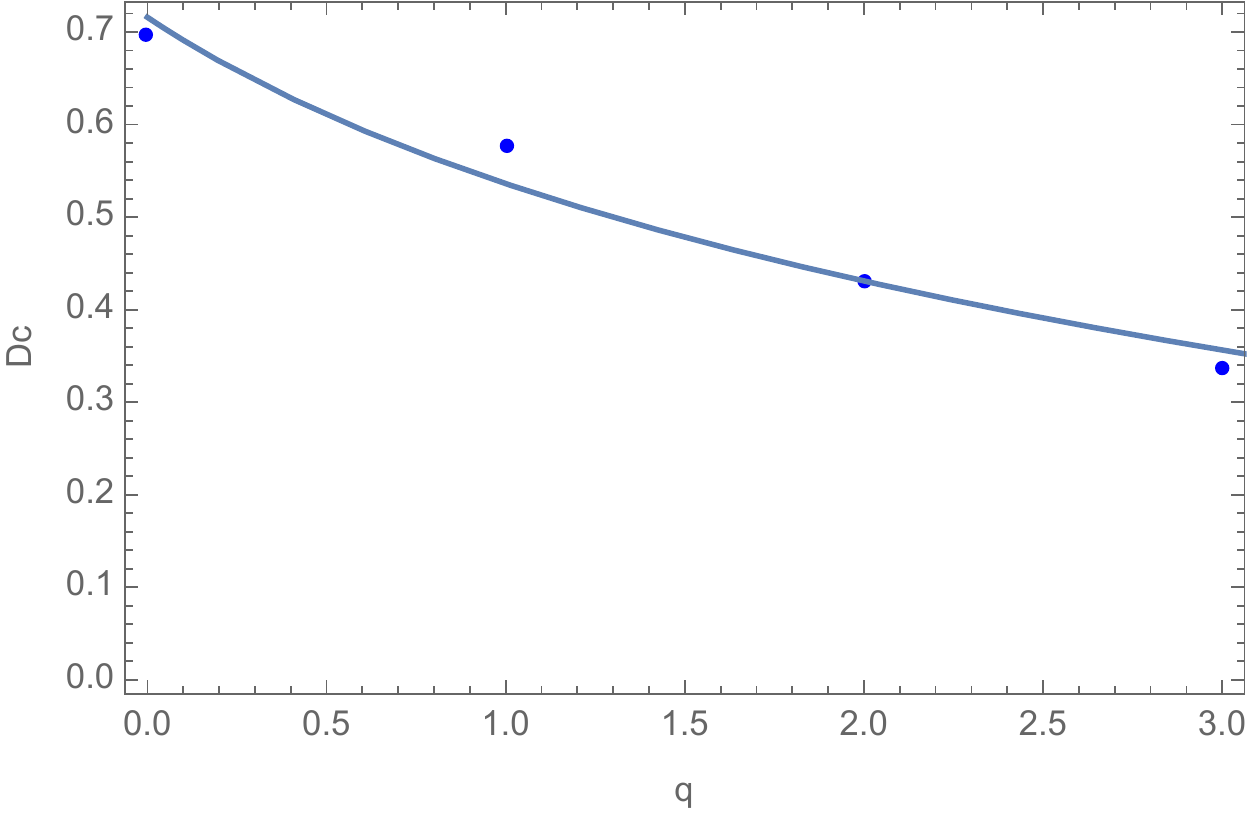}
\caption{Left: The relationship between $1/D_{c}$ and $q$. Right: The relationship between $D_{c}$ and $q$.}\label{fig-Dcfit1}
\end{figure}

In figure \ref{fig-LDeop00}, we show the regions with non-vanishing EoP. As $q$ increases, $D_{c}(q, l\to\infty)$ decreases. Then, we fit $D_{c} (q,l\to\infty)$ as a function of $q$ which satisfies the following relation
\begin{equation}
D_{c}^{-1}(l\to\infty)\simeq1.33+0.521 q,
\end{equation}
or
\begin{equation}
D_{c}(l\to\infty)\simeq0.716-0.259\log q.
\end{equation}
The fitting line is shown in figure \ref{fig-Dcfit1}.

For the cases that $D$ and $l$ are smaller than $D_{c}(d, l)$ and $l_c$, and therefore EoP is non-zero, the area of the surface $\Gamma$ in charged BTZ background could be derived as
\begin{equation}
\Gamma = \int _ { z _ { D } } ^ { z _ { 2 l + D } } \frac { d z } { z \sqrt { 1 - z ^ { 2 } + \frac{q^{2}}{2}z^{2}\ln{z}} }.
\end{equation}

\begin{figure}[ht!]
\centering
\includegraphics[width=0.3\textwidth]{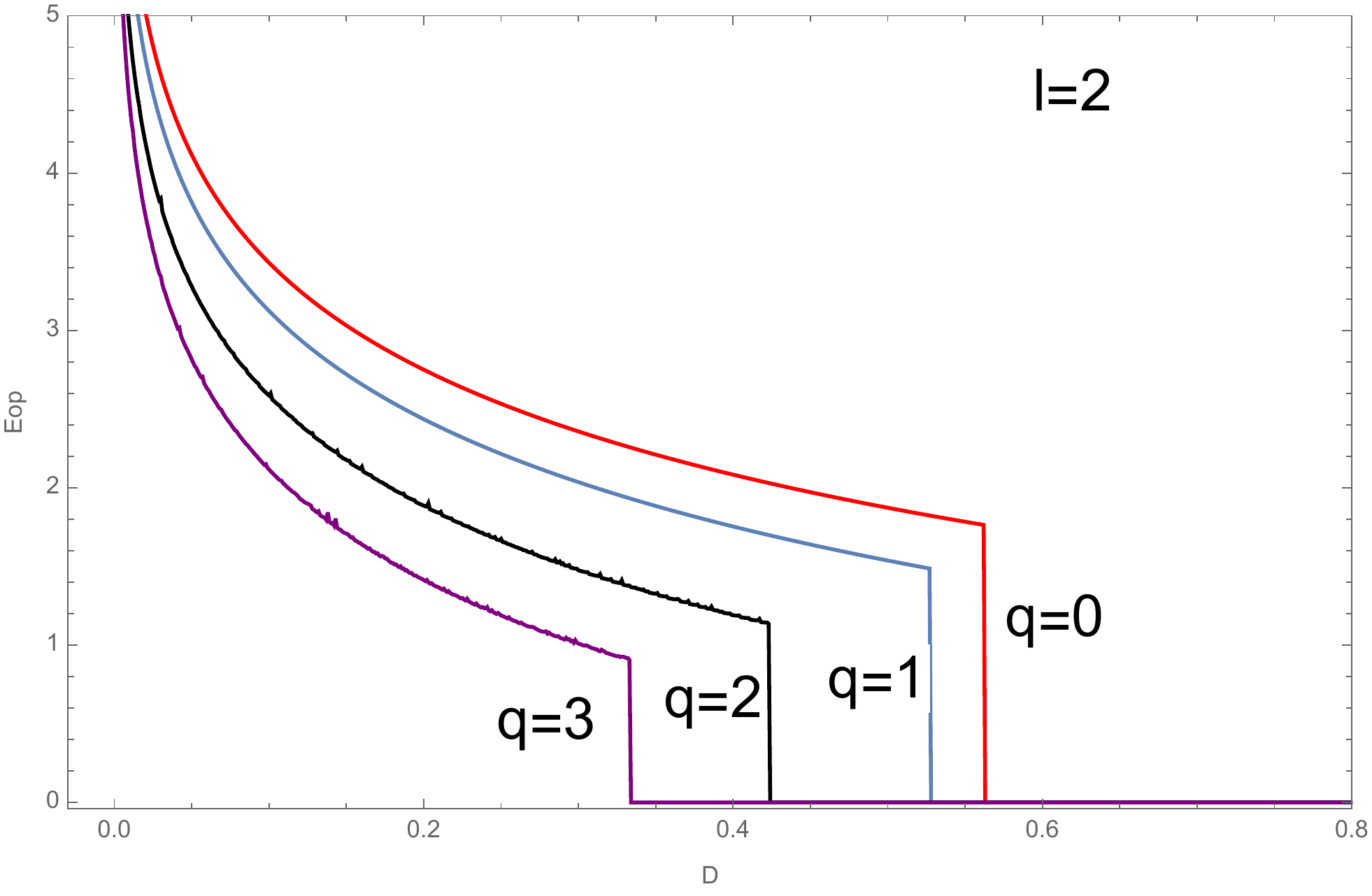}
\includegraphics[width=0.3\textwidth]{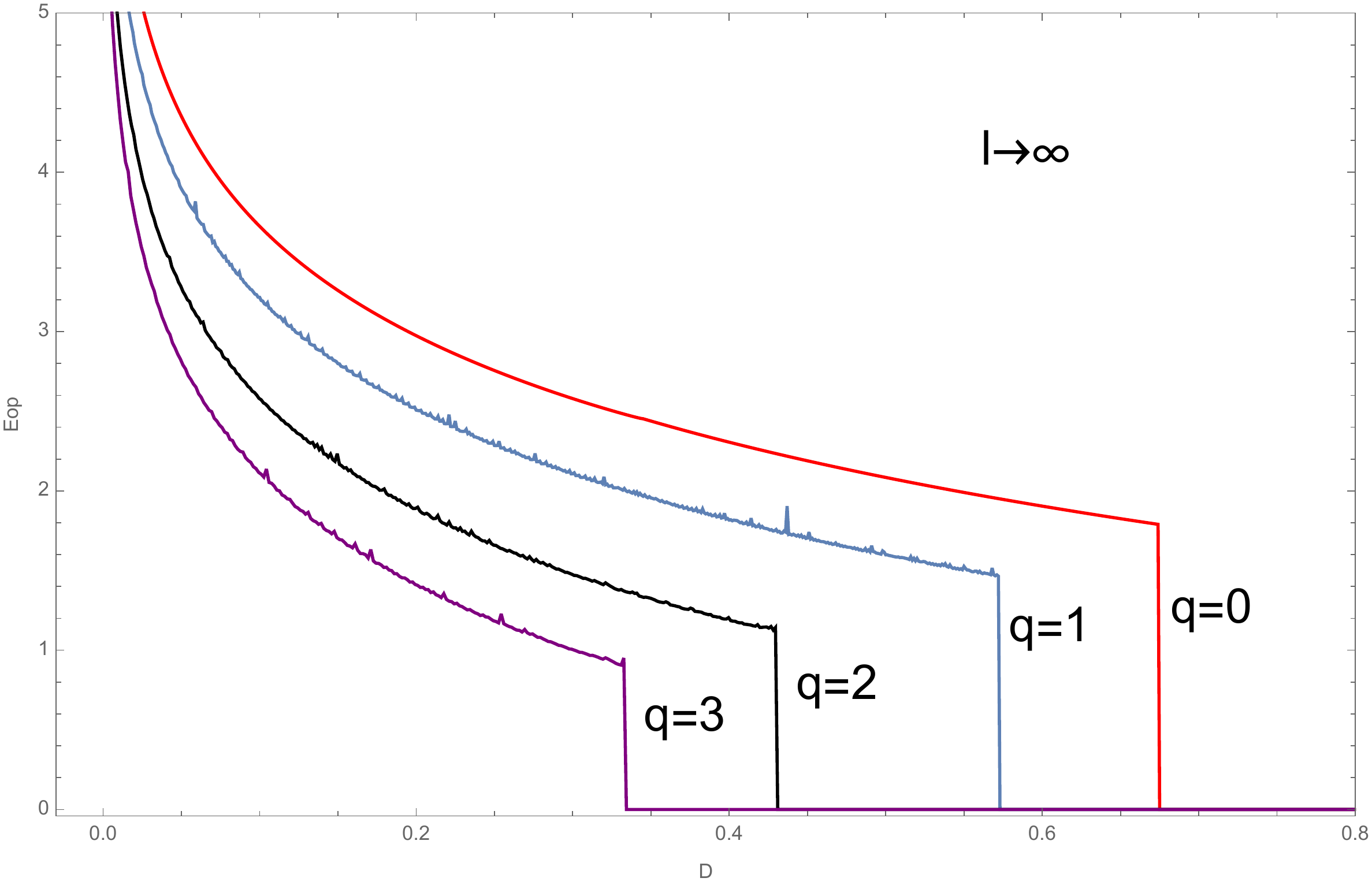}
\includegraphics[width=0.3\textwidth]{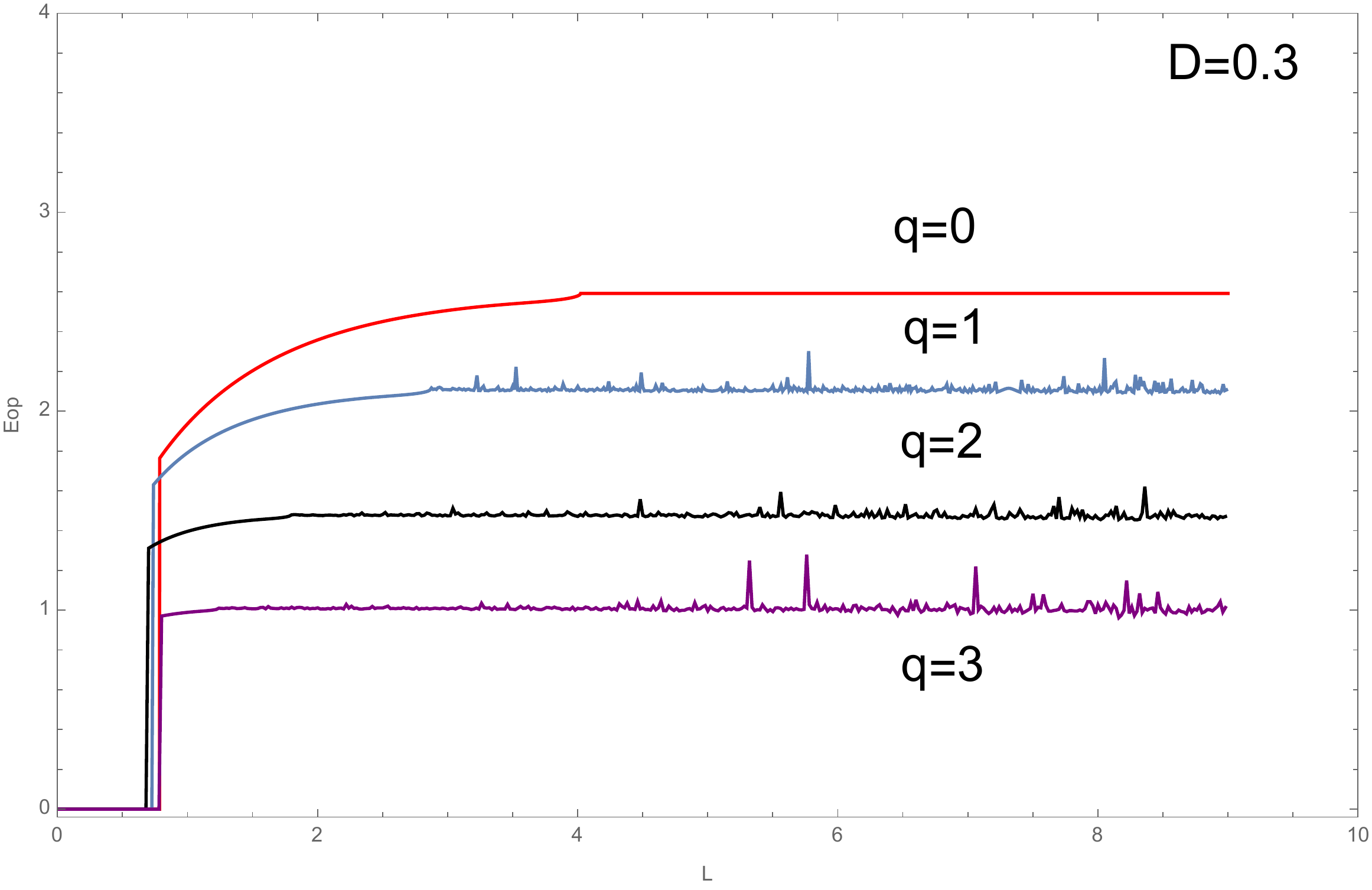}
\caption{The relationship between EoP and $D$ with $l=2$ (left) and l=$\infty$ (middle). The relationship between EoP and $l$ with $D=0.3$ (right).}\label{fig-D-EOP2}
\end{figure}

The  EoP as a function of $D$ for a fixed and finite $l$, and then as a function of $l$ for fixed $D=0.3$ are shown in figure \ref{fig-D-EOP2}. One could also see that relative to the mass parameter $m$, charge has bigger effect on decreasing EoP, as the same sign charges on the two sides of the boundary could greatly limit the correlations among the bit threads and therefore could decrease EoP or CoP a greatly.  

\begin{figure}[ht!]
\centering
\includegraphics[width=0.4\textwidth]{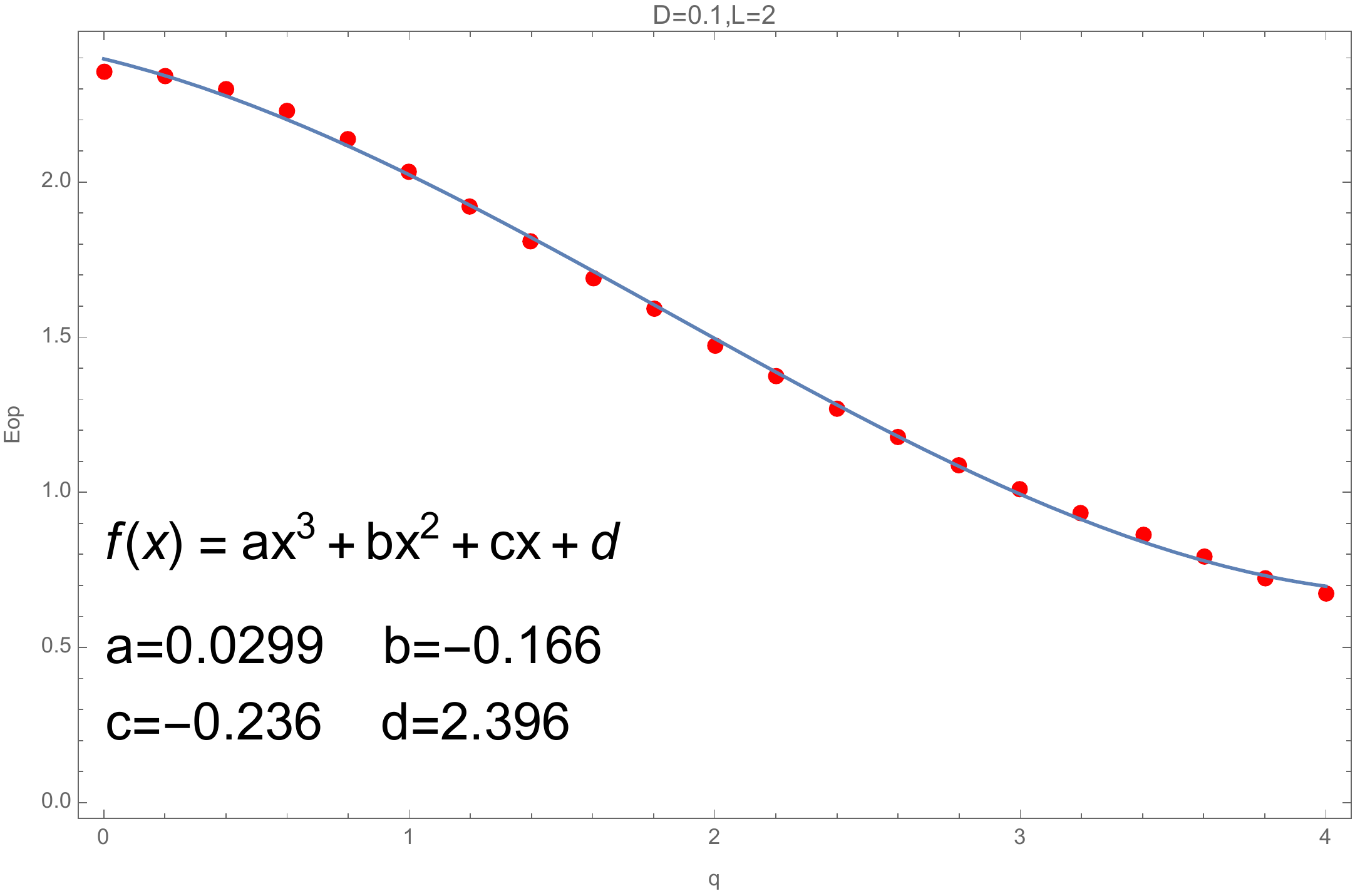}\hspace{0.5cm}
\includegraphics[width=0.4\textwidth]{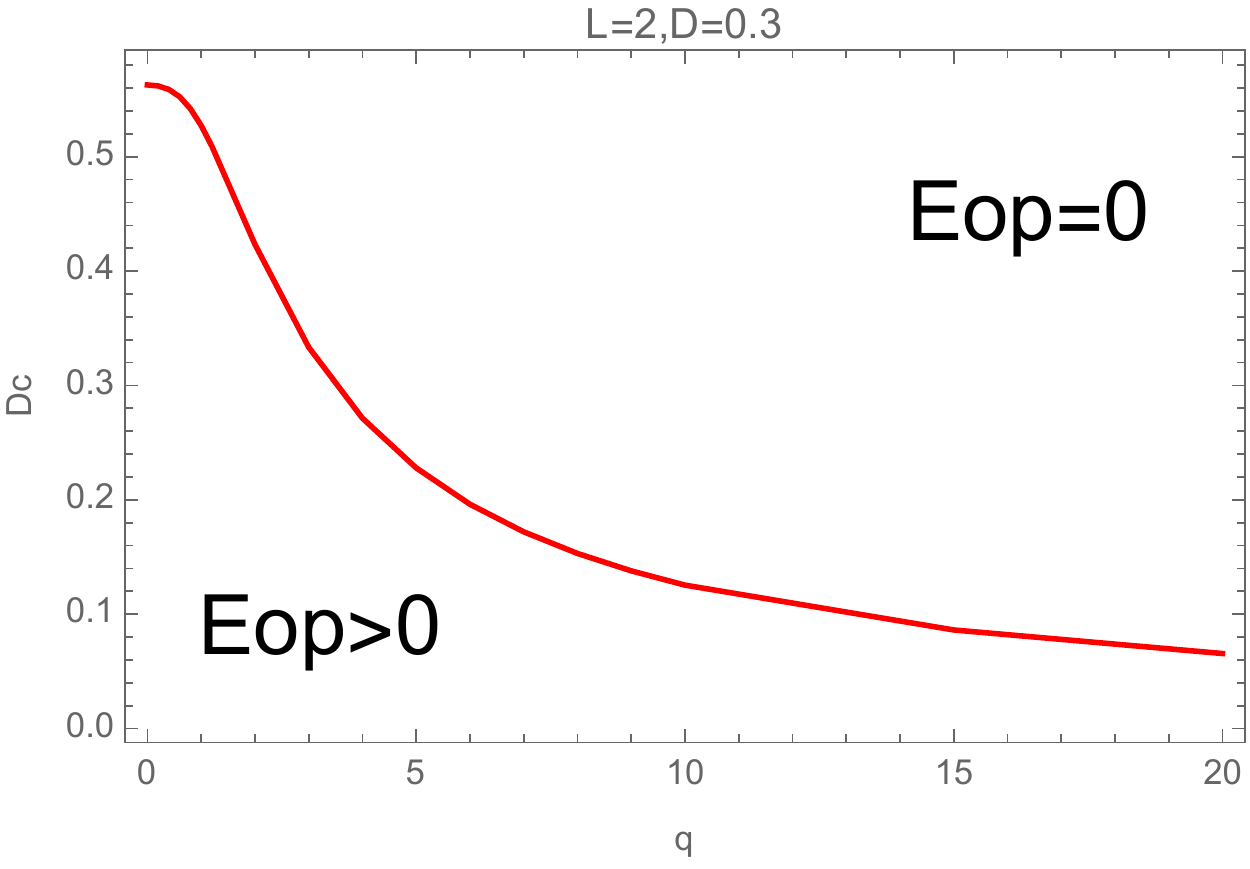}
\caption{Left: EoP as function of $q$ with fixed $D=0.1$ and $l=0.8$. Right: Phase diagram in the $q-D_{c}$ plane.}\label{fig-q-EOP}
\end{figure}

The effect of $q$ on EoP with fixed $D$ and $l$ are shown in the left plot of figure \ref{fig-q-EOP}. From that it would be obvious that EoP decreases as $q$ increases as we have expected. In the right plot of figure \ref{fig-q-EOP}, we figure out the phase diagram in $q-D_{c}$ plane, namely, above the line, we have $\text{EoP}=0$ and below the line, EoP is positive.

\subsubsection{CoP in charged BTZ}

\begin{figure}[ht!]
\centering
\includegraphics[width=0.4\textwidth]{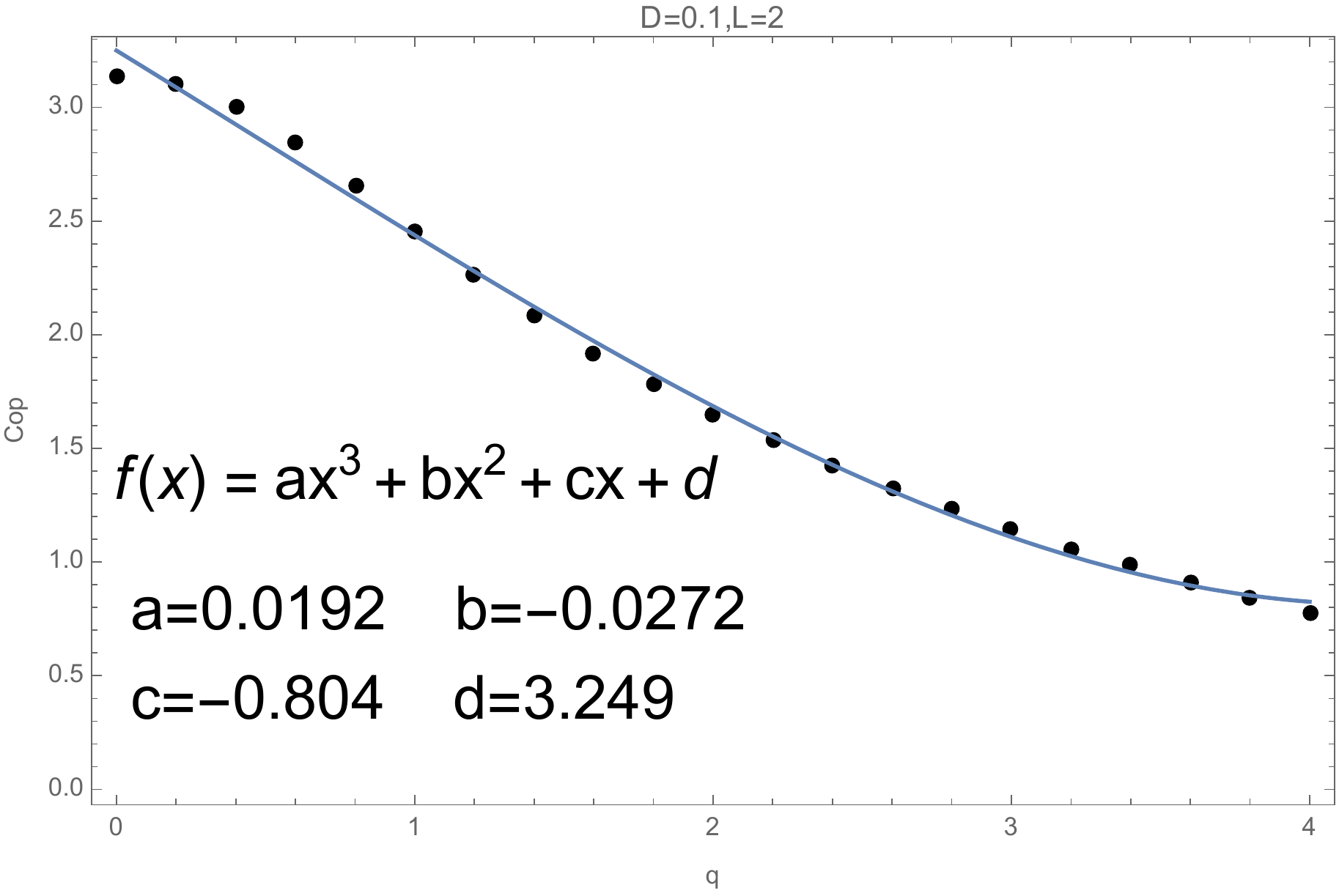}
\caption{CoP as function of $q$ with fixed $D=0.1$ and $l=2$.}\label{fig-q-COP}
\end{figure}
Next, the CoP in charged BTZ black hole could be evaluated as
\begin{eqnarray}
CoP&=&\int_\delta^{z_{2l+D}}\frac{dz}{z^2    \sqrt{1 - z ^ { 2 } + \frac{q^{2}}{2} z^{2}\ln{z}    }}      \int_z^{z_{2l+D}}
\frac{dZ}{\sqrt{(1 - z ^ { 2 } + \frac{q^{2}}{2} z^{2}\ln{z})(z_{2l+D}^2/Z^2-1)}}\nonumber\\
&-&\int_\delta^{z_{D}}\frac{dz}{z^2\sqrt{1 - z ^ { 2 } + \frac{q^{2}}{2} z^{2}\ln{z}}}\int_z^{z_{D}}
\frac{dZ}{\sqrt{(1 - z ^ { 2 } + \frac{q^{2}}{2} z^{2}\ln{z})(z_{D}^2/Z^2-1)}}\nonumber\\
&-&2\int_\delta^{z_{l}}\frac{dz}{z^2\sqrt{1 - z ^ { 2 } + \frac{q^{2}}{2}  z^{2}\ln{z}}}\int_z^{z_{l}}
\frac{dZ}{\sqrt{(1 - z ^ { 2 } + \frac{q^{2}}{2} z^{2}\ln{z})(z_{l}^2/Z^2-1)}}.
\end{eqnarray}

\begin{figure}[ht!]
\centering
\includegraphics[width=0.4\textwidth]{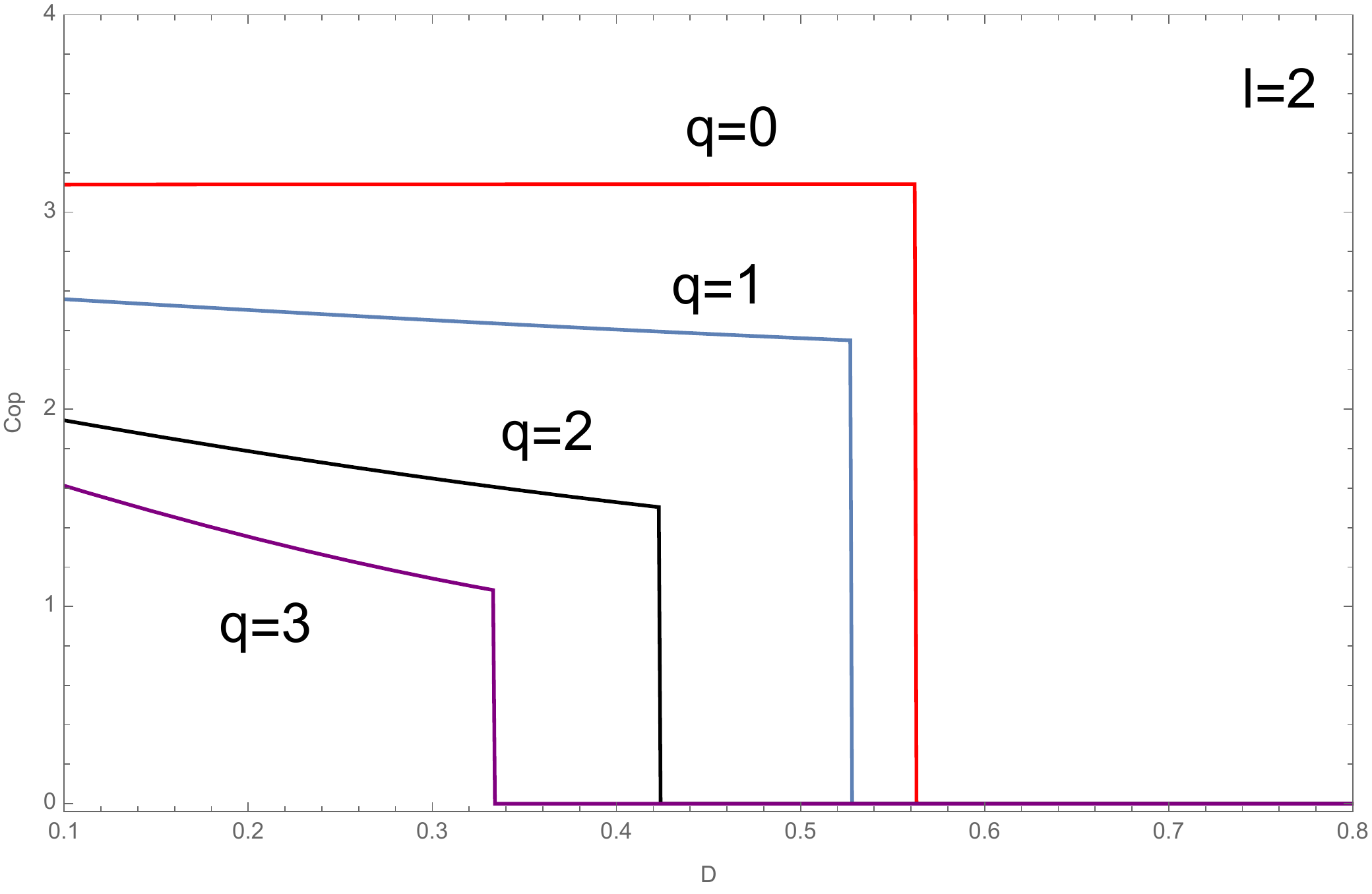}\hspace{1cm}
\includegraphics[width=0.4\textwidth]{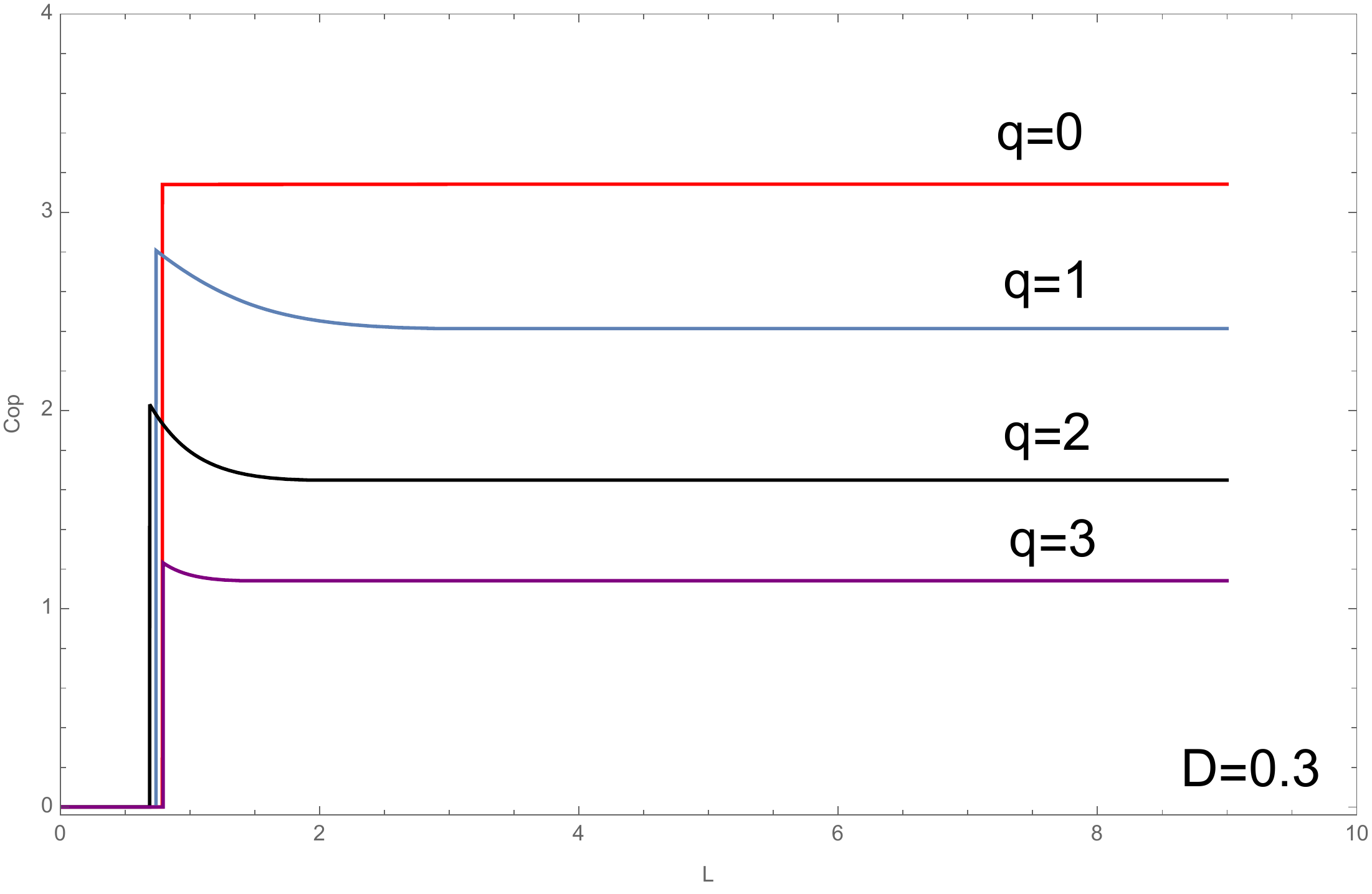}
\caption{Left: The relationship between CoP and $D$ with $l=2$. Right: The relationship between CoP and $l$ with $D=0.3$.}\label{fig-D-COP2}
\end{figure}

The numerical results for the behavior of CoP as a function of $D$ for fixed finite $l$ and as a function of $l$ for fixed $D$ are shown in figure \ref{fig-D-COP2}. Similar to the massive BTZ  black hole, CoP in charged BTZ black hole would not be a constant $2\pi$. The effect of $q$ on CoP with fixed $D$ and $l$ is shown in figure \ref{fig-q-COP}.

\subsection{Purification of multipartite systems}\label{pureMultipartite}
Another class of geometries that one could study is the $\text{AdS}_3$ black hole with $n$ sides and genus $g$. These are the extension of BTZ black holes by quotienting pure $\text{AdS}_3$ by a discrete group of isometries. The entanglement of purification for these black holes has been studied in \cite{Caceres:2018blh}.

In \cite{Umemoto:2018jpc}, the multipartite entanglement of purification $\Delta_P$ has been defined as
\begin{gather}
\Delta_P(\rho_{A_1:...:A_n}) := \underset{\ket{\psi}_{A_1 A'_1 ... A_n A'_n}} {min} \sum_{i=1}^n S_{A_i A'_i},
\end{gather}
where the minimization is over all purification of $\rho_{A_1...A_n}$.
In \cite{Bao:2018gck}, the conditional mutual information for multi-partite states have also been studied. 

\begin{figure}[ht!] 
\centering
\includegraphics[width=4.5cm]{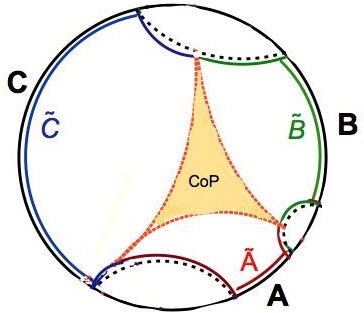}
  \caption{The three orange dashed lines are the entanglement wedge dual to the multipartite entanglement of purification and the corresponding volume is dual to the complexity of purification for the three-partite system.}\label{threeCoPconstrict}
\end{figure}

Note that in the dual bulk holographic definition, the multipartite entanglement wedge cross section $ \Delta_W$ could also be defined. For instance, as shown in figure \ref{threeCoPconstrict}, for the three subsystems of $A$, $B$, and $C$ on a boundary of $\partial M$, the entanglement wedge $M_{ABC}$, shown as orange lines, could be defined as a region of $M$ with the boundary $A$, $B$, $C$ and the Ryu-Takayanagi surface $\Sigma_{ABC}^{min}$ (for $\rho_{ABC}$) \cite{Umemoto:2018jpc} such that

\begin{gather}
\partial M_{ABC}= A \cup B \cup C \cup \Sigma_{ABC}^{min}.
\end{gather}

Note that as for the two-partite case, similar measures such as tripartite information for three (or more)-partite systems could also be defined as \cite{Hayden:2011ag}
\begin{gather}
\tilde{I}_3(A:B:C) := S_A+S_B+S_C-S_{AB}-S_{BC}-S_{CA}+S_{ABC}\nonumber\\
=I(A:C)+I(A:B)-I(A:BC),
\end{gather}
which is actually the generalization of the mutual information.

However, one should note that this quantity could be positive, zero or negative and therefore another quantity, ``the relative entropy" between the original state and its local product state has been defined as
\begin{gather}
I(A:B:C):= S(\rho_{ABC} || \rho_A \otimes \rho_B \otimes \rho_C)= S_A+S_B+S_C-S_{ABC},
\end{gather}
which could also be generalized to the n-partite states. This quantity is always positive and therefore it is a better measure to use for studying multipartite correlations \cite{Umemoto:2018jpc}. So this is the quantity that we should work with while studying the CoP of n-partite systems.

For the general n-partite state, one could also write \cite{Umemoto:2018jpc}
\begin{gather}
I(A_1: ... : A_n) = I(A_1: A_2)+I (A_1 A_2: A_3)+ ...I(A_1 ... A_{n-1}: A_n).
\end{gather}

Here we only consider more than two strips in the background of ``Schwarzchild AdS black brane", similar to the previous section.

\begin{figure}[ht!] 
\centering
\includegraphics[width=5cm]{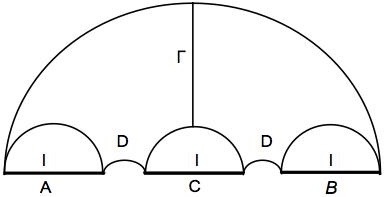}
  \caption{Three strips and the entanglement wedge between the furthest regions, $A$ and $B$. Here $n=2$.}
 \label{fig:threeregion}
\end{figure}

So for $n$ strips in the arrangement of figure \ref{fig:threeregion} where we had $n=1$, the CoP could be found as
\begin{gather}
CoP_{A,B} \left((n+1) l + n D\right)=2 n \pi-\frac{1}{\delta} (n i \pi),
\end{gather}

One can see that the universal part is always an even factor of $\pi$. However, note that if $(n+1)l+nD >D_c$, for the two furthest regions where we call $A$ and $B$, EoP and as the result, CoP would be zero. So the above result is valid only when $(n+1)l+nD <D_c$. These calculations could be done for other backgrounds specifically those multi-boundary wormholes and therefore the effect of higher genus on CoP could be examined further to distinguish further between various definitions of holographic complexity of purification.

\section{Operational and bit thread interpretations}\label{pureOperational}

Here we would like to provide more explanations of what we have observed in the behavior of EoP and CoP in the previous sections, specifically for the case of massive and charged solutions. For doing that, we could use different holographic tools and quantum information concepts such as operational studies in \cite{doi:10.1063/1.1498001}, and also ideas from convex optimization such as bit thread and max-flow, min-cut theorems \cite{Headrick:2017ucz}.

To understand the nature of correlations in each example one could study the problem using operational perspectives, specifically from the point of view of \textit{resource theories}. One of these resource theories is the ``Local Operation and Classical Communication" (LOCC). 

The LO (Local Operations) are in the following form
\begin{gather} 
\rho \to \sum_{i,j} (A_i \otimes B_j). \rho . (A_i^ \dagger \otimes B_j^\dagger ), 
\end{gather}
where
\begin{gather}
\sum_i A_i ^\dagger A=1, \ \ \ \ \sum_j B_j^\dagger B_j=1.
\end{gather}
These operations include projection measurements and unitary transformations. 

One then could add classical communications (CC) between $A$ and $B$, where the combination of these operations are called LOCC. One example of LOCC is the quantum teleportation.  

For the operational definition of EoP, however, one should use the LOq (Local Operations plus some small number of communications). The entanglement of purification is equal to the ``Entanglement Cost" for the LOq process. It would be equal to the number of EPR pairs needed to create $\rho_{AB}$ via LOq. For the definition of CoP one then could take the ``Computational Cost for the LOq process" and similarly connect it with the number of gates or EPR pairs.

Note that the difference between the entanglement cost $E_C (\rho_{AB})$ and entanglement distillations $E_D( \rho_{AB})$ is that the first one is the number of EPR pairs needed to create $\rho_{AB}$ via LOCC while the later one is the number of EPR pairs we can create from $\rho_{AB}$ via LOCC \cite{Takayanagi:2017knl}. Then, the complexity of distillation would be the minimum necessary number of operations via Loq processes or the EPR pairs which are shared between the two regions and contribute to the computations. The distinct characteristics of this measure for different field theory models could be examined numerically similar to the case done for EoP \cite{Bhattacharyya:2018sbw}. The effects of charge and mass on this computational cost for the LOq process and therefore their effects on CoP could be comprehended this way.

Note that in quantum information studies, there would be two sets of states, the ``separable states" which could be prepared by LOCC and are available for ``free" and then the "entangled" or ``precious" operations or states which would be used as resources for various computations \cite{Hubeny:2018ijt}. For purification process and then calculating CoP, we need the ``precious" operations (corresponding to the uncomplexity \cite{Brown:2017jil} of the system).

As the connection between EoP, CoP and LOCC would deserve further studies, here we provide additional statements. First, note that we observed in various examples, such as charged or massive BTZ case, that the behavior of EoP and CoP showed always similar behavior, which could imply the same states would participate in both quantities, which could also be distinguished using LOCC.

The connections between EoP and CoP could become more clear by getting back to the definitions of formation cost and subregion complexity. The main question is how difficult would be to create a bipartite quantum state $\rho$ in the asymptotic regime, from an initial supply of EPR-pairs by local operations and asymptotically vanishing communication (LOq) which are local operations with $o(n)$ communication in the asymptotic regime.

Strictly speaking the \textit{formation cost}, $E_{LOq}$ is \cite{Terhal:2002uua}
\begin{gather}\label{ELOQ}
E_{LOq}(\rho)= \lim_{\epsilon \to 0} \text{inf} \Big \{ \frac{m}{n} \Big | \exists  \ \mathcal{L}_{LOq},  D( \mathcal{L}_{LOq} (\ket{ \Psi_-}  \bra{ \Psi_-} ^ {\otimes m}), \rho^{\otimes n} ) \leqslant \epsilon \Big\}
\end{gather}
where $ \ket{\Psi}$ is the singlet state in $\mathcal{H}_2 \otimes \mathcal{H}_2$, $\mathcal{L}_{LOq}$ is a local superoperator using $o(n)$ quantum communication, $D$ is the Bures distance $D(\rho, \rho')= 2 \sqrt{1-F(\rho, \rho')}$ and the square-root-\textbf{fidelity} is $F(\rho, \rho')= \text{Tr} ( \sqrt{\rho^{1/2} \rho' \rho^{1/2}   })$. 

Then the entanglement of purification, $E_p(\rho)$, would be
\begin{gather}
E_{LOq}= \lim_{n\to \infty} \frac{E_p(\rho ^{\otimes n} ) }{n}. 
\end{gather}

Also in section III of \cite{Alishahiha:2015rta}, the direct relationship between ``fidelity susceptibility" and subregion complexity, specifically for a thermal mixed state has been shown. Specifically, for a subsystem with spherical symmetry in the ground state of a CFT, the thermal fidelity could be written as
\begin{gather}
F(2\pi \ell, \lambda_1, \lambda_2)=1-2\pi \ell \chi_\lambda \frac{\delta \lambda^2}{8}+ \mathcal{O}(\delta \lambda^3),
\end{gather}
where $\chi_\lambda= \partial^2_\lambda \mathcal{F}_{\text{th}}$ is the fidelity susceptibility in terms of the thermal free energy $\mathcal {F_{\text{th}} }$. In \cite{Alishahiha:2015rta}, it was shown that $\chi_\lambda$ behaves as
\begin{gather}
\chi_\lambda \sim \frac{R^d}{\epsilon^d} \left ( 1+c_2 \frac{\epsilon^2}{\ell^2}+c_4 \frac{\epsilon^4}{\ell^4}+.... \right),
\end{gather}
which agrees with the behavior of subregion complexity. Therefore, from \ref{ELOQ} and the connection between fidelity susceptibility and subregion complexity, one could see the direct connection between EoP and CoP too. This is another argument to sketch the connection between EoP and CoP.

Using average fidelity, the connection between EoP, CoP and LOCC could be studied even further. For a fixed set $\mathcal{S}={p_i, \ket{\psi}}$, and for a measurement (POVM) $M={M_a}$ with the guessing strategy $G: a \to \ket{\phi_a}$, the average fidelity is defined as \cite{PhysRevLett.100.070503,PhysRevA.97.022314}
\begin{gather}
\mathbb{F} (\mathcal{S} \big | \mathbf{M}, \mathbf{G}) = \sum_{i,a} p_i \bra{\psi_i} M_a \ket{\psi_i}  |\braket{\psi_i | \phi_a}|^2,
\end{gather}
which measures the ability to prepare a new quantum system in a state which is close to the original state $\ket{\psi_i}$. Then the optimal fidelity which by definition is related to the complexity of purification is defined as
\begin{gather}
\mathbb{F}_{\text{opt}} (\mathcal{S})= \sup_{M \in \text{ALL,G} } \mathbb{F} (\mathcal{S} | \mathbf{M}, \mathbf{G}).
\end{gather}

Here, the optimization would be over all quantum measurements and all guessing strategies.  In the dual bulk, this means one needs to probe ``all" the entanglement wedge cross section, which leads to our region $D$ and the complexity of purification CoP.

Then the LOCC protocol $\mathcal{P}$, comes into play for defining the \textit{optimal local fidelity}, $\mathbb{F}_{\text{local} (\mathcal{S}) } $ as
\begin{gather}
\mathbb{F} (\mathcal{S})= \sup_{\mathbf{P} \in \text{LOCC} } \mathbb{F} (\mathcal{S} | \mathbf{P}) \le \mathbb{F}_{\text{opt}} (\mathcal{S}),
\end{gather}
where the allowed measurements belong to the LOCC class and the optimization is over ``all LOCC protocols". This is a sub-class of average and optimal fidelity and the dual bulk region would be a sub-region of entanglement wedge. Note also that computing the local fidelity would be sufficient for determining how well a given set of states could be locally distinguished using a resource state \cite{PhysRevA.97.022314}.

It worths to note that the connection between local and optimal fidelity would be
\begin{gather}
\mathbb{F} (\psi \otimes \mathcal{S} \big | \mathbf{P}) = p \mathbb{F}_{\text{opt}} (\mathcal{S})+(1-p) \mathbb{F}_{\text{local} } (\mathcal{S}) > \mathbb{F}_{\text{local}} (\mathcal{S})
\end{gather}
where $p$ is the probability of being successful in conversion of any bipartite pure state $\ket{\Psi}$ of Schmidt rank $r$ to maximally entangled state $\ket{\Psi '}$ of Schmidt rank $d_1 \le r$ by LOCC.
If we consider the fidelity/subregion duality \cite{Alishahiha:2015rta}, then the left hand side would be dual to the whole bulk volume, while $\mathbb{F}_{\text{opt}}$ would be dual to CoP and $\mathbb{F}_{\text{local} }$ to some other local subregion, probably region $A$ and $B$ in figure \ref{fig:CoPregion2}. 

Other fidelities, such as maximal achievable fidelity \cite{PhysRevLett.90.097901} which is connected to the robustness of entanglement, fidelity of recovery \cite{PhysRevA.92.042321} which is related to conditional quantum mutual information and geometric squashed entanglement, fidelity of teleportation \cite{Ming:2010Li} which is related to the noise, entanglement fidelity \cite{PhysRevA.74.050302} which is related to how well a channel preserves entanglement, etc, have been defined in the literature where their exact dual holographic bulk volumes would be different. The dualities between any subregion in the bulk and  different fidelities would deserve further investigations.

\begin{figure}[ht!]
 \centering
  \includegraphics[width=5.9cm] {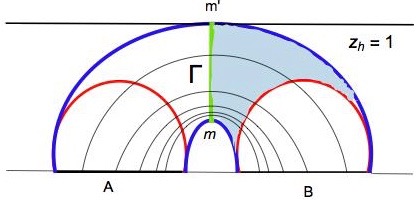} \hspace{1cm}
    \includegraphics[width=5.57cm] {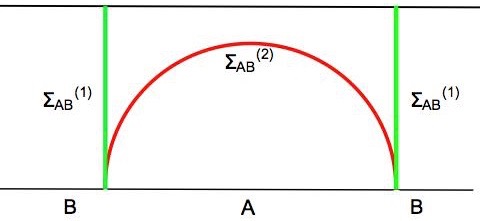}
  \caption{In the left figure, entanglement wedge cross section which captures the quantum correlations and also the bit thread interpretation of complexity of purification for Bell states are shown. In the right figure the two bottlenecks for a strip $A$ are shown which capture the classical correlations.}
 \label{fig:Bitthread2}
\end{figure}

In our bulk setup, for two infinite strips with the same width, in each strip where the gates are located, arbitrary operations could be performed locally. Then, they classically could communicate through a confined region of the bulk, as shown in the left figure of \ref{fig:Bitthread2}, where we called it the complexity of purification. Using this picture, one could intuitionally deduce how any factor that ``locally" decreases the ease of local operations or classical communications through the region, could affect the correlations between the mixed states and therefore EoP and CoP, as comparing to entanglement entropy or complexity.

As we have observed in our diagrams, using this picture, one could expect that the momentum dissipations, through introducing the mass term $m$, or decreasing correlations by adding the same charge on the two sides of the boundary would decrease greatly both EoP and CoP, more than entanglement entropy and complexity.  Also, breaking down the classical communications in some way, for instance by increasing the distance $D$ between the strips, could also decrease or even make the EoP or CoP to vanish faster as we have observed. It also worths to mention here that EoP and CoP would be an entanglement monotones in all cases \cite{PhysRevA.99.022338}. 

\begin{figure}[ht!]
 \centering
  \includegraphics[width=5cm] {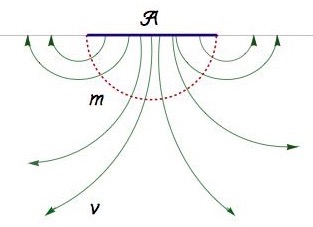}
  \caption{Max Flow-Min Cut theorem pointing out the maximizing flow through the bottleneck gives RT surface and also the entanglement of purification.}
 \label{fig:max2}
\end{figure}

Another method to visualize and study the correlations between the two strips is using the bit thread formalism introduced in \cite{Headrick:2017ucz}. In \cite{Freedman:2016zud}, the authors reformulated entanglement entropy in terms of the flux of some divergenceless vector fields $\vec{v}$ satisfying $\nabla.{\vec{v}}=0$ and $| {\vec{v}}| \le 1$. So as shown in figure \ref{fig:max2}, the entanglement entropy could be written as the maximum of this flux passing through region $\mathcal{A}$ as
\begin{gather}
\mathcal{S} (\mathcal{A}) =\max_{\vec{v}} \int_{\mathcal{A}} \vec{v}
 \ge \int_\mathcal{A} \vec{v}.
\end{gather}

It has been proved in \cite{Headrick:2017ucz} that the ``max flow-min cut" theorem would be equivalent to the Ryu-Takayanagi (RT) prescription, as the bottleneck for the flow is equal to the minimized surface.  

Therefore, we conjecture that both EoP and CoP could be written in terms of these flows as well. In fact, just recently, after our work, in \cite{Du:2019emy} the bit thread formalism for EoP has been worked out. Note that, EoP could be interpreted as the number of threads that pass through the minimal wedge cross section, the surface $\Gamma$ in any construction. For instance consider figure \ref{fig:Bitthread2}. If a thread starts from the region $A$ and one would have the condition that it definitely should end on the region $B$, then it has to pass through the surface $\Gamma$. Therefore, EoP could be written in terms of the number of such threads.

Similar to the entanglement entropy and EoP, for the complexity and CoP one could find a bit thread formalism by considering a small volume for each thread, for instance by modeling them as tubes with a radius of Planck length.  Then complexity or CoP could be calculated through the volumes of such threads. This picture of bit thread led us to propose the best volume one could consider for CoP is the one shown in figure \ref{fig:strips}.  

\begin{figure}[ht!]
 \centering
  \includegraphics[width=7.9cm] {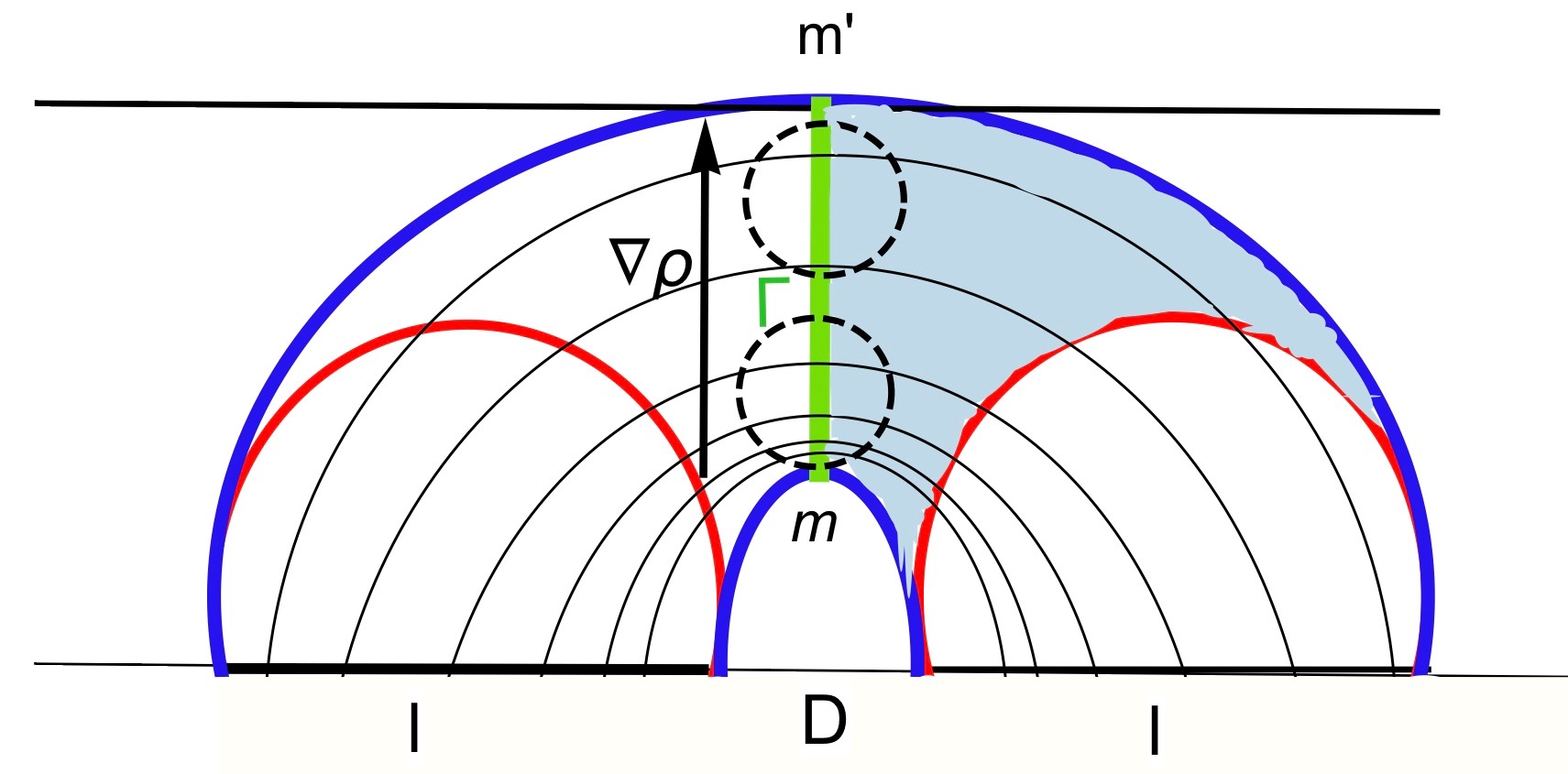}
  \caption{For the symmetric case, the density of the flux of bit threads at each point could be approximated by a linear gradient function and  therefore could be considered to be $\rho(z)= \rho_{2\ell+D}(1-\frac{z}{\Gamma})+\rho_D \frac{z}{\Gamma} $. Then using the flux, the length of the thread at each point and infinitesimal volume for the bit thread, the CoP could be calculated using the bit thread formalism.}
 \label{fig:Bitflux}
\end{figure}

In addition, noting the left part of figure \ref{fig:Bitthread2} where the bit threads for the two strips are shown, we propose the arrangement of bit threads, for the Bell states, considering their densities at each point is actually the pattern which were shown there. This pattern is formed since the density of flow is higher around the point $m$ and it decreases until reaching to the point $m'$, which could also be deduced from figure \ref{fig:Bitflux}. So the gradient of the flux is a decreasing function along $\Gamma$, moving deep into the bulk and for instance could be modeled by a linear function as $\rho(z) = \rho_{2\ell+D}-\left (\frac{\rho_{2\ell+D} -\rho_{D} }{\Gamma}  \right) z$.  This could also be explained by the fact that the gates which are closer together have much stronger correlations among themselves, creating a denser flows of bit threads there. Note also that the density of bit threads at each point is at most $1/4 G_N \hbar$ and also in total we have the relation $E_W \sim c$ \cite{Brown:1986uua}, where $c$ is the central charge of the theory. Also, all these arguments could be applied in any dimension.

In the tensor network construction, considering the structure of bit threads in figure \ref{fig:Bitflux},  the network in the lower part of $\Gamma$ would then be much denser.  Subsequently, this  could be considered in studying the complexity of purification. For instance in each region of space one could define a density of computation or complexity of purification. So the density of CoP around the point $m$ would have a bigger value than the corresponding ones around the point $m'$.  Also, note that all of these statements are compatible with surface/state correspondence \cite{Miyaji:2015yva}.

Another important observation was the ``jumping" of the Ryu-Takayanagi surfaces under the continuous variations of the region on the boundary and also the sudden drop of the mutual information to zero, which represented a second order phase transition. These phase transitions would point out to the fact that the qubits actually don't sit on the RT surface and they behave completely in a non-local way \cite{Fu:2018kcp}. As suggested in \cite{Freedman:2016zud,Cui:2018dyq,Headrick:2014cta}, using the bit thread picture one could visualize this non-locality, and then this picture could be used to interpret the behavior of EoP and CoP as shown in figure \ref{fig:Bitthread2}. Note that actually one could remove these phase transitions by considering the whole quantum effects in the definition of mutual information, EoP and CoP.

There is also another interpretation of entanglement of purification mentioned in \cite{Agon:2018lwq}.  In the right part of figure \ref{fig:Bitthread2}, a single interval $A$ with its complement $B$ is shown. As one could see there are two bottlenecks, the two lines $\Sigma_{AB}^{(1)}$, and also the RT surface $\Sigma_{AB}^{(2)}$ which measures the entanglement of entropy. The entanglement of purification would be the minimum between these two bottlenecks for each width of strip $l$. For larger $l$ the disconnected one would be the solution. Using the bit thread interpretation the authors of \cite{Agon:2018lwq} concluded that the threads that pass through $\Sigma_{AB}^{(1)}$ are related to the maximum number of Bell pairs which one can distill from $\hat{\rho}_{AB}$ using only local operations and classical communication (LOCC) and then it would be the \textit{maximum} amount of entanglement entropy which is present in $S(A)$ whose source is ``purely quantum mechanical".   

Now, recall our new definition, volume of interval (VI) introduced in section \ref{newBV}. This quantity measures a same volume confined between the entanglement wedge cross sections. Therefore, this functional measures the complexity of correlations whose sources are mainly quantum mechanical and therefore some entirely ``quantum" behavior such as quantum locking effect could only be observed using that measure.

On the other hand, there would be some threads that go into the horizon where the authors of \cite{Agon:2018lwq} interpreted them as the \textit{minimum} amount of correlations which are present in $S(A)$ that are ``thermal or classical". The complexity of purification of these threads and sources could only be obtained through our definition of CoP introduced in section \ref{cop31}.

\section{Discussion}\label{purediscussion}

The aim of this paper was to use entanglement of purification and the conjecture between EoP and the minimal wedge cross section, $E_P=E_W$, to find a new definition for the complexity of purification (CoP), and finding the connections between them. For doing that we specifically considered a setup of two strips in the same side of the boundary similar to \cite{Yang:2018gfq}. We first generalized their studies to various temperatures and scrutinized their results considering various factors. 

Then, from the intuition we got from the behavior of EoP, we defined two different measures for the complexity of mixed states. In the first definition we used the volume of a subregion in the bulk and its connection with the minimal wedge cross section and using ``complexity=volume" conjecture, we defined the complexity of purification (CoP) and studied its behavior for different dimensions, temperatures, width of strips and the distance between the strips. Next, we defined another functional that we called ``interval volume (VI)" which measures the complexity from the mainly quantum mechanical sources where we observed a quantum locking effect in its behavior.

Then, to gain further understanding of the behavior of CoP, we considered it in various more general examples such as massive BTZ, charged black holes and also n-partite solutions. We noted that the mass parameter would decrease both EoP and CoP and it has a higher effect on CoP as we have expected. This could be explained from both operational studies and bit thread picture. The mass term is equivalent to introducing the momentum dissipations on the boundary which could reduce the freedom in the movement of bit threads and therefore decreases EoP and CoP. Also it can decrease the local interactions between the quantum gates in the boundary which could break the correlations.

Similarly, for the charged case, when there are same charges in the boundary, EoP and CoP would also decrease as one could expect. However, in this case the effect would be higher on EoP than on CoP.

Also, for the case of $d=2$ we observed that CoP would be a constant $2\pi$ as in this case the complexity is topological. Introducing the mass parameter $m$ or charge however changes this fact as they can add additional degrees of freedom in the bulk.

 For the multi-partite case, and for $d=2$, CoP is a factor of $2\pi$ which depends on the number of parts and also number of genuses. If the distance between the two strips become bigger than a critical distance $D_c$, in any case, the mutual information and as the result EoP and CoP becomes zero.

We then gave further interpretations and intuitions for our results using resource theory such as LOCC and also bit thread pictures in the final part. 

There are various other points worth to study. For example, it would be interesting to compare the complexity of purification for various purifications with the ``path-integral complexity" similar to the work of \cite{Caputa:2018xuf, Brown:2018bms} and get further concrete evidences with the results from holographic correspondence. For doing this, the first steps recently have been taken in \cite{Caputa:2018xuf}. 

Note that the holographic purification is inspired mainly from the surface/state correspondence \cite{Miyaji:2015yva}. It would be interesting to understand complexity of purification further from this duality as well. 

Also, note that the surface/state correspondence was based on the conjectured relationship between tensor network and AdS/CFT \cite{PhysRevLett.115.200401, Swingle:2009bg}. One then could understand how tensor network renormalization would be related to different volumes in the bulk and specifically to the complexity of purification.

In \cite{Chen:2018mcc}, the holographic complexity under a thermal quench using CV conjecture in the Vaidya-AdS has been worked out. They have studied the effect of quench speed, the strip size, the mass of black holes and the dimension of spacetimes on the evolution of complexity. Then, later in \cite{Ling:2018xpc}, they have studied this quench process for the case of charged Einstein-Born-Infeld theory, again using CV proposal. There, they have also studied the effect of charge on the patterns of evolution. We now could use this setup and instead implement complexity of purification as a probe for studying the quench system which will be the aim of our up-coming work \cite{MG:2019KM}. 

One could also find the relationships between the minimal wedge cross section $\Gamma$ and the complexity of purification through the integral of foliation of different bulk extremal surfaces and further quantify their connections. 

It would also be interesting to study the relationships between entanglement wedge and complexity of purification from the point of view of CFTs using ideas such as modular Hamiltonian and Berry connections similar to \cite{Czech:2017zfq}. One would like to see if a flow could be constructed along $\Sigma$ in region $D$ using modular Hamiltonian and the edge modes, similar to the work in \cite{Czech:2017zfq} or in the neighborhood of RT surface of region $A$ and $B$ \cite{MG:2019MGG}. 

The relationship between complexity of purification and the change of Noether charge and also space of entanglement wedges could also be studied.

 In \cite{Balasubramanian:2018hsu}, the binding complexity (the complexity of connectedness) has been introduced. One could also try to define CoP using such ideas.
 
  Also, the same studies could be done for other geometries such as confining backgrounds \cite{Ghodrati:2015rta}, or Lifshitz or Hyperscaling backgrounds \cite{Ghodrati:2014spa} and then study the effect of various characteristics of each of these geometries on EoP and CoP, \cite{Ghodrati:2014spa,Zhou:2019jlh}.

\section*{Acknowledgement}
MG would like to thank Bartek Czech for useful discussions and Tsinghua University for the warm hospitality during her visit. XMK is supported by the Natural Science Foundation of China (grant No.11705161) and Natural Science Foundation of Jiangsu Province (grant No.BK20170481). C.-Y. Zhang is supported by National Postdoctoral Program for Innovative Talents BX201600005 and Project funded by China Postdoctoral Science Foundation.

 \medskip

\bibliography{complexityPurification}
\bibliographystyle{JHEP}
\end{document}